\documentclass[a4paper,twoside]{report}
\usepackage{amsmath}
\usepackage{amssymb}
\usepackage{amsthm}
\usepackage{bbm}
\usepackage{graphicx}
\usepackage[counterclockwise]{rotating}   % for sideways table

\newcommand{\heth}{${^3}$He}
\newcommand{\hefo}{${^4}$He}
\newcommand{\lisi}{${^6}$Li}
\newcommand{\lise}{${^7}$Li}
\newcommand{\bese}{${^7}$Be}
\newcommand{\beei}{${^8}$Be}
\newcommand{\ctwelve}{${^{12}}$C}
\newcommand{\define}{:=}
\newcommand{\mgut}{M_{GUT}}
\newcommand{\lqcd}{\Lambda_{QCD}}
\newcommand{\mplanck}{M_{\rm P}}
\newcommand{\mplanckred}{\bar{M}_{\rm P}}
\newcommand{\piZero}{$\pi^0$}
\newcommand{\pipm}{$\pi^{\pm}$}
\newcommand{\mev}{\mbox{ MeV}}
\newcommand{\gev}{\mbox{ GeV}}

\newcommand{\mproton}{m_p}
\newcommand{\mnucleon}{m_N}
\newcommand{\mneutron}{m_n}
\newcommand{\alphagut}{\alpha_X}
\newcommand{\alphastrong}{\alpha_S}
\newcommand{\gnewton}{G_{\rm N}}
\newcommand{\dotgnewton}{\dot{G}_{\rm N}}

\newcommand{\ee}{\begin{equation}}
\newcommand{\eee}{\end{equation}}
\newcommand{\beq}{\begin{equation}}
\newcommand{\eeq}{\end{equation}}
\newcommand{\bea}{\begin{eqnarray}}
\newcommand{\eea}{\end{eqnarray}}
\newcommand{\vev}[1]{\langle #1\rangle}
\newcommand{\vp}{\varphi}

\newcommand{\etc}{\textit{etc.}}
\newcommand{\etal}{\textit{et al.\ }}
\newcommand{\ie}{\textit{i.e.\ }}
\newcommand{\eg}{\textit{e.g.\ }}
\newcommand{\chap}{Chap.\ }
\newcommand{\sect}{Sec.\ }
\newcommand{\eqn}{Eq.\ }
\newcommand{\eqns}{Eqs.\ }
\newcommand{\fig}{Fig.\ }
\newcommand{\tab}{Tab.\ }
\newcommand{\BigBangNucleosynthesis}{Big Bang Nucleosynthesis}
\newcommand{\OmegaBaryon}{\Omega_b}

\newtheorem{assumption}{Assumption}

\begin{document}

\pagestyle{headings}

%%%%%%%%%%%%%%%%%%%%%%%%%%%%%%%%%%%%%%%%%%%%%%%%%%%%%%%%%%%%%%%%%%%%%%%%%%%%%
%%%%%%%%%%%%%%%%%%%%%%%%%%%%%      TITLEPAGE      %%%%%%%%%%%%%%%%%%%%%%%%%%%
%%%%%%%%%%%%%%%%%%%%%%%%%%%%%%%%%%%%%%%%%%%%%%%%%%%%%%%%%%%%%%%%%%%%%%%%%%%%%

\begin{titlepage}
\begin{center}

\Huge \bfseries Dissertation\\
\vspace{2cm}
\normalfont
\normalsize submitted to the\\
\hfill\\
\large Combined Faculties for the Natural Sciences and for Mathematics\\
of the Ruperto-Carola University of Heidelberg, Germany\\
\hfill\\
\normalsize for the degree of \\
\hfill\\
\large Doctor of Natural Sciences
\vspace{10cm}

\normalsize Put forward by\\
\vspace{0.2cm}
\large \bfseries Dipl.-Phys. Steffen Stern\\
\normalsize \normalfont
\vspace{0.2cm}
\normalsize
born in Giessen, Germany\\
\vfill
Oral examination: November 19, 2008

\end{center}
\end{titlepage}

%%%%%%%%%%%%%%%%%%%%% empty page %%%%%%%%%%%%%%%%%%%%%%%%%%%%%%%%%
\newpage
\thispagestyle{empty}
\setcounter{page}{0}
\vspace*{\fill}

\newpage

%%%%%%%%%%%%%%%%%%%%%%%%%%%%%%%%%%%%%%%%%%%%%%%%%%%%%%%%%%%%%%%%%%
\begin{center}
\pagenumbering{roman}

\thispagestyle{empty}
\setcounter{page}{1}
\parbox{13cm}{
\linespread{3}
\center \Huge \bfseries Dynamical dark energy and variation of fundamental ``constants''
\normalsize
\normalfont
}
\vfill
\begin{tabbing}
Referees: \hspace{0.5cm} \=Prof. Dr. Christof Wetterich\\
\>Prof. Dr. Carlo Ewerz
\end{tabbing}

\end{center}
\newpage

%%%%%%%%%%%%%%%%%%%%%%%%%%%%%%%%%%%%%%%%%%%%%%%%%%%%%%%%%%%%%%%%%%%%%%%%%%%%%
%%%%%%%%%%%%%%%%%%%%%%%%%%%%%      ABSTRACT      %%%%%%%%%%%%%%%%%%%%%%%%%%%%
%%%%%%%%%%%%%%%%%%%%%%%%%%%%%%%%%%%%%%%%%%%%%%%%%%%%%%%%%%%%%%%%%%%%%%%%%%%%%

\thispagestyle{plain}
\noindent \textbf{\Large Dynamische Dunkle Energie und Variation fundamentaler ``Konstanten''}\\
\ \\
\noindent
Diese Arbeit besch\"aftigt sich mit den Auswirkungen m\"oglicher Variationen fundamentaler ``Konstanten'' auf den Prozess der primordialen Nukleosynthese (BBN). Die gewonnenen Ergebnisse  zur Nukleosynthese werden mit Untersuchungen zu variierenden Konstanten in anderen physikalischen Prozessen kombiniert, um Modelle der gro\ss en Vereinheitlichung (GUT) und Quintessence zu \"uberpr\"ufen. Unsere Untersuchungen ergeben, dass das \lise-Problem der Nukleosynthese stark gemildert werden kann, sofern man Variationen von Konstanten zul\"asst, wobei insbesondere eine Variation der leichten Quarkmassen einen starken Einfluss hat. Weiterhin finden wir, dass aktuelle Messungen zu variablen Konstanten im Rahmen von sechs exemplarischen GUT Modellen nicht miteinander und mit BBN in Einklang gebracht werden k\"onnen, sofern eine monotone zeitliche Variation angenommen wird. Wir folgern, dass aktuelle Messungen nichtverschwindender Variationen in starkem Widerspruch zueinander stehen und entweder selbst revidiert werden m\"ussen, oder in der Natur erheblich komplexere GUT-Zusammenh\"ange (und/oder nicht-monotone Variationen) vorliegen. Die im Rahmen dieser Dissertation vorgestellten Methoden erweisen sich hierbei als m\"achtige Werkzeuge, um per Experiment unzug\"angliche Bereiche weit jenseits des Standardmodells der Teilchenphysik bzw.~des concordance Modells der Kosmologie auf ihre intrinsische Konsistenz sowie auch Vereinbarkeit miteinander zu \"uberpr\"ufen, sofern einmal erste unumst\"o\ss liche Beweise f\"ur Variationen von Naturkonstanten vorliegen sollten.

\vfill

\noindent \textbf{\Large Dynamical dark energy and variation of fundamental ``constants''}\\
\ \\
\noindent
In this thesis we study the influence of a possible variation of fundamental ``constants'' on the process of \BigBangNucleosynthesis\ (BBN). Our findings are combined with further studies on variations of constants in other physical processes to constrain models of grand unification (GUT) and quintessence. We will find that the \lise\ problem of BBN can be ameliorated if one allows for varying constants, where especially varying light quark masses show a strong influence. Furthermore, we show that recent studies of varying constants are in contradiction with each other and BBN in the framework of six exemplary GUT scenarios, if one assumes monotonic variation with time. We conclude that there is strong tension between recent claims of varying constants, hence either some claims have to be revised, or there are much more sophisticated GUT relations (and/or non-monotonic variations) realized in nature. The methods introduced in this thesis prove to be powerful tools to probe regimes well beyond the Standard Model of particle physics or the concordance model of cosmology, which are currently inaccessible by experiments. Once the first irrefutable proofs of varying constants are available, our method will allow for probing the consistency of models beyond the standard theories like GUT or quintessence and also the compatibility between these models.

%%%%%%%%%%%%%%%%%%%%%%%%%%%% CONTENTS %%%%%%%%%%%%%%%%%%%%%%%%%%%%

\tableofcontents

\pagebreak

%%%%%%%%%%%%%%%%%%%%%%%%%%%%%%%%%%%%%%%%%%%%%%%%%%%%%%%%%%%%%%%%%%

%%%%%%%%%%%%%%%%%%%%% empty page %%%%%%%%%%%%%%%%%%%%%%%%%%%%%%%%%
\newpage
\thispagestyle{empty}
\setcounter{page}{0}
\vspace*{\fill}

\newpage

%%%%%%%%%%%%%%%%%%%%%%%%%%%%%%%%%%%%%%%%%%%%%%%%%%%%%%%%%%%%%%%%%%%%%%%%%%%%%
%%%%%%%%%%%%%%%%%%%%%%%%%%%%%    Part 1     %%%%%%%%%%%%%%%%%%%%%%%%%%%
%%%%%%%%%%%%%%%%%%%%%%%%%%%%%%%%%%%%%%%%%%%%%%%%%%%%%%%%%%%%%%%%%%%%%%%%%%%%%

\pagenumbering{arabic}
\setcounter{page}{1}
\part{Introduction and prerequisites}

%%%%%%%%%%%%%%%%%%%%%%%%%%%%%%%%%%%%%%%%%%%%%%%%%%%%%%%%%%%%%%%%%%%%%%%%%%%%%
%%%%%%%%%%%%%%%%%%%%%%%%%%%%%    INTRODUCTION     %%%%%%%%%%%%%%%%%%%%%%%%%%%
%%%%%%%%%%%%%%%%%%%%%%%%%%%%%%%%%%%%%%%%%%%%%%%%%%%%%%%%%%%%%%%%%%%%%%%%%%%%%

\chapter{Introduction}

\section*{The constants of nature}

Since the time of Newton, the constancy of the fundamental laws of nature has been undoubted. Comparing and reproducing experiments have been at the root of the scientific approach: A physical experiment which we perform today will have the same outcome as the same experiment performed tomorrow\footnote{Neglecting experiments which incorporate probabilities, for instance quantum mechanical effects.}. Neglecting local gravitational effects, it should also not matter where we perform the experiment. Hence, it has been unquestionable for a long time that the laws of nature are constant over space and time. Moreover, Einstein formulated this space- and time independence of physics in his strong equivalence principle, making it an essential part of his theory of general relativity.

Today's view of this question is somewhat different, at least from theoretical aspects. Even though compelling evidence for changes in the laws of physics has up to now not been found, we have to admit that we are still lacking a profound test of this constancy. In the past, the laws of physics have only been thoroughly tested on time and length scales accessible by mankind, \ie on timescales of years and on length scales that do not go beyond the size of our solar system\footnote{Note that general relativity has furthermore not been tested on length scales smaller than about 1mm.}. Only recently astrophysics and cosmology have opened a door to test physics on immensely broader scales, reaching out to unimaginable length scales of several gigaparsecs and going back in time to the very beginning of our Universe. 

This thesis will deal with probes of possible variations of constants throughout the whole accessible history of the Universe. In a first part, we will study one of the most distant (in time and space) events where physics can be applied and tested, primordial nucleosynthesis. It is the process during which the light elements of our Universe were formed and which happened when our Universe was only one minute old, extremely hot and dense. If physics was really subject to variations, primordial nucleosynthesis is a prime candidate for any studies of this kind.
In a second step the obtained results will be combined together with further tests of varying constants at later times to derive a ``history of variations''. Finally, we will show how these results can be used to test models beyond standard physics which currently cannot be accessed directly by experiments.

\section*{Outline}

In this thesis I will work out the influence of varying ``constants'' on the process of primordial nucleosynthesis and implied constraints to theories beyond standard physics. In the next chapter, I will give a short introduction to variations of physical constants, some historical remarks and theoretical motivations. Chapter \ref{Chap:Cosmology} will lay the theoretical framework for our understanding of the Universe as a whole, explaining general relativity and the main laws of cosmology. In Chapter \ref{chap:SUSYandGUT} I will introduce the concepts of supersymmetry and grand unified theories (GUTs) which are widely accepted as extensions of the Standard Model of particle physics. Chapter \ref{chap:Quintessence} will introduce quintessence models which can yield variations of constants. 

Part II will focus on the details of one particular process in the history of our Universe, \BigBangNucleosynthesis\ (BBN). Chapter \ref{chap:BBN} will explain the standard process of BBN and the physics behind. Chapter \ref{chap:BBNvaryingNucl} will introduce the possibility of varying constants in the process of BBN, and Chapter \ref{chap:BBNvaryingFund} will demonstrate how one can relate the results to variations of the Standard Model parameters. Finally, in Chapter \ref{chap:BBNobsVStheo}, the observed element abundances will be used to derive constraints on variations of fundamental parameters.

In Part III I will study relevant tests of varying constants from the Big Bang until today. Chapter \ref{chap:ExpTestofVariations} gives an overview over tests of varying constants, and in Chapter \ref{chap:VariationsFromBBNtoTodayinGUT} I will combine these tests within six different GUT models, showing how variations of constants can in principle be used to probe models of grand unification. Using the six GUT models, Chapter \ref{chap:ProbingQuintessence} shows how models of quintessence can be probed under the assumption of grand unification.

Finally, in Chapter \ref{chap:Conclusion}, I will sum up the findings of this thesis and give some final conclusions and outlook.
\\
\\
\\
The work on this thesis has led to four main publications:

\begin{itemize}
\item Michael Doran, Steffen Stern, Eduard Thommes,\\
      Baryon Acoustic Oscillations and Dynamical Dark Energy,\\
      JCAP 0704:015 (2007) \cite{DST06} (not in focus of this thesis)
\item Thomas Dent, Steffen Stern, Christof Wetterich,\\
      Primordial nucleosynthesis as a probe of fundamental physics parameters,\\
      Phys.~Rev.~D 76, 063513 (2007) \cite{DSW07}
\item Thomas Dent, Steffen Stern, Christof Wetterich,\\
      Unifying cosmological and recent time variations of fundamental couplings,\\
      Phys.~Rev.~D 78, 103518 (2008) \cite{DSW08.1}
\item Thomas Dent, Steffen Stern, Christof Wetterich,\\
      Time variation of fundamental couplings and dynamical dark energy,\\
      Preprint arXiv:0809.4628, accepted by JCAP \cite{DSW08.2}
\end{itemize}

\chapter{Variation of ``constants''}

\section{The laws of physics and the constants of nature}
The fundamental laws of physics, represented by the Standard Model of particle physics and Einstein's theory of general relativity, consist of two parts. One part is the mathematical form of the laws (\eg the $1/r$ behavior of Newton's theory of gravity), the other part is the actual strength of the interactions relative to each other. Whilst the first part, the mathematical form of the laws of nature, can be derived from considerations of fundamental symmetries of nature\footnote{For example, the Standard Model of particle physics is obtained when demanding a local $SU(3) \times SU(2) \times U(1)$ symmetry. See \sect \ref{sec:StandardModel}.}, the second part has to be put into the theories ``by hand'' in form of about 27 - from a theory standpoint a priori absolutely arbitrary - numerical values, the constants of nature. Tab.~\ref{Tab:CstOfNature} gives a list of these fundamental constants\footnote{As will be explained in \sect \ref{sec:VariationDimensionfulParams}, only ratios of masses are measurable fundamental parameters. Hence, in fact one can get rid of one the mass terms in Tab.~\ref{Tab:CstOfNature}, for instance by defining all masses with respect to the Planck mass. This would reduce the number of fundamental parameters by one.} for the Standard Model of particle physics and general relativity\footnote{In cosmology some more free parameters turn up which have to be determined by observations, for instance those describing the composition of our Universe. However, it is assumed that these parameters can in principle be obtained from some fundamental laws of physics once the processes in the very early stage of the Universe are better understood.}. Note that the list of fundamental parameters gets much larger when going to theories beyond the Standard Model, \eg supersymmetry (see \sect \ref{sec:Supersymmetry}). Up to now it is unclear where these constants come from and if they are ``real'' constants in the sense that their numerical values are fixed once and for all.

\begin{table}
\center
\begin{tabular}{c|c}
 & Number of \\
Type of constant & parameters\\
\hline
\hline
3 coupling constants & 3  \\
masses of 6 quarks & 6  \\
CKM matrix (3 angles + 1 complex phase) & 4  \\
masses of 3 leptons & 3  \\
Higgs mechanism & 2  \\
strong CP phase & 1  \\
masses of 3 neutrinos & 3  \\
PMNS mixing matrix for neutrinos & 4  \\
\hline
gravitational constant & 1\\
\hline
\hline
in summa & 27
\end{tabular}
\caption[The fundamental constants of nature]{The fundamental constants of nature.} \label{Tab:CstOfNature}
\end{table}

\section{The question of constancy}

The question if the constants of nature are actually constant was probably first raised by Dirac \cite{Dirac37,Dirac38,Dirac79}. In his ``large numbers hypothesis'', he argues that very large (or small) dimensionless constants must not enter in basic laws of physics. Based on his numerological principle, he suggests that very large numbers rather characterize the state of the Universe, specifically the time which has passed since the Big Bang. For instance, he finds that the age of the Universe in atomic time, $H_0 e^2 / m_e c^2 \approx 2 \times 10^{-41}$, is of the same order of magnitude as the ratio of electrostatic to gravitational force between proton and electron, $\gnewton \mproton m_e / \frac{e^2}{4\pi \epsilon_0} \approx 4 \times 10^{-40}$. Consequently, he suggests that also the latter quantity should vary with cosmic time. Attributing the variation to the gravitational sector, the intensity of all gravitational effects would then decrease with a rate of about $10^{-10}$ y$^{-1}$. It was quickly found that this would lead to astrophysical effects \cite{Chandrasekhar37} which could not be detected in the following time. Hence, Dirac's theory was finally abandoned, but the discussion on varying constants had started\footnote{See for instance \cite{Uzan02} for a more complete review of the history of varying constant theories.}.

In 1961, Brans and Dicke \cite{BransDicke61} used Mach's principle\footnote{There are different formulations of Mach's principle. In Brans' and Dicke's argument it states \cite{Brans05} that the gravitational constant should be a function of the mass distribution in the universe.} to derive what we now call a ``scalar-tensor theory''. In their model, the gravitational constant is replaced by a scalar field which can vary in space and time. Besides others, models of this kind are still being considered as theoretical arguments for variation of constants.

\section{Theoretical arguments for variation of constants}
\label{sec:TheoretArgumentsForVariation}
In high-energy theories such as string theory, which unifies gravity with the Standard Model of particle physics, our low-energy laws of physics appear as an effective theory whose parameters are set dynamically by vacuum expectation values which break the ``higher'' symmetry. In particular string theory offers a plethora of possibilities to introduce variations of constants, for instance due to the fact that it is formulated with 10 (or 11) spacetime dimensions which need to be compactified in order to arrive at the 4 spacetime dimensions of the Standard Model (we will give some more details in \sect \ref{sec:Stringtheory}). Similar considerations also apply to other theories with extra dimensions, for instance the possibility of varying constants in Kaluza-Klein theories has been studied in \cite{Marciano83}. Hence, both temporal and spatial variations of constants are from a theoretical standpoint well founded, even though those high-energy theories mostly do not give any hint on the actual size of the variations.

Also, ``low-energy'' theories, for instance theories which extend the concordance model of cosmology by introducing a cosmological scalar field, allow variations of constants. In this thesis we will concentrate on theories of coupled quintessence in which constants can depend on cosmic time and the environment.

This thesis will examine the possibility of variations of constants from today back to the time of \BigBangNucleosynthesis\ (BBN). During BBN, the composition of the Universe was quite different from today's composition (concerning temperature and pressure). Hence, composition dependent effects which might cause spatial variations today might have caused variations at BBN time. However, since the Universe was almost homogeneous at BBN, these variations can effectively be treated as a time-dependent effect. This thesis will not evoke the question of space-dependence of constants but treat possible variations at BBN as purely temporal effects.

\section{Equivalence principles and possible violations}
\label{sec:ViolationOfEquivalencePrinciple}
Particle theory is based on Poincar\'{e} covariance. In quantum field theory (QFT), we demand that each of the fundamental fields is a representation of the Poincar\'{e} group. Hence, amongst others, invariance under spacetime translations is automatically built in. However, we can still implement spacetime variations by introducing additional dynamical fields, whose values are determined by the fields' own actions and their couplings to the rest of the theory. While the theory as a whole remains Poincar\'{e} invariant, variations in measurable quantities can still arise if the solution for the additional fields has a nontrivial spacetime dependence.

This discussion can be extended to general relativity (GR), which is also based on symmetry principles that are apparently violated by variations of constants. In particular, GR is based on the strong equivalence principle, which can be decomposed into the following symmetries
\begin{itemize}
\item Weak equivalence principle: The trajectory of a freely falling test body only depends on its initial position and velocity and is independent of its composition.
\item Local Lorentz invariance: The outcomes of any experiments (whether gravitationally or not) in a laboratory moving in an inertial frame of reference are independent of the velocity of the laboratory.
\item Local position invariance: Outcomes of experiments (whether gravitationally or not) do not depend on their position in space and time.
\end{itemize}
A space or time variation of fundamental constants obviously violates local position invariance. Also, as the gradient of any varying fundamental parameter defines a direction in spacetime, local Lorentz invariance is violated. Finally, it has been shown (see \eg \cite{Nordtvedt02}) that any space-time variation of fundamental constants will necessarily lead to an additional gravitational force, hence also the weak equivalence principle will be violated\footnote{We will work out the relation between violation of the weak equivalence principle and variation of constants in \sect \ref{sec:WEP}.}. As probes of general relativity so far do not find any violation of the theory, we can immediately conclude that variations of constants must be extremely tiny. Note, however, that GR has only been tested on relatively small time scales and also only on length scales from 1mm to the size of our solar system.

\newpage

\section{Variation of dimensionful parameters}
\label{sec:VariationDimensionfulParams}
When measuring or estimating possible variations of constants, one always has to keep in mind that the variation of any dimensionful quantity is not physically well-defined, as one always has to specify how the dimension (\eg the unit [Energy]) is defined. In general, a dimensionful quantity can only be measured by comparison with another dimensionful quantity, so in fact only dimensionless ratios are measurable. For example, measurements of variations of the electron mass $m_e$ are only well-defined if one states how the mass unit is defined, for instance by choosing a system of units where the mass scale is kept fixed. Popular system of units are the ``Einstein frame'' where the Planck mass $\mplanck$ is kept constant, or the ``Jordan frame'' where some particular particle physics scale is kept fixed. Considering the electron mass in the Einstein frame, the actually measured varying quantity (without system of units ambiguities) is then rather $m_e / \mplanck$.

In the part of this thesis which deals with \BigBangNucleosynthesis, we use a system of units where the QCD invariant scale $\lqcd$ is kept constant. This is convenient for dealing with nuclear reactions, where the energy scales are mainly determined by the strong interaction. Thus the variations of dimensionful parameters include implicitly some power of $\lqcd$. For example, if we take the electron mass $m_e$ as a varying parameter we are implicitly considering a variation of $m_e/\lqcd$. In the last part of this thesis we will work with theories of grand unification. There, the grand unified scale $\mgut$ enters as natural scale which we choose to be constant. The appropriate conversion from a constant $\lqcd$ to a constant $\mgut$ system of units is explained in \sect \ref{sec:ConversionUnits}.

\subsection{The chiral limit}

Many studies on varying constants work with the chiral limit, \ie they assume that all quarks are massless \cite{Epelbaum02,BeaneSavage02,Donoghue06}. Then all dimensionful QCD parameters are simply proportional to a power of $\lqcd$, which ameliorates their treatment considerably. For instance, QCD masses like the proton mass simply scale like
\beq
\Delta \ln \mproton = \Delta \ln \lqcd
\eeq
and any other dimensionful QCD parameter according to its mass scale (for instance, cross sections with $[\sigma]$ = [Energy]$^{-2}$ scale like $\Delta \ln \sigma = -2 \Delta \ln \lqcd$).  Switching on the quark masses, one obtains a finite range for pion-mediated interactions, which may greatly affect static and dynamical properties of nuclei. Also, the masses of all hadrons are affected at some order in chiral perturbation theory \cite{Gasser82}. 
In this thesis we will work with the full quark contributions, which are for most QCD parameters known at least in first order chiral perturbation theory, \ie to terms linear in the quark masses.

\section{Probes of varying constants}
A possible variation of constants can be tested in various ways. Common tests are laboratory based measurements, for instance of atomic transitions. Also, a multitude of astrophysical and cosmological effects can be studied under the question of constancy, which allow to probe physics over a timescale unreachable with laboratory measurements. In recent years probes of variations in the constants of nature have been performed with increasingly high accuracy. Whilst direct laboratory measurements do not point towards any variation, some astrophysical tests yield slight variations. In part III (\sect \ref{chap:ExpTestofVariations}) we will list all recent relevant probes of varying constants, followed by detailed studies on how one can combine the different outcomes in unified scenarios. BBN as a probe of varying constants will be examined in detail in part II of this thesis.

\section[Fine-tuning of constants and the anthropic principle]{Fine-tuning of constants and the anthropic\\ \mbox{principle}}

Connected to the question of constancy of fundamental constants is the question of fine-tuning of these constants. Even though this question can be seen as a rather philosophical one, we will shortly comment on it.

As far as we know today, the value of most of the 27 fundamental constants is extremely fine-tuned in order to allow life to appear. It has been argued \cite{Tegmark97} that even small deviations (less than or order of 1\%) will make the appearance of any life impossible. For example, if the strong force was slightly weaker, multi-proton nuclei would not be stable, and if it was slightly stronger, hydrogen could fuse into helium-2. Similar arguments can be found for the electromagnetic and weak force and for many other natural constants.

This fine-tuning problem can be ameliorated, like all problems of this kind, by evoking the anthropic principle\footnote{The concept of the anthropic principle was systematically introduced by Brandon Carter in a contribution to a symposium honoring Copernicus' 500th birthday in 1973 \cite{Carter74}, even though the idea of the anthropic principle has already been used long before.}. In short, this principle states that the Universe which we observe has to be capable to develop intelligent life like us. Otherwise we would not be here and could not ask the question why the Universe has exactly the laws of nature which it has. The final outcome is that the question why we are living in such a highly fine-tuned, \ie extremely improbable, universe has simply disappeared, because the actual probability we have to discuss is rather the probability under the condition of our existence, which is no longer vanishingly small.

In recent years scientists have come up with the idea of ``multiverses'', stating that universes with many different kinds of physical properties are constantly formed \cite{Linde86}. This is supported by candidate theories of everything (like string theory) which ab initio do not seem to have hard constraints which would exclusively select our physics. Rather, they allow an extremely high number of different physical configurations. In the framework of those theories, universes with many different physical configurations bubble out constantly, and the anthropic principle states that our universe is the one of these many universes which allowed us to appear.

These considerations are not directly connected to the investigations which are subject of this thesis, except the fact that varying constants would lead to an even more fine-tuned universe: Not only the values of the constants today, but also their whole time evolution would need to be tuned such that we could appear. We will not comment on the point of fine-tuning in the following, but it has become clear that the problems we are tackling have some deeper connection to philosophy and the question of why we are actually here.

%%%%%%%%%%%%%%%%%%%%%%%%%%%%%%%%%%%%%%%%%%%%%%%%%%%%%%%%%%%%%%%%%%%%%%%%%%%%%
%%%%%%%%%%%%%%%%%%%%%%%%%%%%%%%  Cosmology  %%%%%%%%%%%%%%%%%%%%%%%%%%%%%%%%%
%%%%%%%%%%%%%%%%%%%%%%%%%%%%%%%%%%%%%%%%%%%%%%%%%%%%%%%%%%%%%%%%%%%%%%%%%%%%%

\chapter{Cosmology}
\label{Chap:Cosmology}

In this thesis we will consider probes for varying constants from today back to the first minute after the Big Bang. Hence it is essential to understand the evolution of our Universe from the Big Bang until today. This chapter gives a short review of our current picture of the Universe, its history and present status, and the important equations that govern its evolution.

\section{General relativity and the basics of cosmology}
\label{SectionGR}

\subsection{General relativity}

General relativity is an extension of the theory of special relativity, which states that gravity is a purely geometric effect, generated by the curvature of spacetime. The relation between curvature and stress-energy is given by the Einstein field equations
\beq
\label{eqnGR}
R_{\mu \nu} - \frac{R}{2}g_{\mu \nu} = \frac{8 \pi \gnewton}{c^4} T_{\mu \nu}\, ,
\eeq
where $R_{\mu \nu}$ is the Ricci tensor, $R$ the Ricci scalar, $g_{\mu \nu}$ the metric tensor and $T_{\mu \nu}$ the stress-energy tensor. Equation \eqref{eqnGR} is a complicated differential equation which can in general only be solved if one makes simplifying assumptions and/or uses numeric techniques.

\subsection{The basics of cosmology}
\label{sec:BasicsOfCosmology}
In cosmology one is interested in the evolution of the Universe as a whole. Thus, one usually confines oneself to physics on large scales which allows to make some simplifying assumptions that dramatically reduce the complexity of equation \eqref{eqnGR}.

\begin{assumption}\label{cosmol_assumption}
The main assumption of cosmology is that the Universe is homogeneous and isotropic on large scales.
\end{assumption}

Of course, the existence of objects like the earth, sun \etc\ contradicts this assumption locally. However, if one averages over distances ($> 1000$ Mpc), it turns out that Assumption \ref{cosmol_assumption} is observationally well-justified\footnote{The biggest known structure is the Sloan great wall which is 1.37 billion lightyears long.}. Demanding all quantities to be homogeneous and isotropic, one can show \cite{WeinbergGRT} that the metric takes the simple form\footnote{There are theories claiming that the averaged Einstein tensor $G_{\mu \nu} = R_{\mu \nu} - \frac{R}{2}g_{\mu \nu}$ which enters in \eqn \eqref{eqnGR} is not equivalent to the Einstein tensor derived from an averaged metric as given in \eqn \eqref{eqn:FRWmetric}. Since the actual outcome of these considerations is still unclear, we will not consider those theories in this thesis. See \cite{Buchert07} for a recent review.}
\beq \label{eqn:FRWmetric}
ds^2 = dt^2 - a(t)^2 \left( \frac{dr^2}{1-kr^2} + r^2 d \Theta^2 + r^2 \sin^2 \Theta d \phi^2 \right) \, ,
\eeq
where k describes the curvature and $a(t)$ is the scale parameter, related to the redshift $z$ via
\beq
a = \frac{1}{1+z}  \, .
\eeq
The metric \eqref{eqn:FRWmetric} is called Friedmann-Robertson-Walker metric (FRW metric). The scale parameter fulfills the Friedmann equations
\beq \label{eqn:Friedmann1}
H^2 \define \left(\frac{\dot{a}}{a}\right)^2 = \frac{8\pi \gnewton}{3} \rho 
\eeq
\beq \label{eqn:Friedmann2}
3 \frac{\ddot{a}}{a} = -4\pi \gnewton \left(\rho + \frac{3p}{c^2} \right) \, , 
\eeq
where $H$ is the Hubble constant and $\rho$ and $p$ denote the total energy and pressure density. These two densities are usually split up into the different components which are assumed to be present in today's Universe, baryonic and dark matter, dark energy (denoted with the symbol $\Lambda$), photons, neutrinos and curvature\footnote{In the very early Universe, also electrons will make a substantial contribution to the expansion rate. This applies to the epoch of BBN and will be explained in more detail in chapter \ref{chap:BBN}.},
\beq
\rho = \rho_{B} + \rho_{DM} + \rho_{\Lambda} + \rho_{\gamma} + \rho_{\nu} + \rho_K \, . 
\eeq
The pressure is related to the energy via an equation of state,
\beq
p_i = w_i \rho_i \; ,
\eeq
where the equation-of-state parameter $w_i$ depends on the composition of the components as shown in Table~\ref{tab:EqnOfStateParams}.
\begin{table}
\center
\begin{tabular}{|c|c|}
\hline
Composition & $w$ \\
\hline
non-relativistic matter & 0  \\
ultra-relativistic matter (radiation) & 1/3  \\
curvature & -1/3  \\
cosmological constant & -1  \\
\hline
\end{tabular}
\caption{Equation-of-state parameters for different types of cosmological components}\label{tab:EqnOfStateParams}
\end{table}

With the critical density defined as
\begin{equation}\label{eqn:DefRhoCrit}
\rho_C \define \frac{3H^2}{8\pi \gnewton} \; , 
\end{equation}
all densities are usually given as fractional densities
\begin{equation}\label{eqn:DefOmega}
\Omega_i \define \frac{\rho_i}{\rho_C}\, . 
\end{equation}
Note that equation \eqref{eqn:Friedmann1} yields
\begin{equation}
\Omega_{B} + \Omega_{DM} + \Omega_{\Lambda} + \Omega_{\gamma} + \Omega_{\nu} + \Omega_K \equiv 1 
\end{equation}
at all times. 

\begin{figure}
\begin{center}
\includegraphics[width=11cm]{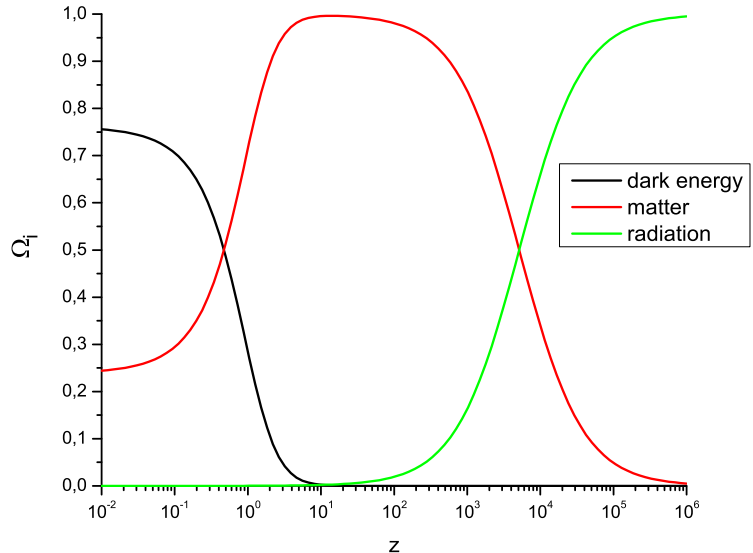}
\end{center}
\caption{Evolution of the components in a $\Lambda$CDM Universe}\label{Fig:evolutionOfComponents}
\end{figure}

In the course of the evolution of the Universe, the energy densities scale like
\beq \label{eqn:ScalingBehaviour}
\rho \propto a^{-3(1+w)} \, ,
\eeq
which means that the values of $\Omega_i$ do not stay constant over time since we have different $w$ for different kinds of energy densities (Tab.~\ref{tab:EqnOfStateParams}). As baryons and dark matter follow the same equation of state, one can combine these to the matter energy density 
\begin{equation}
\Omega_M \define \Omega_{DM} + \Omega_{B} \, .
\end{equation}
Given today's values $\Omega_i^0$, one can combine \eqns \eqref{eqn:Friedmann1}, \eqref{eqn:DefRhoCrit}, \eqref{eqn:DefOmega} and \eqref{eqn:ScalingBehaviour} and use Tab.~\ref{tab:EqnOfStateParams} to derive the time evolution of the Hubble constant,
\beq
H^2(a) = H_0^2 \left[ \Omega_{\gamma}^0 a^{-4} + \Omega_M^0 a^{-3} + \Omega_K^0 a^{-2} + \Omega_{\Lambda}^0 \right] \, , 
\label{eqn:HubbleEvolution}
\eeq
where we have neglected the neutrinos which have no substantial contribution to today's content of the Universe\footnote{Further note that due to the tiny but non-vanishing mass of the neutrinos, the neutrino equation of state might change during the evolution of the universe.} (see Tab.~\ref{Tab:CosmologicalParams}). \eqn \eqref{eqn:HubbleEvolution} shows that at early times ($a \ll 1$) non-relativistic and relativistic matter become dominant and any cosmological constant component irrelevant, whilst at late times ($a \gg 1$) $\Omega_{\Lambda}$ dominates. The flow of cosmological components in the $\Lambda$CDM concordance model (see \sect \ref{sec:concordanceModel}) is depicted in Fig.~\ref{Fig:evolutionOfComponents}, where the time evolution of the fractional components is given by
\beq
\Omega_i = \frac{ \Omega_i^0 a^{-3(1+w)}}{\Omega_{\gamma}^0 a^{-4} + \Omega_M^0 a^{-3} + \Omega_K^0 a^{-2} + \Omega_{\Lambda}^0} \, .
\eeq
As can be seen in Fig.~\ref{Fig:evolutionOfComponents}, today's Universe ($z=0$) is dominated by dark energy ($\Omega_{\Lambda}$) but did undergo 2 transitions, from radiation dominated to matter dominated and from matter to dark energy dominated:

\label{PhaseTransitions}
\begin{itemize}
\item In the early Universe, the expansion was almost completely due to relativistic particles $\Rightarrow$ radiation-dominated era.
\item At $z \approx 5000$, about 70,000 years after the Big Bang, we have matter-radiation-equality and the Universe becomes matter dominated.
\item At $z \approx 0.4$, about $4.3$ Gyrs ago, the Universe becomes dominated by dark energy (in a $\Lambda$CDM model).
\end{itemize}

\section[The concordance model]{The concordance model: Our current picture of the Universe}
\label{sec:concordanceModel}

\subsection{Historical development}

Presumably, the question of where we come from and where we will go is as old as mankind. As a first modern physical approach to questions of origin, evolution and fate of the Universe, one usually considers Einstein's paper ``Cosmological Considerations in the General Theory of Relativity'' from 1917 \cite{Einstein17}. One might say that high-precision observational cosmology started with the Hubble space mission in 1990. It was followed by further astrophysical and cosmological investigations, and basically all of these (mainly observational) tests point towards a coherent picture of our Universe, which is called the \textit{``concordance model''}.

\subsection{Our current picture of the Universe}

\begin{figure}[th]
\begin{center}
\includegraphics[width=13cm]{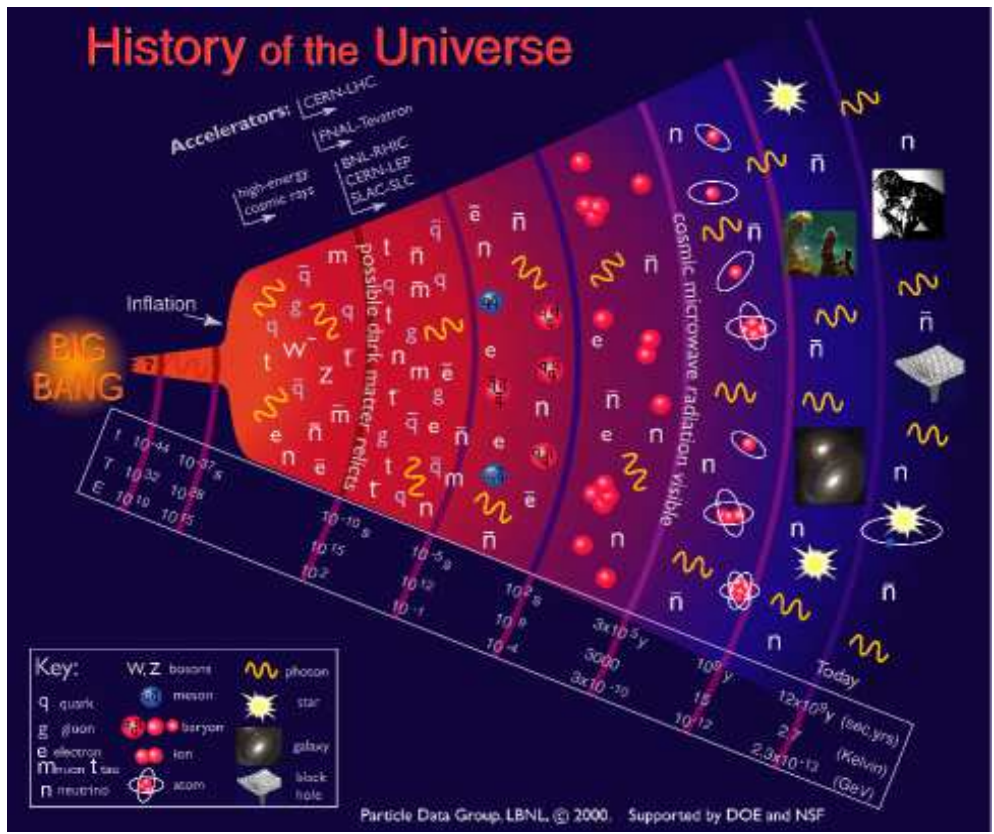}
\end{center}
\caption{History of our Universe from Particle Data Group 2000}\label{Fig:historyUniverse}
\end{figure}

According to the concordance model, the Universe started in a Big Bang\footnote{Even though it is hoped that physics will once be able to explain the actual origin of this singular event, one is lacking an accepted theory of quantum gravitation which would allow to go beyond the time of the Planck epoch.} and has been expanding since then. All observational evidence points towards a so-called $\Lambda$CDM cosmology, stating that the Universe is geometrically flat ($\Omega_K \equiv 0$) and consists besides known baryonic matter, leptons and photons of an unknown ``dark matter'' component which has the property of non-relativistic, only gravitationally interacting heavy particles, and a ``dark energy'' component, in the simplest version described by a cosmological constant $\Lambda$. See for instance \cite{Bartelmann06} for a compilation of the major observational evidences for the Big Bang and \cite{WMAP5} for recent parameter determinations including all major probes of the concordance model.

As can be seen in Fig.~\ref{Fig:evolutionOfComponents}, we live in a dark-energy dominated universe just now but have undergone both radiation- and matter-dominated phases. The question why dark energy is taking over ``just now'' is unclear; this problem is called the ``coincidence problem'' or ``why now problem'' (see \sect \ref{sec:ProblemsWithCosmologicalConstant}).

Looking back to the very beginning of our Universe, the concordance model suggests the following history of our Universe which is depicted in \fig \ref{Fig:historyUniverse}. Here, $t$ is the age of the Universe, \ie the time after the Big Bang, and $T$ is the temperature of the Universe, defined by the photon temperature\footnote{See Sec.~\ref{sec:BBNbackgroundEqns} for details on the photon temperature.}.

\begin{itemize}
\item $t \approx 10^{-43}$ seconds ($T \approx 10^{19}$ GeV): \\
      Planck epoch, needs to be described by a quantum theory of gravity.
\item $t \approx 10^{-35}$ seconds ($T \approx 10^{15}$ GeV):\\
      Inflation (exponentially fast expansion of the Universe), \\
      Baryogenesis (production of matter-antimatter asymmetry) 
\item $10^{-6}$ seconds $\le t \le 10^{-2}$ seconds ($T \approx 0.1$ GeV): \\
      Quark-hadron transition: protons and neutrons form
\item $1\, {\rm second} \le t \le 3\, {\rm minutes}$, ($T \approx 1$ MeV, $z \approx 10^{10}$): \\
      Nucleosynthesis: light elements (D, He, Li) form 
\item $t \approx 70,000$ years ($T \approx 1$ eV, $z \approx 5000$): \\
      Beginning of matter dominated era
\item $t \approx 300,000$ years ($T \approx 0.25$ eV, $z \approx 1100$): \\
      Recombination, the cosmic microwave background (CMB) forms
\item After that:\\
      Galaxy/star formation.
\end{itemize}

Experimental tests of high-energy physics currently do not go beyond the TeV region. The processes which happened during the Planck epoch, inflation and also baryogenesis go beyond the physics of the Standard Model and are hence subject to a lot of speculation. Also the quark-hadron transition is hard to describe since it involves QCD at low energies, in the regime where QCD effects cannot be evaluated perturbatively. Hence, the oldest cosmological events which can be reasonably described quantitatively with known physics are primordial nucleosynthesis and CMB formation. 

\section{Cosmological parameter values}

\begin{figure}[th]
\begin{center}
\includegraphics[width=6cm]{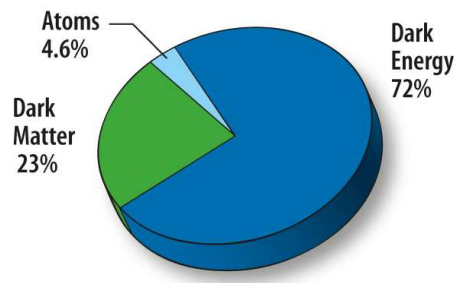}
\end{center}
\caption[Today's energy content of our Universe]{Today's energy content of our Universe. From NASA / WMAP Science Team}\label{fig:contentUniverse}
\end{figure}

In recent years, important and quite extensive missions have been undertaken to deepen our understanding of cosmological relations. In particular, WMAP, SDSS and Supernovae Ia \cite{WMAP5, HinshawWMAP5}(see also references therein) have yielded a coherent set of cosmological parameters of a precision which had been inconceivable 10 years ago. However, compared to the Standard Model of particle physics, the concordance model of cosmology is rather new and by far less tested. The set of cosmological parameters for a $\Lambda$CDM cosmology is given in Tab.~\ref{Tab:CosmologicalParams}. It turns out that only $4.6\%$ of our Universe is made of ``known'' ordinary baryonic matter, the rest of the Universe is dark matter and dark energy. The energy composition of our Universe today is shown in Fig.~\ref{fig:contentUniverse}.

Using the evolution equations of \sect \ref{SectionGR}, the present-day values of cosmological parameters allow to deduce the content of our Universe in the past and also in the future\footnote{Cosmological models allow to extrapolate cosmology into the future, however models are not tested sufficiently to allow a definite prediction of what the ultimate fate of the Universe will be.}. As observations also allow us to look back in time, the picture for the past is nowadays quite clear and observationally probed. The composition history of our Universe in a $\Lambda$CDM model is shown in Fig.~\ref{Fig:evolutionOfComponents}.

{\renewcommand{\baselinestretch}{1.3}
\begin{table}[h!]
\begin{tabular}{l|l|l}
Quantity & Symbol & Value\\
\hline
Hubble expansion rate & $H_0$ & $100\, h\, \mbox{km/s Mpc}^{-1}$ \\
normalized Hubble expansion rate & $h$ & $0.701 \pm 0.013$ \\
\hline
baryon density & $\OmegaBaryon \equiv \rho_b/\rho_c$ & $0.0462 \pm 0.0015 $ \\
dark matter density & $\Omega_{dm} \equiv \rho_{dm}/\rho_c$ & $0.233 \pm 0.013 $ \\
matter density & $\Omega_m \equiv \OmegaBaryon + \Omega_{dm}$ & $0.279 \pm 0.013 $ \\
dark energy density & $\Omega_{\Lambda} \equiv \rho_{\Lambda}/\rho_c$ & $0.721 \pm 0.015$ \\
radiation density & $\Omega_{\gamma} \equiv \rho_{\gamma}/\rho_c$ & $(5.0 \pm 0.2)\cdot 10^{-5}$ \\
neutrino density & $\Omega_{\nu} \equiv \rho_{\nu}/\rho_c$ & $<0.013 \; (95\% \, \mbox{CL}) $ \\
\hline
baryon to photon ratio & $\eta \equiv n_b/n_{\gamma}$ & $(6.21 \pm 0.16) \cdot 10^{-10}$ \\
CMB temperature & $T$ & $2.725$ K 
\end{tabular}
\caption[Parameters describing our Universe]{Parameters describing our Universe. WMAP ``recommended parameter values'' \cite{WMAP5, HinshawWMAP5} from WMAP5, BAO and SN for a $\Lambda$CDM cosmology}\label{Tab:CosmologicalParams}
\end{table}
}

\chapter{The Standard Model and beyond}
\label{chap:SUSYandGUT}

\section{The Standard Model of particle physics}
\label{sec:StandardModel}
The Standard Model (SM) of particle physics describes the elementary particles and three of the four fundamental interactions, the strong, weak and electromagnetic interaction. This section will only give a rough overview over some particular aspects of the Standard Model which will be of relevance later. For a more comprehensive introduction, see for instance \cite{HalzenMartin} or any other textbook on modern particle physics.\\
\\
The list of the fundamental particles of the Standard Model comprises 
\begin{itemize}
\item six leptons ($e, \mu, \tau, \nu_e, \nu_{\mu}, \nu_{\tau}$),
\item six quarks ($u,d,s,c,b,t$), 
\item the gauge bosons as mediators of the fundamental interactions,
\item a Higgs boson.
\end{itemize}
The matter particles enter as pointlike massless fermions, and the interactions are introduced by demanding a local $SU(3) \times SU(2) \times U(1)$ gauge symmetry. Here, the group $SU(3)$ is responsible for quantum chromodynamics (QCD), with 8 massless gauge bosons called \textit{gluons} as mediators of the strong force. At small momenta, the strong coupling constant becomes large (see Sec.~\ref{section:RunningOfCouplings}), which is thought to be the explanation for \textit{confinement}, \ie the fact that only color-neutral particles are observed in nature.

The theory of electroweak interaction goes back to the seminal work of Glashow, Salam and Weinberg \cite{Weinberg67, Salam69, Glashow70} (Nobel Prize 1979). It describes the electroweak interaction by a $SU(2) \times U(1)$ gauge symmetry which is broken by the Higgs mechanism into the weak interaction (with massive gauge bosons $W^{\pm}$ and $Z^0$) and the electromagnetic sector with the photon as massless mediator of the electromagnetic force. In particular, the $SU(2) \times U(1)$ gauge symmetry implies four massless gauge bosons, written as $W_{\mu}^{\pm}$, $W_{\mu}^3$ and $B_{\mu}$. Additionally, one introduces a scalar Higgs field $\phi$ (as a weak doublet under $SU(2)$ which has altogether 4 real components) and gives it a potential $V(\phi)$ which results in a vacuum that is not symmetric under the $SU(2) \times U(1)$ gauge symmetry. This leads to a spontaneous symmetry breakdown of the electroweak symmetry, and the Higgs field obtains a nonzero vacuum expectation value $\vev{\phi} \approx 246 \gev$. One of the four Higgs components becomes a massive scalar particle, which is the only particle of the SM which has not yet been observed. The $W_{\mu}^{\pm}$ and a combination of $W_{\mu}^3$ and $B_{\mu}$ obtain masses proportional to $\vev{\phi}$ and become the massive mediators of the weak force, $W^{\pm}$ and $Z^0$, where the three remaining components of the Higgs form the longitudinal modes of the $W^{\pm}$ and $Z^0$. The coupling constant $\alpha_{em}$ of the remaining electromagnetic symmetry group $U(1)_{em}$ can be obtained from the coupling constants of the original $SU(2) \times U(1)$ coupling constants $\alpha_1$, $\alpha_2$ (see \eg \cite{HalzenMartin}),
\beq
\alpha_{em}^{-1} = \alpha_1^{-1} + \alpha_2^{-1} \; . 
\label{eqn:ElectroweakCouplingRelation}
\eeq
Also, the SM fermions obtain masses via the Higgs mechanism, their mass is a product of Higgs v.e.v.\ and a Yukawa coupling $h_i$, for instance for the electron
\beq
m_e = h_e \vev{\phi} \; .
\eeq

\section{Running of couplings}
\label{section:RunningOfCouplings}
The influence of fluctuations with different momenta leads to scale dependent coupling constants. See for instance \cite{Wilson71,Wegner72,Wilson73} or any good textbook on quantum field theory for details of this process. Generally, physical systems at slightly different scales are described by the similar laws of physics, with slightly changed parameters. In quantum field theory, this behavior is described by the famous beta function, which describes the behavior of the coupling parameter $g$ under slight changes of the energy scale $\mu$,
\beq \label{eqn:RG}
\mu \frac{ \partial g}{\partial \mu} = \beta(g) \; .
\eeq
Using the coupling constant $\alpha \define \frac{g^2}{4\pi}$ instead, one can also define a $\beta$ function for $\alpha$,
\beq \label{eqn:RG2}
\mu \frac{ \partial \alpha}{\partial \mu} = \beta(\alpha) \, . 
\eeq

The mathematical apparatus to investigate these changes of physical systems under scale transformations is called the renormalization group (RG). In quantum field theory, the renormalization group equation \eqref{eqn:RG} can only be computed perturbatively as the exact RG equation would in principle include an infinite order of loop corrections. For our purpose the first-order (one-loop) RG equations are sufficient, and these are known for all three interactions of the Standard Model. In particular, the beta function for QED (with only photons and electrons present) at first order is given by
\beq
\label{eqn:BetaEm}
\beta_{em}(\alpha_{em}) = \frac{2 \alpha_{em}^2}{3\pi} \; , 
\eeq
which is solved by
\beq \label{eqn:RunningOfAlphaEm}
\alpha_{em}(\mu) = \frac{ \alpha_{em}(\mu_0)}{1 - \frac{2 \alpha_{em}(\mu_0)}{3\pi} \ln \left( \frac{\mu}{\mu_0} \right)} \, . 
\eeq
The fine structure constant $\alpha$ is defined in the limit of zero momentum transfer, \ie for $\mu \le m_e$. For QCD, the beta function is
\beq
 \label{eqn:BetaStrong}
\beta_{strong}(\alphastrong) = -\left(11 - \frac{2 n_f}{3} \right) \frac{\alphastrong^2}{2\pi} \; , 
\eeq
solved by
\beq \label{eqn:RunningOfAlphaStrong}
\alphastrong(\mu) = \frac{ \alphastrong(\mu_0)}{1 + \frac{\alphastrong(\mu_0)}{6\pi} (33 - 2n_f) \ln \left( \frac{\mu}{\mu_0} \right)} \, . 
\eeq
Here $n_f$ is the number of quark flavors present, \ie the number of quarks with mass $m_q \le \mu$. As $n_f \le 6$ in the Standard Model, the beta function $\beta_{strong}$ is negative.

\begin{figure}
\begin{center}
\includegraphics[width=6.2cm]{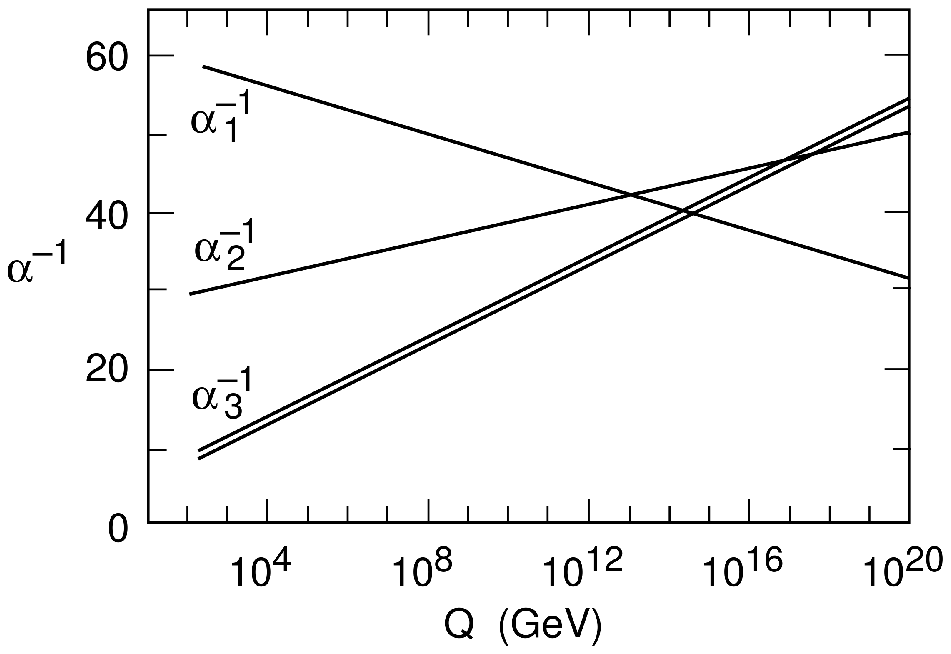}
\includegraphics[width=6.2cm]{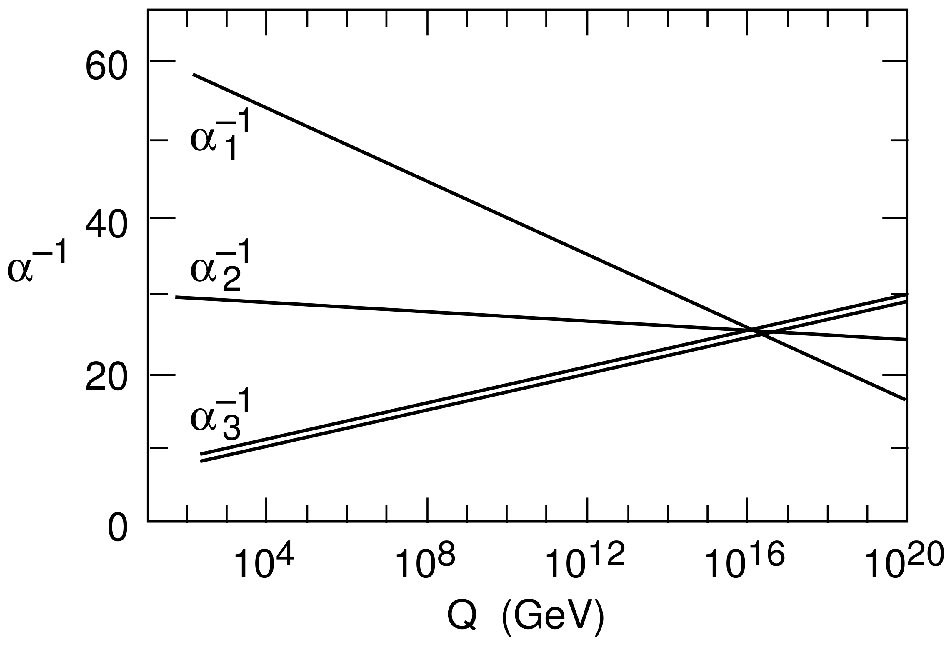}
\end{center}
\caption[Running of coupling constants in the Standard Model and in SUSY]{Running of coupling constants of the three gauge groups $SU(3) \times SU(2) \times U(1)$ in the Standard Model (left) and in SUSY (right). $\alpha_1$ is scaled by a factor $\frac{5}{3}$, see \eqn \eqref{eqn:GUTCouplingRelation}. From \cite{Peskin97}.} \label{fig:RunningCouplings}
\end{figure}

The running of coupling constants is shown in Fig.~\ref{fig:RunningCouplings}.
As the beta function for QCD is negative, the QCD coupling diverges when going to low energies. This effect, which was found by Wilczek, Politzer and Gross (Nobel price 2004), is thought to be the reason for confinement. As opposed to the electroweak theory, QCD thus has an intrinsic energy scale induced by the RG equation, the scale where $\alphastrong$ becomes formally infinite. Choosing $m_s < \mu_0 < m_c$ such that $n_f$ remains constant ($n_f \equiv 3$), this happens when the denominator in equation \eqref{eqn:RunningOfAlphaStrong} becomes zero, \ie 
\beq
\frac{\alphastrong(\mu_0)}{6\pi} (33 - 2n_f) \ln \left( \frac{\mu}{\mu_0} \right) = -1 \; , 
\eeq
which happens at the \textit{QCD invariant scale} $\mu \equiv \lqcd$, defined by
\beq \label{eqn:defLambdaQCD}
\lqcd \define \mu_0 \exp \left( \frac{-6\pi}{ (33-2n_f) \alphastrong(\mu_0)} \right) \,. 
\eeq
Hence \eqn \eqref{eqn:RunningOfAlphaStrong} can be rewritten as ($\mu < m_c$)
\beq
\alphastrong(\mu) = \frac{ 6\pi} {(33 - 2n_f) \ln ( \mu / \lqcd)} \, . 
\eeq
Note that when the energy $\mu$ becomes of the order of $\lqcd$, perturbation theory breaks down, and a world of quarks and gluons becomes a world of pions, protons and so on. This is revealed in the fact that $\lqcd$ is of the order of light meson masses \cite{Berger06}, 
\beq
\lqcd \approx 200 \mev \;. 
\eeq

The beta functions given in \eqns \eqref{eqn:BetaEm} and \eqref{eqn:BetaStrong} are simplified versions. In the first case, \eqn \eqref{eqn:BetaEm} only holds for one particle present, and in the second case one has to note that $n_f$ is not constant at all energies $\mu$. In the full functions, particles only contribute when energies are above the particle's threshold energies, which are typically the corresponding particle masses. Hence every particle contributes one threshold term to the renormalization group equation. We will give the full renormalization group equations, including extra terms coming from additional supersymmetric particles (see \sect \ref{sec:Supersymmetry}) in \sect \ref{sec:GUT}.

\section{The necessity of a ``theory beyond''}

From the point of a theorist, the established Standard Model of particle physics cannot be the end of the story. One can definitely say that at latest at the Planck energy scale $\mplanck \approx 10^{19} \gev$ quantum gravity effects will become important, demanding a quantized description of gravity. 
A further hint that the Standard Model of particle physics might not be the end of the story is the running of the coupling constants. They seem to meet at a energy scale of $\mgut \approx 10^{16} \gev$ as depicted in Fig.~\ref{fig:RunningCouplings}. Hence, it appears likely that electroweak and strong interaction can be unified in a grand unified theory (GUT). Within such grand unified theories, it is most likely that if any coupling constant of the Standard Model varies, all coupling constants will vary. In the later chapters, we will assume that some kind of GUT is realized, and hence the electroweak and strong coupling constants are related to each other. Further details on grand unified theories are given in \sect \ref{sec:GUT}. 

\begin{figure}
\begin{center}
\includegraphics[width=11cm]{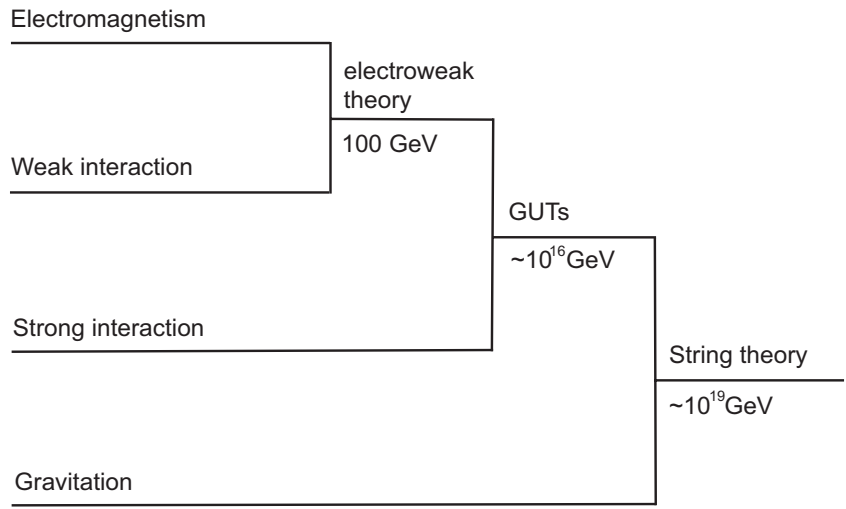}
\end{center}
\caption{Unification of the four forces in the string theory picture} \label{fig:UnificationInStringtheory}
\end{figure}

\newpage

Actually, one can even go further with the idea of unification. String theory, for instance, implements an unification of the interactions of the Standard Model and also gravity. The unification of the four forces including gravity in a string theory picture is shown in Fig.~\ref{fig:UnificationInStringtheory}. In such theories, parameters related to the mass of the SM particles (\eg the Yukawa couplings) should also derive from some sort of unified greater theory, whereas at the level of the Standard Model and also at the level of simple GUTs, there is no direct relation to the gauge coupling sector\footnote{Due to renormalization group effects, also the Yukawa couplings get contributions from coupling parameters. However, we will show in Sec.~\ref{sec:GUTrels} that these effects are small and can be neglected when studying variations of parameters.}. In this thesis, we will assume constant Yukawa couplings.

\section{Supersymmetry and the MSSM}
\label{sec:Supersymmetry}
Many high-energy theories (\eg string theory) contain supersymmetry as an essential part of the theory. Supersymmetry establishes a symmetry between bosons and fermions. Every boson gets a fermionic partner and vice versa with the same quantum numbers. For this thesis the motivation and theoretical framework of supersymmetry are not needed, so I will refrain from going into too much detail. Introductions to supersymmetry can be found in many textbooks and reviews, \eg \cite{NillesSUSY}.

Within the Standard Model of particle physics, no supersymmetric partner can be found, hence in a supersymmetric extension of the Standard Model one has to introduce additional supersymmetric partners for every single particle of the SM. These supersymmetric partners are assumed to be heavier than the current experimentally tested mass region ($\approx 100 \gev$). 

The minimal supersymmetric extension of the Standard Model is called the \textit{MSSM (minimal supersymmetric standard model)}. It contains an additional supersymmetric partner for every SM particle. Furthermore, with supersymmetry, a single Higgs doublet would result in a gauge anomaly, so a second Higgs doublet is introduced. Hence the MSSM contains 2 additional (heavy) neutral Higgs scalars and two charged Higgs scalars, supplemented by the appropriate superpartners.

The complete particle spectrum of the MSSM is given in Tab.~\ref{TabMSSM}. When working with supersymmetric theories in this thesis, we will always assume that the MSSM particle spectrum is realized.

\begin{table}
\center
\begin{tabular}{ccc|cc}
SM particle & El. charge & Spin & SUSY partner & Spin\\
\hline
quarks u,c,t & 2/3 & 1/2 & squarks $\tilde{u}, \tilde{c}, \tilde{t}$ & 0\\
quarks d,s,b & -1/3 & 1/2 & squarks $\tilde{d}, \tilde{s}, \tilde{b}$ & 0\\
charged leptons e, $\mu, \tau$ & -1 & 1/2 & sleptons & 0\\
neutrinos $\nu_e, \nu_{\mu}, \nu_{\tau}$ & 0 & 1/2 & sneutrinos ($\tilde{\nu}$) & 0\\
photon $\gamma$ & 0 & 1 & photino $\tilde{\gamma}$ & 1/2\\
$Z^{0}$ & 0 & 1 & Zino ($\tilde{Z}$) & 1/2\\
$Z^{0}$ neutral Higgs scalar & 0 & 0 & Zino Higgs & 1/2\\
$W^{\pm}$ & $\pm 1$ & 1 & Wino ($\tilde{W}$) & 1/2\\
$W^{\pm}$ charged Higgs scalar & $\pm 1$ & 0 & Wino Higgs & 1/2\\
8 gluons & 0 & 1 & 8 gluinos ($\tilde{g}$) & 1/2\\
neutral Higgs $H^0$ & 0 & 0 & higgsino ($\tilde{H}^0$) & 1/2\\
2 MSSM neutral Higgs & 0 & 0 & 2 neutral higgsinos & 1/2\\
2 MSSM charged Higgs $H^{\pm}$ & $\pm 1$ & 0 & 2 charged higgsinos $\tilde{H}^{\pm}$ & 1/2\\
\end{tabular}
\caption{The MSSM particle spectrum} \label{TabMSSM}
\end{table}

\section{Grand unification}
\label{sec:GUT}

In grand unified theories, the gauge group of the Standard Model $SU(3) \times SU(2) \times U(1)$ with coupling constants $g_S, g_2, g_1$ is unified into a bigger Lie group (\eg $SU(5)$ or $SO(10)$) with a single coupling constant $g_X$ at a certain energy scale,
\beq
\mgut \approx 10^{16} \gev \;, 
\eeq
which is assumed an independent parameter and can also vary with time. Its actual value depends on the specific form of the grand unified theory (\eg SUSY/non-SUSY). 

\newpage

It can be shown (see \eg \cite{WeinbergQFT2}) that for any unification of $SU(3) \times SU(2) \times U(1)$ with couplings $g_3 \equiv g_S$, $g_2$ and $g_1$ into a simple Lie group with coupling $g_X$, one obtains the relation
\beq
g_X^2 = g_S^2 = g_2^2 = \frac{5}{3} g_1^2 
\label{eqn:GUTCouplingRelation}
\eeq
which holds at $E=\mgut$. For the electromagnetic interaction, \eqns \eqref{eqn:ElectroweakCouplingRelation} and \eqref{eqn:GUTCouplingRelation} yield
\beq
 \label{eqn:AlphaEMinGUT}
 \alpha_{em}^{-1} = \alpha_2^{-1} + \alpha_1^{-1} = \alphagut^{-1} + \frac{5}{3} \alphagut^{-1} = \frac{8}{3} \alphagut^{-1} \; , 
\eeq
which actually only holds at $E = \mgut$ but will be of relevance when studying the running of $\alpha_{em}$ in a GUT framework.

The value of the unified coupling $\alphagut$ can roughly be estimated from the RG running as displayed in Fig.~\ref{fig:RunningCouplings}, showing that $\alphagut^{-1}$ is of the order $38-45$ in the non-SUSY case and $23-29$ in the SUSY case \cite{Amaldi91}. We take as representative values \cite{Dent03}
\begin{align}
\alphagut &= 1/40 \qquad \mbox{(non-SUSY)} \\
\alphagut &= 1/24 \qquad \mbox{(SUSY).} 
\end{align}

At lower energies, the GUT symmetry is broken and the relation \eqref{eqn:GUTCouplingRelation} does not hold any longer. The coupling constants of the SM evolve separately according to the renormalization group equations (see section \ref{section:RunningOfCouplings}). Generalizing the running of couplings as given in equations \eqref{eqn:RunningOfAlphaEm} and \eqref{eqn:RunningOfAlphaStrong} to the full SM/MSSM particle spectrum, we obtain for QCD
\beq \label{eqn:FlowStrong}
\alphastrong^{-1}(\mu) = \alphagut^{-1} - \frac{1}{2\pi} \sum_{i} b_i \ln \frac{\mu}{\mgut} - \frac{1}{2\pi} \sum_{th} b^{th} \ln \left( \frac{m^{th}}{\mgut} \right) 
\eeq
where the first sum goes over all particles $i$ with threshold mass $m^{th} < \mu$ and the second sum goes over all particles with  $m^{th} > \mu$. For the $b_i$ and $b^{th}$ the values from \tab \ref{tab:bthFactors} have to be applied.

\begin{table}
\center	
\begin{tabular}{c|c}
Type of particle & $b^{(th)}$\\
\hline
quarks & 2/3 \\
gluons & -11/8 \\
squarks & 1/3 \\
gluinos & 1/4 \\
\end{tabular}
\caption{Renormalization group coefficients for the strong interaction}
\label{tab:bthFactors}
\end{table}

\noindent The corresponding expression for the fine-structure constant $\alpha \define \alpha_{em}(m_e)$ is
\beq \label{eqn:FlowElm}
\alpha^{-1} = \frac{8}{3}\alphagut^{-1} - \frac{1}{2\pi} \sum_{th} Q^{th ^2}_{em} f^{th} \ln \left( \frac{m^{th}}{\mgut} \right) \, , 
\eeq
where the factor $\frac{8}{3}$ derives from \eqn \eqref{eqn:AlphaEMinGUT}, $Q_{em}$ denotes the electric charge and for $f^{th}$ the values given in Tab.~\ref{tab:fthFactors} are applied. An analogous equation also holds for the weak coupling, where the electric charge is replaced by the weak isospin\footnote{For the SU(2) weak interaction, the weak isospin effectively acts like a multiplicative charge factor, hence it can be treated analogously to the electric charge.}. When dealing with weak interactions in this thesis, we will only be working with terms that contain the weak coupling in terms of the Fermi constant\footnote{See for instance the weak decay of the neutron, \sect \ref{sec:NeutronLifetime}.},
\beq
G_F = \frac{\sqrt{2}}{8} \frac{g_w^2}{M_W^2} \, . 
\eeq
As $M_W = \frac{g_w \vev{\phi}}{2}$ with $\vev{\phi}$ the Higgs v.e.v., the weak coupling $g_w$ drops out and the Fermi constant can be expressed only in terms of the Higgs v.e.v.,
\beq
\label{eqn:RelationFermiConstHiggs}
G_F = \frac{1}{\sqrt{2} \vev{\phi}^2} \; .
\eeq

\begin{table}
\center	
\begin{tabular}{c|c}
Type of particle & $f^{th}$\\
\hline
chiral (or Majorana) fermion & 2/3 \\
complex scalar & 1/3 \\
vector boson & -11/3 
\end{tabular}
\caption{Renormalization group coefficients for the electromagnetic interaction} \label{tab:fthFactors}
\end{table}

\section{Variations in a GUT framework}
\label{sec:VariationsInGutFramework}
The GUT relations which were introduced in the preceding section show that within a GUT framework, the coupling constants are usually related to further fundamental parameters, in particular the GUT coupling $\alphagut$ and threshold masses. In this section we will derive the equations which relate variations in the GUT coupling constant $\alphagut$ and particle masses to variations in the SM coupling constants.

\subsection{Variation of the electromagnetic coupling}

For the MSSM particle spectrum, we obtain for the fine-structure constant from \eqn \eqref{eqn:FlowElm}
\begin{multline}\label{eqn:FlowElmMSSM}
\alpha^{-1} = \frac{8}{3}\alphagut^{-1} - \frac{1}{2\pi} \left[ \frac{4}{3} \cdot 3 \cdot \left( \left(\frac{2}{3}\right)^2 + 2 \left(\frac{1}{3}\right)^2 \right) \ln \frac{\lqcd}{\mgut} 
+ \frac{4}{3} \cdot 3 \left( \frac{2}{3} \right)^2 \ln \frac{m_c m_t}{\mgut^2} \right.\\
\left. + \frac{4}{3} \cdot 3 \left( \frac{1}{3} \right)^2 \ln \frac{m_b}{\mgut} + \frac{4}{3} \left(1\right)^2 \ln \frac{m_e m_{\mu} m_{\tau}}{\mgut^3} + \left( -\frac{11}{3} \cdot 2 + \frac{1}{3} \right) \ln \frac{M_W}{\mgut} \right.\\
\left. + \frac{2}{3} \cdot 3 \cdot (1)^2 \ln \frac{m_{\tilde{l}}}{\mgut} + \frac{2}{3} \left( 3 \cdot 3 \cdot \left( \frac{2}{3} \right)^2 + 3 \cdot 3 \cdot \left( \frac{1}{3} \right)^2 \right) \ln \frac{m_{\tilde{q}}}{\mgut} \right.\\
\left. + \frac{2}{3} \cdot 2 \cdot (1)^2  \ln \frac{m_{\tilde{W}}}{\mgut} + \frac{2}{3} \cdot 2 \cdot (1)^2  \ln \frac{m_{\tilde{H}}}{\mgut} + \frac{1}{3} \ln \frac{m_{H^{\pm}}}{\mgut} \right] 
\end{multline}
where it has been used that
\begin{itemize}
\item The light quarks $u,d,s$ decouple at $m^{th} = \lqcd$.
\item The quarks enter in 3 different colors
\item The charged leptons enter as both left- and righthanded particles (2 chiral fermions)
\item The massive gauge bosons $W^{\pm}$ have to be supplemented by a charged complex Higgs scalar (longitudinal DOFs)
\item $m_{H^{\pm}}$ is the mass of the additional charged Higgs scalars which have to be introduced in MSSM.
\end{itemize}
Taking the linear variation of \eqn \eqref{eqn:FlowElmMSSM}, we obtain for the variation of the fine structure constant, including the MSSM particles,
\begin{multline}
\label{eqn:dlnAlpha}
\frac{\Delta \ln \alpha}{\alpha} = +\frac{8}{3} \frac{\Delta \ln \alphagut}{\alphagut} + \frac{1}{2\pi} \left( \frac{8}{3}\Delta \ln \frac{\lqcd}{\mgut} + \frac{16}{9} \Delta \ln \frac{m_c m_t}{\mgut^2} + \frac{4}{9} \Delta \ln \frac{m_b}{\mgut} \right.\\
\left. + \frac{4}{3} \Delta \ln \frac{m_e m_{\mu} m_{\tau}}{\mgut^3} - \frac{21}{3} \Delta \ln \frac{M_W}{\mgut} + 8 \Delta \ln \frac{\tilde{m}}{\mgut} + \frac{1}{3} \Delta \ln \frac{m_{H^{\pm}}}{\mgut} \right) \, . 
\end{multline}
Not knowing the actual mass or mass generating mechanism for the superpartners, we assume that the mechanism is the same for all superpartners and define $\tilde{m}$ as the average superpartner mass. To obtain the corresponding relation in nonsupersymmetric models, one simply has to leave out the terms with $\tilde{m}$ and $m_{H^{\pm}}$.\footnote{Note that the RG equations \eqref{eqn:FlowElmMSSM}, \eqref{eqn:dlnAlpha} and also \eqref{eqn:FlowStrongMSSM} and \eqref{eqn:dlnLambdadlnMgut} only hold under the condition that all threshold masses are smaller than $\mgut$. Hence, the nonsupersymmetric case is \textit{not} obtained in the limit $m_{\rm susy} \rightarrow \infty$.} When later dealing with variations in supersymmetric models, we will further assume $\Delta \ln \tilde{m} = \Delta \ln m_{H^{\pm}}$, so the last two terms can be combined into one.

\newpage

\subsection{Variation of the QCD scale}

As $\alphastrong$ diverges at $\mu = \lqcd \approx 200 \mev$, we are interested in the value of $\alphastrong(\mu)$ in the regime $\lqcd < \mu < m_c$ (in this regime, $n_f = 3$). For the MSSM particle spectrum, we obtain
\beq \label{eqn:FlowStrongMSSM}
\alphastrong^{-1}(\mu) = \alphagut^{-1} + \frac{9}{2\pi} \ln \left( \frac{\mu}{\mgut} \right) - \frac{1}{2\pi} \left( \frac{2}{3} \ln \frac{m_c m_b m_t}{\mgut^3} + \frac{6}{3} \ln \frac{m_{\tilde{q}}}{\mgut} + \frac{8}{4} \ln \frac{m_{\tilde{g}}}{\mgut}\right) \; . 
\eeq
Inserting \eqn \eqref{eqn:FlowStrongMSSM} into \eqn \eqref{eqn:defLambdaQCD} ($\lqcd < \mu_0 = \mu < m_c$, $n_f=3$) yields
\beq
\frac{\lqcd}{\mgut} = e^{-2\pi/9 \alphagut} \left( \frac{m_c m_b m_t}{\mgut^3} \right)^{2/27} \left( \frac{m_{\tilde{q}} m_{\tilde{g}}}{\mgut} \right)^{2/9} 
\eeq
and the linear variation gives
\beq
\label{eqn:dlnLambdadlnMgut}
\Delta \ln \frac{\lqcd}{\mgut} = \frac{2\pi}{9 \alphagut} \Delta \ln \alphagut + \frac{2}{27} \Delta \ln \frac{m_c m_b m_t}{\mgut^3} + \frac{2}{9} \left( \Delta \ln \frac{m_{\tilde{q}}}{\mgut} + \Delta \ln \frac{m_{\tilde{g}}}{\mgut} \right) \, . 
\eeq
When later dealing with variations in supersymmetric models, we will further assume $\Delta \ln m_{\tilde{q}} = \Delta \ln m_{\tilde{g}}$, so the last two terms can be combined into one.

\subsection{Conversion of units}
\label{sec:ConversionUnits}
As has been explained in \sect \ref{sec:VariationDimensionfulParams}, we will work with two different systems of units. During the discussion of BBN processes, we choose units with $\lqcd = const.$~as the BBN energy scale is of roughly the same order of magnitude. When applying grand unified theories, $\mgut$ is the more appropriate energy scale to keep constant. However, we usually neglect the reference scale and write $\Delta \ln m_e$ instead for $\Delta \ln \frac{m_e}{\lqcd}$, for instance. The conversion to a different system is then performed by keeping track of all reference scales,
\beq
\Delta \ln \frac{m_e}{\mgut} = \Delta \ln \frac{m_e}{\lqcd} + \Delta \ln \frac{\lqcd}{\mgut} \, . 
\eeq
Obviously, the term $\Delta \ln \frac{\lqcd}{\mgut}$ naturally enters when converting from the $\lqcd=const.$ to the $\mgut=const.$ system of units, its explicit dependence on the unified coupling and particle masses is given in \eqn \eqref{eqn:dlnLambdadlnMgut}. Keep in mind, however, that the reference scale has to enter in the correct power, for instance the gravitational constant has units $[\gnewton] = {\rm [Energy]}^{-2}$, hence it enters as $\Delta \ln \gnewton \lqcd^2$.

\chapter{Models of quintessence}
\label{chap:Quintessence}

\section{Problems of the cosmological constant}
\label{sec:ProblemsWithCosmologicalConstant}
Recent observations show that roughly 75\% of the energy content of our Universe is made from dark energy (see \sect \ref{sec:concordanceModel}). However, the nature of dark energy is still far from being clear. The assumption that the cosmological constant derives from a vacuum energy density sufferes from a severe fine-tuning problem. In particular, the oberseved dark energy density evaluates to \cite{Copeland06}
\beq
 \rho_{\Lambda} = \frac{\Lambda \mplanck^2}{8\pi} \approx 10^{-47} \gev^4 \, , 
\eeq
while the vacuum energy density of particle physics which is evaluated by summing up the zero-point energies of the present quantum fields gives \cite{Copeland06}
\beq
 \rho_{vac} \approx \frac{\mplanck^4}{16 \pi^2} \approx 10^{74} \gev^4 \, . 
\eeq
Here we have chosen $\mplanck$ as a natural cut-off scale where we assume that the known quantum field theory is no longer applicable. Obviously, there is a discrepancy of the order of $10^{121}$. Assuming that the dark energy comes from a particle physics origin, one would have to introduce counter terms which have to be extremely fine-tuned. Hence this problem is called the ``finetuning'' problem. 

A further problem which is related with dark energy can be seen in Fig.~\ref{Fig:evolutionOfComponents}. In a $\Lambda$CDM model, the dark energy is only recently becoming important, and the time when the universe switched from a matter-dominated to a dark energy dominated epoch is only 4.3 billion years ago. There is no natural reason why these two presumably completely independent constituents of our Universe are of about the same order of magnitude and / or why we live in a period of time where this is the case. This problem is called the ``coincidence problem'' or ``why now problem'', and typically cosmological models with a cosmological constant (like the $\Lambda$CDM model) fail to address this issue.

\section{Basics of quintessence}
\label{sec:BasicsOfQuintessence}
Models of quintessence \cite{Wetterich88.1,RatraPeebles88} can offer an explanation to the issues mentioned in the previous section. A good review on dark energy models can be found in \cite{Copeland06}. 

In quintessence theories, one introduces a scalar field $\vp$ (called the cosmon) which is coupled to gravity and, most times, also to matter and gauge fields. A typical Lagrangian for a quintessence theory including couplings to matter and the electromagnetic gauge field looks like \cite{Wetterich02.1, Copeland06}
\beq
 {\cal L} = \mplanckred^2 R + \frac{1}{2} (\partial \vp)^2 + V(\vp) - V_{\vp m} + {\mathcal L}_{\rm em}
\eeq
with gauge field coupling
\beq
{\cal L}_{\rm em} = -\frac{1}{4} (1 + \lambda_{\rm em} \vp)^{-1} F_{\mu \nu} F^{\mu \nu} 
\eeq
and matter term
\beq
V_{\varphi m} = m_e(\vp) \bar{e}e + m_u(\vp) \bar{u}u + ... + m_{\rm dark}(\vp) \overline{\Psi_{\rm dark}} \Psi_{ \rm dark} + ...
\eeq
Here $\mplanckred$ denotes the reduced Planck mass,
\beq
 \mplanckred = \frac{1}{\sqrt{8\pi}} \mplanck = \frac{1}{\sqrt{8\pi \gnewton}} \, .
\eeq
If there are only slight changes in the cosmon field, the dependence of the mass, $m(\vp)$, can be linearized, \ie
\beq
 m(\vp) = (1+\lambda \vp) m_0  
\eeq
with some coupling $\lambda$. However, there are also models with significant changes in the masses, for instance in the models of growing neutrinos in \sect \ref{sec:GrowingNeutrinoModels}, where we apply a more advanced nonlinear expression.

In order to derive one of the main properties of quintessence, its capability of producing accelerated expansion, we can neglect couplings to matter and gauge fields and work instead with the action
\beq \label{eqn:QuintessenceAction}
{\cal S} = \int d^4 x \sqrt{-g} \left[-\frac{1}{2} (\partial \vp)^2 - V(\vp) \right] \, . 
\eeq
In the background of a flat FRW cosmology (\sect \ref{sec:BasicsOfCosmology}), and assuming that $\vp$ is homogeneous, \ie it only depends on time, a variation of the action \eqref{eqn:QuintessenceAction} with respect to $\vp$ yields the equation of motion
\beq
 \ddot{ \vp } + 3H \dot{ \vp } + \frac{ {\rm d} V}{{\rm d} \vp} = 0 \, . 
\eeq
The corresponding energy momentum tensor 
\beq
T_{\mu \nu} = \frac{-2}{\sqrt{-g}} \frac{\delta {\cal S}}{\delta g^{\mu \nu}}  
\eeq
yields the energy and pressure densities
\beq
\label{eqn:RhoDark}
\rho = - T^0_0 = \frac{1}{2} \dot{\vp}^2 + V(\vp)  
\eeq
\beq
p = T^i_i = \frac{1}{2} \dot{\vp}^2 - V(\vp) \, .  
\eeq
Then \eqns \eqref{eqn:Friedmann1} and \eqref{eqn:Friedmann2} yield the relations
\bea
 H^2 &=& \frac{8 \pi \gnewton}{3} \left[\frac{1}{2} \dot{\vp}^2 + V(\vp) \right] \\
 \frac{\ddot a}{a} &=& - \frac{8 \pi \gnewton}{3} \left[ \dot{\vp}^2 - V(\vp) \right] \; , 
\eea
showing that we get an accelerating universe when $\dot{\vp}^2 < V(\vp)$. Introducing the kinetic energy $T \define \dot{\vp}^2 / 2$, we define the equation of state parameter of quintessence,
\beq
\label{eqn:DefineWh}
w_h \define \frac{p}{\rho} = \frac{T - V}{T + V} \, .  
\eeq
In the next section we introduce two specific models of quintessence, crossover quintessence which has been introduced 20 years ago \cite{Wetterich88.1} and a very recent model, where quintessence is strongly coupled to neutrinos.

As the cosmon also couples to other fields and matter, one question one might ask is whether the cosmon evolution decouples in a local cluster with high ``cosmon charge density'' from the cosmological evolution. It has been shown \cite{Wetterich02.1, Mota03, Shaw05} that for a very light field weakly coupled to matter the local perturbations are generally small relative to the cosmological evolution. In other words, the evolution of the scalar field in a cluster of galaxies or on Earth does not decouple from the cosmological evolution (in distinction to the gravitational field), such that its cosmological time evolution is reflected in a universal variation of couplings, both on Earth and in the distant Universe.

\section{Crossover quintessence models}
\label{sec:CrossoverQuintessence}
Our first class of models is ``crossover quintessence'' \cite{Hebecker00,Doran07,Wetterich02.2}. Here the scalar field follows tracking solutions \cite{Wetterich88.1, RatraPeebles88} at large redshift. In this early epoch the equation of state $w_h$ is equal to that of the dominant energy component (matter or radiation). One particular difference to cosmological constant models is that this type of quintessence models yields a non vanishing amount of early dark energy, $\Omega_{h,e}$ \footnote{The effects of early dark energy on the measurements of baryon acoustic peaks have been studied by the author in \cite{DST06}. However, these considerations are not subject of this thesis.}.
Typically, such models have an exponential potential, for instance
\beq
\label{eqn:CrossoverPotential}
V(\vp) = \mplanckred^4 e^{- \alpha \frac{\vp}{\mplanckred}} \; . 
\eeq
In this specific case, $\alpha$ is related to the early dark energy fraction \cite{Wetterich88.1, Amendola08},
\beq
\Omega_{h,e} = \frac{n}{\alpha^2}  
\eeq
with $n=3 (4)$ for the matter (radiation) epoch. Late-time acceleration can be achieved, for instance, by slight modifications in the cosmon potential or, equivalently, in the kinetic term \cite{Hebecker00}. 

\begin{figure}
\begin{center}
 \includegraphics[width=8cm]{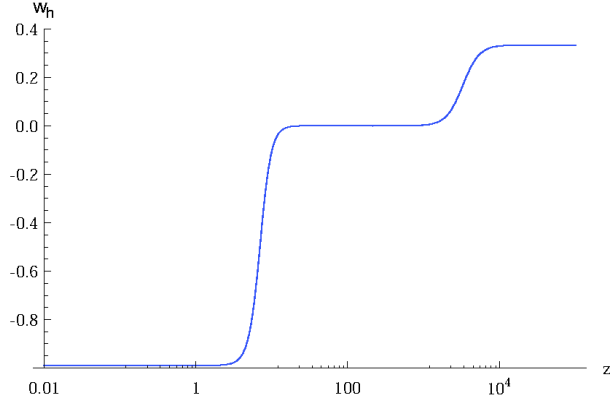}
\caption[Equation of state of quintessence in the crossover quintessence model]{Equation of state of quintessence in the crossover quintessence model ($w_{h0} = -0.99, z_+ = 7$)} \label{fig:CrossoverEqnOfState}
\end{center}
\end{figure}

\begin{figure}
\begin{center}
 \includegraphics[width=8cm]{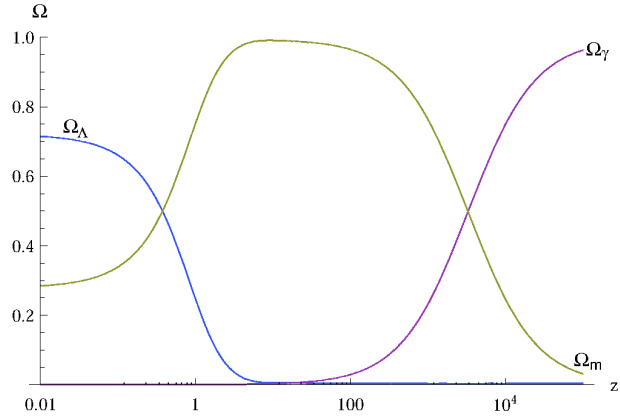}
\end{center}
\caption[Energy components of our Universe in the crossover quintessence model]{Energy components of our Universe in the crossover quintessence model\\($w_{h0} = -0.99, z_+ = 7$)} \label{fig:CrossoverComponents}
\end{figure}

\begin{figure}
\begin{center}
 \includegraphics[width=8cm]{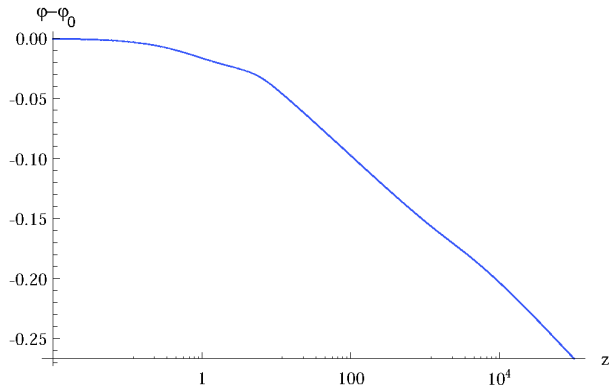}
\end{center}
\caption[Dimensionless cosmon field $\vp$ in the crossover quintessence model]{Dimensionless cosmon field $\vp$ in the crossover quintessence model\\ ($w_{h0} = -0.99, z_+ = 7$)} \label{fig:CrossoverPhi}
\end{figure}

At some intermediate redshift before the onset of acceleration, the time evolution of the cosmon slows down. In consequence, there is a crossover to a negative equation of state and the fraction of energy density due to the scalar begins to grow. In recent epochs the field has an effective equation of state $w_h \gtrsim -1$. The aim of this thesis is not building and solving models of this type in detail, but rather estimating general properties of the scalar evolution. Hence we will not start with appropriate potentials for crossover quintessence and evolve the cosmon over time, but rather simulate the behavior of the field by defining the quintessence equation of state by hand. We set the dark energy equation of state to constant at late times with value $w_{h0}$. Above some given redshift $z_+$ the equation of state crosses over to the scaling condition $w_h=0$ in the matter dominated era; then for $z>z_{eq}$, before matter-radiation equality, we again have scaling through $w_h=1/3$, where $z_{eq}$ can be obtained via
\beq
  z_{eq} = \frac{\Omega_M}{\Omega_{\gamma}} - 1 \; .
\eeq
Then the general relation ($a = (1+z)^{-1}$) 
\beq
	\frac{d \ln\rho}{d \ln a} = -3(1+w(a)) 
\eeq
may be used to find the matter, radiation and dark energy densities over cosmological time. Combining \eqn \eqref{eqn:DefineWh} with \eqn \eqref{eqn:RhoDark}, we can estimate the scalar kinetic energy via 
\beq
\dot{\vp}^2/2=\rho_h(1+w_h)/2
\eeq
and thus integrate $d\vp/da=\dot{\vp}/aH$ from the present back to any previous redshift. The initial conditions are set by specifying the present densities of matter, radiation and dark energy and the model parameters $w_{h0}$ and $z_+$. For illustration, we set $w_{h0} = -0.99$ and $z_+ = 7$. The resulting equation of state is displayed in Fig.~\ref{fig:CrossoverEqnOfState}, the corresponding evolution of energy components in Fig.~\ref{fig:CrossoverComponents} and the dimensionless cosmon field $\vp/\mplanckred$ in Fig.~\ref{fig:CrossoverPhi}.
As can be seen from Fig.~\ref{fig:CrossoverPhi}, in this type of models the scalar field has a monotonic evolution. Assuming a constant coupling $\delta$ to the fundamental varying parameter, usually $\alphagut$, the variation is given by
\beq \label{eq:DalphaXgrowing}
	\Delta \ln \alphagut(z) = \delta( \vp(z) - \vp(0)) \; . 
\eeq
Hence this ansatz implies a monotonic evolution of variations.

\section{Growing neutrino mass models}
\label{sec:GrowingNeutrinoModels}

Growing neutrino models \cite{Amendola08, Wetterich08} explain the value of today's dark energy density by the ``principle of cosmological selection''. The present fraction of dark energy, $\Omega_h^0$, is set by a dynamical mechanism. The essential ingredient of this class of models is a neutrino mass that depends on the cosmon field $\vp$ and grows in the course of the cosmological evolution. As soon as the neutrinos become non-relativistic, their coupling to the cosmon triggers an effective stop (or substantial slowing) of the evolution of the cosmon. Before this event, the quintessence field follows the tracking behavior described in the preceding section. In the models which we will study, the cosmon is assumed to have the potential from \eqn \eqref{eqn:CrossoverPotential},
\[
V(\vp) = \mplanckred^4 e^{- \alpha \frac{\vp}{\mplanckred}} \; . 
\]
The present dark energy density, $\rho_{h0}$, can be expressed in terms of the average present neutrino mass, $m_{\nu}(t_0)$, and a dimensionless parameter $\zeta$ of order unity \cite{Amendola08},
\beq
(\rho_{h0})^{1/4} = 1.07 \left( \frac{\zeta m_{\nu}(t_0)}{eV} \right)^{1/4} 10^{-3} {\rm eV} \; .
\eeq

We follow again our simple proportionality assumption, namely that the cosmon coupling to a typical fundamental parameter is given by \eqn \eqref{eq:DalphaXgrowing},
\[ 
	\Delta \ln \alphagut(z) = \delta (\varphi(z)-\varphi(0)) \; , 
\]
with a proportional variation for other couplings according to the unification scenario that we will study. This is the only contribution to the variation of the unified coupling $\alphagut$ and $\mgut / \mplanck$. However, a new ingredient is an additional variation of the Higgs v.e.v.~$\vev{\phi}$ with respect to $\mgut$, which only becomes relevant at late time \cite{Wetterich08}. It is due to the effect of a changing weak triplet operator on the v.e.v.~of the Higgs doublet. If the dominant contribution to the neutrino mass arises from the ``cascade mechanism'' (or ``induced triplet mechanism'') via the expectation value of this triplet, this changing triplet value is directly related to the growing neutrino mass \cite{Wetterich08}. To understand this mechanism, we start with the most general mass matrix for the light neutrinos,
\beq
m_{\nu} = M_D M^{-1}_R M_D^T + M_L \; . 
\eeq
The first term is responsible for the seesaw mechanism \cite{Minkowski77} with the mass matrix for heavy ``right handed'' neutrinos $M_R$ and a Dirac mass term $M_D$. The second term accounts for the ``induced triplet mechanism'' \cite{Magg80}
\beq
M_L \propto \frac{\vev{\phi}^2}{M_t^2} \; ,
\eeq
where a heavy $SU(2)_L$-triplet field $t$ with mass $M_t$ enters the equation (see \cite{Wetterich08} for details). It is assumed that the mass of the triplet depends on the cosmon field, $M_t = M_t(\vp)$.

The $\vp$-dependence of the Higgs v.e.v.\ $\vev{\phi}$ is introduced by assuming a general effective potential $U(\vp, \phi, t)$. Solving the field equations for the Higgs doublet field $\phi$ and the triplet field $t$, $\partial U / \partial \phi = 0$, $\partial U / \partial t = 0$, the cosmon potential is then obtained as
\beq
 V(\vp) = U(\vp, \phi(\vp), t(\vp)) \; .
\eeq
In \cite{Wetterich08} the simple potential
\beq
  U = U_0(\vp) + \frac{\lambda}{2} (\phi^2 - \phi^2_0)^2 + \frac{1}{2} M_t^2(\vp) t^2 - \gamma \phi^2 t 
\eeq
is assumed, with $\gamma$ and $\lambda$ some coupling parameters. Solving the field equations for the Higgs doublet, it is found \cite{Wetterich08}
\beq
 \frac{\vev{\phi}^2}{\mgut^2} (\vp) = \frac{\vev{\phi}^2_0}{\mgut^2} \left( 1 - \frac{\gamma^2}{\lambda M_t^2(\vp)} \right)^{-1} \; , 
\eeq 
where $\vev{\phi}^2_0$ has to be chosen such that the measured Higgs v.e.v.\ is obtained today.

In the following we consider two models, with slightly different functional dependence of the Higgs v.e.v.\ and neutrino mass on the scalar field. 

\subsection{Stopping growing neutrino model}
\label{sec:StoppingGrowingModel}
\begin{figure}
\begin{center}
 \includegraphics[width=8cm]{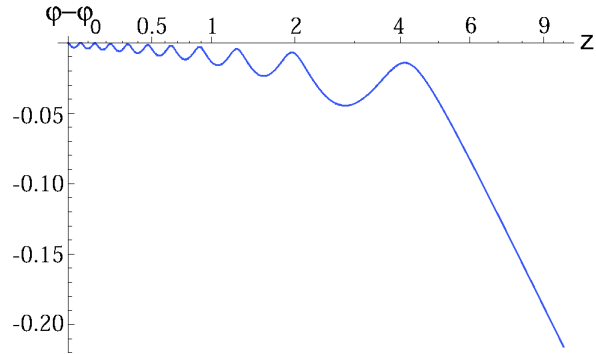} 
\end{center}
\caption[Evolution of the cosmon field in the stopping growing neutrino model]{Evolution of the dimensionless cosmon field in the stopping growing neutrino model of \cite{Wetterich08}.} \label{fig:VariationFixedPointPhi}
\end{figure}

\begin{figure}
\begin{center}
 \includegraphics[width=8cm]{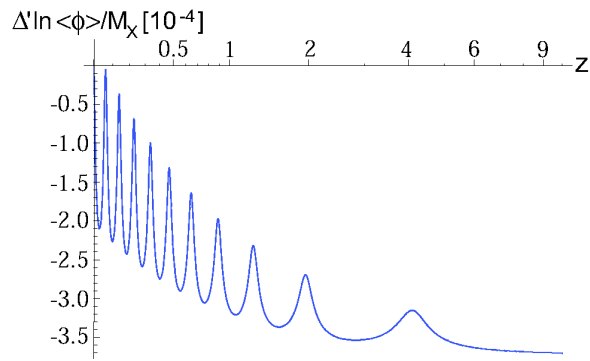}
\end{center}
\caption[Additional variation of the Higgs v.e.v.\ in the stopping growing neutrino model]{Additional variation of the Higgs v.e.v.\ according to \eqn \eqref{eq:phiRofz} in the stopping growing neutrino model of \cite{Wetterich08}.} \label{fig:VariationFixedPointHiggs}
\end{figure}

In the first model, the cosmon asymptotically approaches a constant value (``stopping growing neutrino model'') \cite{Wetterich08} and the neutrino mass is given by
\beq
  m_{\nu}(\vp) = \bar{m_{\nu}} \left \lbrace 1 - \exp \left[ -\frac{\epsilon}{\mplanckred} (\vp - \vp_t )  \right] \right \rbrace^{-1} \; . 
\eeq
With a triplet mass dependence
\beq
M_t^2(\vp) = \bar{M_t^2} \left[ 1 - \exp \left( - \frac{\epsilon}{\mplanckred} (\vp - \vp_t)  \right) \right] \; ,
\eeq
the additional Higgs variation is given according to
\beq \label{eq:phiRofz}
	\frac{\vev{\phi}}{\mgut}(z) = \bar{H} \left(1 - R(z)\right)^{-0.5},
\eeq
where
\beq \label{eq:RofzphiWett}
	R(z) = \frac{R_0}{1-\exp(-\frac{\epsilon}{\mplanckred}(\vp(z) - \vp_t))}.
\eeq
Here, $\vp_t \approx 27.6$ is the asymptotic value (choosing the parameter $\alpha=10$ in the exponential potential \cite{Wetterich08}). For illustration we take the set of parameters given in \cite{Wetterich08}, $\epsilon = -0.05$, and $\bar{H}$ is set by demanding the Higgs v.e.v.\ being consistent with measurements today, $\vev{\phi}(z=0) = 175\,$GeV. We set $R_0 = 10^{-7}$, however in general we only require $R(z=0) \ll 1$. The resulting variations are shown in Fig.~\ref{fig:VariationFixedPointPhi} and Fig.~\ref{fig:VariationFixedPointHiggs}.

The stopping growing neutrino model has an oscillation in $\vev{\phi}$ that grows both in frequency and amplitude at late times as $\vp$ approaches its asymptotic value. Such oscillations must not be too strong as measurements between $z=2$ and today would measure a high rate of change. The oscillation may be made arbitrarily small by choosing small $R_0$. However, the linear variation \eqref{eq:DalphaXgrowing} is independent of $R_0$.

\newpage

\subsection{Scaling growing neutrino model}
\label{sec:ScalingGrowingModel}
The second growing neutrino model \cite{Amendola08} does not lead to an asymptotically constant $\vp$. Now the coupling of the neutrino to the cosmon $\vp$ is given by a constant $\beta$, according to
\beq
	m_\nu = \tilde{m}_\nu e^{-\beta \vp}.
\eeq
This ``scaling growing neutrino model'' leads in the future to a scaling solution with a constant ratio between the neutrino and cosmon contributions to the energy density. 

\begin{figure}
\begin{center}
 \includegraphics[width=8cm]{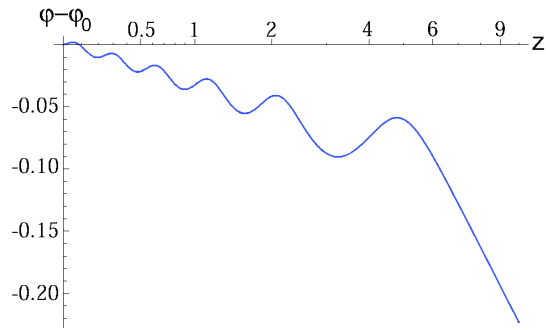} 
\end{center}
\caption[Evolution of the cosmon field in the scaling growing neutrino model]{Evolution of the dimensionless cosmon field in the scaling growing neutrino model of \cite{Amendola08}.} \label{fig:VariationConstantBetaPhi}
\end{figure}

\begin{figure}
\begin{center}
  \includegraphics[width=8cm]{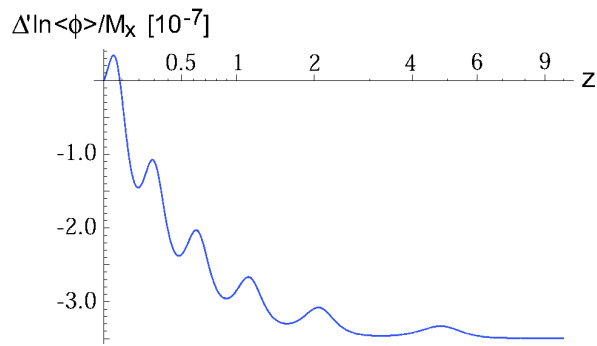}
\end{center}
\caption[Additional variation of the Higgs v.e.v.\ in the scaling growing neutrino model]{Additional variation of the Higgs v.e.v.\ according to \eqn \eqref{eq:phiRofz} in the scaling growing neutrino model of \cite{Amendola08}.} \label{fig:VariationConstantBetaHiggs}
\end{figure}

With the choice of parameters $\beta=-52$, $\alpha=10$ and $m_{\nu,0}=2.3\,$eV \cite{Amendola08}, and given the triplet mechanism of \cite{Wetterich08}, the Higgs v.e.v.\ varies as \eqn \eqref{eq:phiRofz}, where now $R$ is given by 
\beq \label{RofzphiAmen}
	R(z) = R_0 e^{-\beta \vp(z)}.
\eeq
Here the Higgs oscillations remain comparatively small in amplitude, while the absolute value of $\vev{\phi}$ grows overall with time: see Figs.~\ref{fig:VariationConstantBetaPhi} and \ref{fig:VariationConstantBetaHiggs} with the parameter choice $R_0=10^{-6}$. Compared to the ``stopping growing neutrino model'', this model has milder oscillations at late time.

\section{A short note on string theory}
\label{sec:Stringtheory}
String theory is a candidate for a ``theory of everything''. The underlying concept is that fundamental particles are one dimensional objects (strings) which live in a 10 dimensional spacetime. As the world as we know it has only four spacetime dimensions, the remaining 6 dimensions are usually compactified in a way that only 4 ``large'' spacetime dimensions remain. During compactification, several scalar fields appear which can in principle have all sorts of couplings to the Standard Model fields. A further scalar field, the dilaton, is present in string theory from the very beginning. Its v.e.v.~sets the string theory coupling constant. 

With a multitude of possible fields and couplings, one should in principle be able to model any quintessence scenario within the framework of string theory. However, there are strong arguments against it, in particular based on the small mass and potential of the quintessence field (see \eg \cite{Banks01}). A deeper study of variations implied by string theory or possible tests of string theory is out of scope of this thesis, but in principle our methods will also apply to any string theory induced variations of constants, as long as they can be described with effective field theories. Then our methods also allow to constrain the allowed regions in the landscape of string theory.

%%%%%%%%%%%%%%%%%%%%%%%%%%%%%%%%%%%%%%%%%%%%%%%%%%%%%%%%%%%%%%%%%%%%%%%%%%%%%
%%%%%%%%%%%%%%%%%%%%%%%%%%%%%    Part 2     %%%%%%%%%%%%%%%%%%%%%%%%%%%
%%%%%%%%%%%%%%%%%%%%%%%%%%%%%%%%%%%%%%%%%%%%%%%%%%%%%%%%%%%%%%%%%%%%%%%%%%%%%

\part{Big Bang Nucleosynthesis}

\chapter{Big Bang Nucleosynthesis}
\label{chap:BBN}

\section{Why BBN?}

As we have mentioned in \sect \ref{sec:TheoretArgumentsForVariation}, changes in fundamental constants are most likely to appear over large time scales and / or different environments. Big Bang Nucleosynthesis is the earliest process in the history of the Universe which can be both reasonably described with standard physics and astrophysically probed. Also, the Universe was much denser at this time\footnote{The energy density at BBN was roughly the same as the the density of normal water.}, even though locally the environment one finds for instance in heavy stars, supernovae or black holes is much more extreme than at BBN. Hence, it is very reasonable to carefully study the influence of varying parameters on the process of primordial element production.

\section{How will we study BBN?}
BBN is a complex process involving a lot of nuclear reactions and particles. There are two main approaches to a theoretical prediction of primordial element abundances. The first one is purely analytical and only gives rough estimates, whereas the second one numerically simulates the whole process of BBN and gives high-precision abundance predictions.

For our purpose, analytical estimates will turn out to have insufficient accuracy, hence we will utilize a numerical procedure to obtain our findings on the influence of varying constants on BBN. Numerical codes for the simulation of primordial nucleosynthesis go back to the late 60s and 70s \cite{Peebles66, Wagoner69, Wagoner72}. Our BBN code is based on the Kawano 1992 code \cite{Kawano88,Kawano92}, with updated nuclear reaction cross sections as given in \cite{NACRE99} and \cite{NETGEN}. We have improved the code in terms of numerical accuracy in order to be able to derive high precision parameter dependences (see \sect \ref{sec:NumericalAspectsofBBN} for details). In the following sections we explain the process of BBN and the underlying physics in greater detail. As has been explained in \sect \ref{sec:VariationDimensionfulParams}, we will work in a system of units where $\lqcd$ is kept constant.

\section{The process of BBN}
\label{sec:ProcessBBN}

\begin{figure}
\begin{center}
\includegraphics[width=8cm]{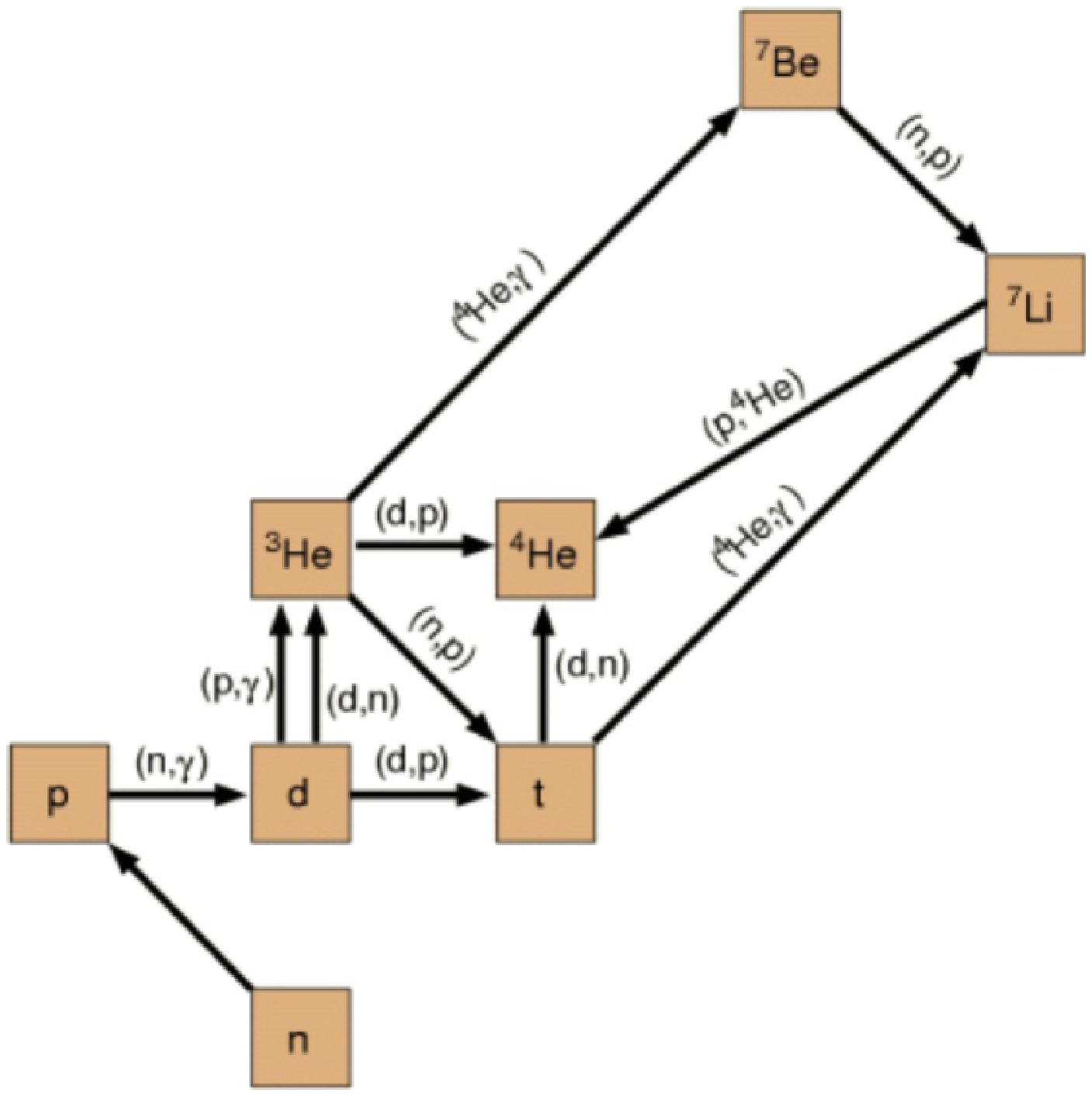}
\end{center}
\caption[Network of main reactions responsible for primordial nucleosynthesis]{Network of main reactions responsible for primordial nucleosynthesis. From \cite{Bartelmann06}.} \label{fig:BBNReactionNetwork}
\end{figure}

In the standard BBN process (SBBN), the element synthesis does not depend on any pre-BBN phase. At a temperature of, for instance, $T= 10^{11}$K, baryogenesis and quark condensation have ended and the Universe only contains protons and neutrons in the baryonic sector. These are held in equilibrium via the weak reactions
\bea\label{eqn:NPEquilibrium}
p + e^- &\longleftrightarrow& n + \nu_e \nonumber \\
n + e^+ &\longleftrightarrow& p + \bar{\nu_e} \, .
\eea
As the Universe expands and cools down, these reactions freeze out at $T \approx 800$ keV, and the neutrons decay freely. The next step in nucleosynthesis is the fusion of deuterium, which happens when reaction
$$ n + p \longleftrightarrow D + \gamma $$
drops out of equilibrium. Due to the large photodissociation cross section, any produced deuterium is immediately destroyed by photons from the background radiation until their temperature has dropped considerably below the binding energy of deuterium, $T \approx B_D = 2.2$ MeV. In fact, since photons are much more abundant than baryons, the high-energy photons in the Maxwell tail keep the reaction in equilibrium until the temperature reaches $T \approx 80$ keV. Once deuterium fusion sets in, elements up to mass number A=7 are synthesized via a network of 11 main nuclear reactions displayed in Fig.~\ref{fig:BBNReactionNetwork}. 
As the Universe continues expanding and cooling down, the temperature and density will fall below the level required for the nuclear reactions at some point. This happens a few minutes after the Big Bang, when the reaction rates become slower than the expansion rate of the Universe, 
\beq
n \vev{\sigma v} < H \; ,
\eeq
with $n$ the particle number density and $\vev{\sigma v}$ the Maxwell-Boltzmann averaged cross section (see Sec.~\ref{sec:ElementSynthesisProcess}). At the end of the BBN process, the element composition of the Universe is roughly 75\% hydrogen, 25\% helium (numbers w.r.t.~mass) and small amounts of deuterium and \lise. Two elements produced during BBN, \bese\ and tritium, decay via a slow $\beta$-decay to \lise\ and \heth\ respectively after BBN has ended. In fact, the primordial \lise\ abundance primarily derives from \bese\ produced during BBN. 

\begin{figure}
\begin{center}
\includegraphics[width=11cm]{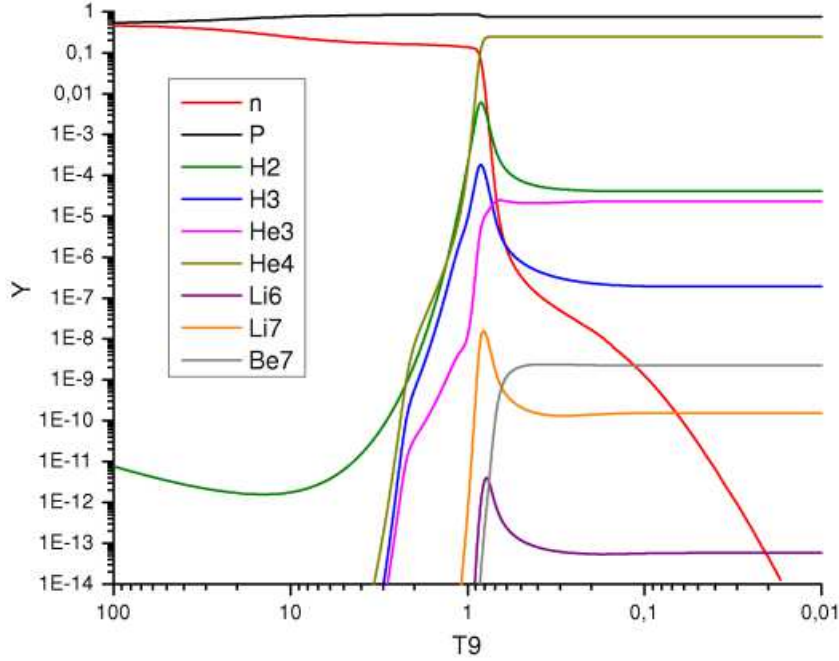}
\end{center}
\caption[Element abundances as a function of the temperature of the universe]{Element abundances $Y$ as a function of the decreasing temperature of the universe $T_9 = T \times 10^{-9} \rm{K}^{-1}$ } \label{fig:BBNAbundanceEvolution}
\end{figure}

The evolution of element abundances during the process of BBN is shown in Fig.~\ref{fig:BBNAbundanceEvolution}, where we define the mass abundance $Y_i$ for a nucleus $i$ as
\beq
   Y_i \define A_i \frac{n_i}{n_B} 
\eeq
with $n_B$ the baryon number density, $n_i$ the number density of nucleus $i$ and $A_i$ its mass number.

\subsubsection{Beyond $A=8$ elements}
During primordial nucleosynthesis, essentially no elements are produced beyond the $A=7$ mass limit. Heavier elements, including carbon and oxygen which are ubiquitous on earth, have been produced later in stars and were thrown back into space during supernova explosions. The process leading to those elements proceeds over the unstable nucleus \beei\ via the ``triple-alpha process''
\bea
 ^4\mbox{He} + \mbox{\hefo} &\rightarrow& \mbox{\beei} + \gamma \nonumber\\
 ^8\mbox{Be} + \mbox{\hefo} &\rightarrow& \mbox{\ctwelve} + \gamma \; . \nonumber
\eea
Due to the low density and low temperature at BBN (compared to stars), the probability that a \beei\ nucleus meets a \hefo\ nucleus during its lifetime of $67 \times 10^{-18} \rm{s}$ is extremely low. Hence, the absence of stable $A=8$ nuclei prohibits any element fusion beyond \lise\ during BBN.

\section{The physics of BBN}

In this section we will describe the underlying processes of BBN in greater detail and introduce the relevant equations.

\subsection{Cosmological background equations}
\label{sec:BBNbackgroundEqns}
BBN is described in the framework of the FRW metric. The basic equations describing the expansion of the Universe are the two Friedmann equations \eqref{eqn:Friedmann1} and \eqref{eqn:Friedmann2}. The process of BBN happened at the a redshift of $z \sim 10^{10}$, a time when the Universe was radiation dominated (see Fig.~\ref{Fig:evolutionOfComponents}), \ie the expansion rate of the Universe is completely controlled by relativistic particles. The temperature $T$ of the Universe is defined by the photon temperature $T_{\gamma}$, and the Stefan-Boltzmann law can be used to express energy and number density of the photons in terms of $T_{\gamma}$,
\bea\label{eqn:StefanBotzmann}
\rho_{\gamma} &=& \frac{\pi}{15} \frac{k_B^4}{(c \hbar)^3} T_{\gamma}^4 \\
\label{eqn:PhotonNumberDensity}
n_{\gamma} &=& \frac{2 \zeta(3)}{\pi^2} \left( \frac{k_B T_{\gamma}}{\hbar c} \right)^3 \; . 
\eea

BBN is taking place in the temperature regime $3 \times 10^9 \rm{K} < T < 0.1 \times 10^9 \rm{K}$. Combining \eqn \eqref{eqn:StefanBotzmann} with the Friedmann equation \eqref{eqn:Friedmann1} yields a relation between temperature and time after the Big Bang,
\beq
t = \left( \frac{16 \pi^3 \gnewton }{5 \hbar^3 c^5} \right)^{-1/2} (k_B T)^{-2} \, . 
\label{eqn:TempTimeRelation}
\eeq
This relation can be applied in the early Universe until the number of photons in the Universe is increased by $e^+ e^-$ annihilation.

Before BBN, neutrinos, photons and electrons were in equilibrium, for instance via
\beq
\nu + \bar{\nu} \longleftrightarrow e^+ + e^- \longleftrightarrow 2\gamma \; . 
\eeq
This reaction freezes out at $T \approx 1.5 \mev$ \cite{Fornengo97}, but as also the neutrinos are ultra relativistic, both $T_{\gamma}$ and $T_{\nu}$ depend on the scale factor\footnote{Equation \eqref{eqn:Tarelation} also gives a relation between temperature and redshift, $T \sim (1+z)$. Note however that this relation does not hold during reheating.}
\beq \label{eqn:Tarelation}
T \sim \frac{1}{a} 
\eeq
and consequently should evolve equally over time. Due to $e^+ e^-$ annihilation, the photon temperature increases whilst the neutrino temperature strictly follows the $T \sim \frac{1}{a}$ law. 
With $T_{\gamma}$ and $T_{\nu}$ evolving differently, the two temperatures have to be tracked separately during BBN. For the neutrino temperature we get an expression equivalent to the Stefan-Boltzmann law,
\beq
\rho_{\nu} = N_{\nu} \frac{7}{8} \frac{\pi}{15} \frac{k_B^4}{(c \hbar)^3} T_{\nu}^4 \; ,  
\eeq
where $N_{\nu}$ is the number of neutrino generations (in SBBN $N_{\nu} = 3$).

Note that temperatures can be given both in units of Kelvin (K) and energy (MeV) via $E = k_B T$, with the conversion factor
\beq
1 \mev \simeq 11.6 \cdot 10^{9} \mbox{K} 
\eeq
\subsubsection{$e^+ e^-$ annihilation}
\label{sec:Reheating}
At $T \approx 10^{10}$K ($E \approx 1$ MeV) the reaction $e^+ + e^- \leftrightarrow 2\gamma$ drops out of thermodynamic equilibrium. Electrons and positrons annihilate to photons, the number of photons and hence also the temperature of the photon gas rises (this is called ``reheating''). Entropy conservation yields that the number of photons, and hence the photon energy- and number density, rise by a factor 11/4. This is also the reason for today's difference in photon and neutrino temperature, $T_{\gamma}^0 = 2.73$ K, $T_{\nu}^0=1.95$ K ($T_{\gamma}^0 / T_{\nu}^0 = (11/4)^{1/3}$).

\subsection{Initial conditions}
\label{sec:InitialConditions}
We start our simulation of the BBN process at a temperature of $T= 10^{11}$ K. As has been explained in \sect \ref{sec:ProcessBBN}, protons and neutrons are in thermodynamic equilibrium via the reaction \eqref{eqn:NPEquilibrium}, hence the neutron-to-proton ratio is given by the thermodynamic relation
\beq
\frac{n}{p} = e^{-(\mneutron - \mproton)c^2/k_B T} \, . 
\eeq
Knowing the starting temperature (which has to be well above the freezeout temperature of the reaction \eqref{eqn:NPEquilibrium}), the initial ratio of neutrons to protons is well defined.
Via \eqn \eqref{eqn:TempTimeRelation} the initial temperature can be translated into the initial time $t_{init}$.

The only additional cosmological parameter necessary for BBN is the initial baryon number density\footnote{A further cosmological parameter often considered in BBN studies is the `number of neutrino species' at BBN, \ie the number of relativistic particles present at BBN. Since the presence of those additional particles is equivalent to a change in expansion rate, we treat this parameter effectively as a change in the gravitational constant $\gnewton$. In SBBN, $\eta$ is the only parameter which enters, and $N_{\nu}$ is set to 3.},
\beq
n_B^{initial} = n_n^{initial} + n_p^{initial} \,. 
\eeq
In principle, this parameter should drop out of a theory of baryogenesis. However, lacking this theory, we have to take $n_B$ as a cosmological parameter which has to be plugged into the BBN simulation by hand. Since $n_B$ scales like the photon number density $n_{\gamma}$ during cosmological expansion,
\beq
n_B \propto n_{\gamma} \propto a^{-3} \; , 
\eeq
the parameter
\beq
\label{eqn:EtaOmegaB}
\eta \define n_B / n_{\gamma} = 2.74 \cdot 10^{-8} \OmegaBaryon h^2 
\eeq
stays constant throughout the later evolution of the Universe, unless the number density of the photons is not significantly increased by later events, \eg by the decay of particles. This is exactly what happens during $e^+ e^-$ annhililation as described in section \ref{sec:Reheating}. The effect on the photon temperature could be computed, giving that $\eta$ at $T > 10^{10}$ K was higher than the current $\eta_0$ by a factor $\frac{11}{4}$, $\eta_{T>10^{10}K} = \frac{11}{4} \eta_0$. Including this correction factor, we can directly deduce the baryon density at BBN (for a given temperature $T \approx 10^9$ K) from today's value of $\eta$ by applying \eqn \eqref{eqn:PhotonNumberDensity},
\beq
  n_B^{BBN} = \frac{11}{4} \eta_0 \cdot \frac{2 \zeta(3)}{\pi^2} \left( \frac{k_B T}{\hbar c} \right)^3 \; .
\eeq
As this quantity determines the frequency of nuclear collisions, it is clear that a modified $n_B$ will change the ``speed'' of the BBN process.

\subsection{The element synthesis process}
\label{sec:ElementSynthesisProcess}

During BBN, element production is dominantly driven by three types of processes\footnote{Due to the comparably low temperature and density at BBN, three-particle interactions which are important in the stellar nucleosynthesis process do not play a role at BBN \cite{Aprahamian05} and are hence not considered in BBN simulations.},
\begin{itemize}
\item $2 \rightarrow 1$ fusion processes $A+B \rightarrow C$, 
\item $2 \rightarrow 2$ scatterings $A+B \rightarrow C+D$, and 
\item particle decays, $A \rightarrow B+C$. 
\end{itemize}
The parameters describing these processes are the reaction cross sections $\sigma_{A+B \rightarrow C}$, $\sigma_{A+B \rightarrow C+D}$ and the decay width $\lambda_{A \rightarrow B+C} \equiv \tau^{-1}$ (inverse mean lifetime). Given these quantities, one derives simple differential equations that describe the time evolution of the number densities of the nuclei $A,B,C,D$, $n_A, n_B, n_C, n_D$. 
\begin{itemize}
\item For a decay $A \rightarrow B+C$,
\bea
\frac{d n_A}{dt} = - \lambda n_A \\
\frac{d n_B}{dt} = + \lambda n_A \\
\frac{d n_C}{dt} = + \lambda n_A 
\eea
\item For a $2 \rightarrow 1$ process $A+B \rightarrow C$
\bea
\frac{d n_A}{dt} = - n_A n_B \vev{\sigma v} \\
\frac{d n_B}{dt} = - n_A n_B \vev{\sigma v} \\
\frac{d n_C}{dt} = + n_A n_B \vev{\sigma v} 
\eea
\item For a $2 \rightarrow 2$ process $A+B \rightarrow C+D$
\bea
\frac{d n_A}{dt} &=& - n_A n_B \vev{\sigma v} \\
\frac{d n_B}{dt} &=& - n_A n_B \vev{\sigma v} \\
\frac{d n_C}{dt} &=& + n_A n_B \vev{\sigma v} \\
\frac{d n_D}{dt} &=& + n_A n_B \vev{\sigma v} \; . 
\eea
\end{itemize}
Here $\vev{\sigma v}$ is the Maxwell-Boltzmann averaged cross section,
\beq
\label{eqn:MaxwellBoltzmannCrossSection}
\vev{\sigma v} = \frac{(8/\pi)^{1/2}}{\mu^{1/2} (k_B T)^{3/2}} \int_0^{\infty} \sigma(E) E \exp(-E/k_B T) dE \; , 
\eeq
where $\mu$ is the reduced mass of the reactants, and $E$ is the reaction energy in the center of mass system.

\section{Nuclear reaction rates and the $Q$ value}
For nuclear reactions, one of the most important quantities is the reaction $Q$ value, which gives the amount of energy released (or absorbed) in a reaction. Working with positive binding energies\footnote{The positive binding energy is defined via $B(A,Z) \define Z \mproton + (A-Z) \mneutron - m(A,Z)$.} $B_i$, the $Q$ value for a reaction is defined as
\beq
Q \define \sum_{outgoing} B_{out} - \sum_{incoming} B_{in} \, .
\eeq
Hence each reaction $Q$-value is determined by the masses of reactants and products. The $Q$-value of each reaction mainly affects the abundances via the reverse thermal reaction rate relative to the forward rate, and via phase space and radiative emission factors in the reaction cross-sections. 

The reverse reaction rate is simply related to the forward rate via statistical factors, due to time reversal invariance (see for example \cite{NACRE99}): the relevant dependence for a $12 \rightarrow 34$ reaction is
\beq
\frac{\vev{\sigma v}_{34\rightarrow 12}}{\vev{\sigma v}_{12\rightarrow 34}} \propto e^{-Q/T} \; . 
\eeq
The $Q$-dependence of radiative capture reactions $A+B \rightarrow C + \gamma$, assuming a dominant electric dipole, is 
\beq \label{radiative}
\sigma(E) \propto E_\gamma^3 \sim (Q+E)^3 \; ,
\eeq
whereas for $2\rightarrow 2$ inelastic scattering or transfer reactions the dependence is
\beq \label{2to2}
\sigma(E) \propto \beta \sim (Q+E)^{1/2} \; , 
\eeq
where $\beta$ is the outgoing channel velocity. In the current treatment we assume $E\ll Q$ at relevant temperatures, and simply scale rates by the appropriate power of $Q$.\footnote{Clearly this breaks down when $Q$ approaches zero, and we have not considered varying any binding energy to the point where this happens. A more accurate treatment would involve applying the phase space dependences directly to the cross-sections, for example in the S-factor description (see \sect \ref{sec:Sfactor}) of charged particle reactions, which involves an expansion in $E$; the dependence on $(Q+E)$ can then be applied order by order.}

A variation in binding energies can have two kinds of effect. The $Q$ value can change the time when a reaction drops out of equilibrium, for instance at the $n+p \rightarrow d+\gamma$ reaction. Or it can change the absolute rate of a reaction, and thus the production rate of a given species, for example at the \bese-producing reaction whose cross-section varies with $Q^3$.

Whether the reaction matrix elements have a dependence on the binding energies and on $Q$ is in general not clear because there is no systematic effective theory for multi-nucleon reactions. The exception is the $np \rightarrow d\gamma$ reaction, for which we have implemented a nuclear effective theory result (see \sect \ref{sec:npdgamma}).

\section{Nuclear reactions important for BBN}

Simulations of the BBN process usually track a large amount of elements and reactions. However, only a few of them turn out to be important for BBN. In the next section, we will be interested in the behavior of BBN under variations of parameters, hence we will be concentrating on those reactions where a variation of parameters will result in a variation in final abundances. In order to estimate which reactions are more or less important for the variation of the final abundances, we varied each thermal averaged cross-section $\vev{\sigma v}$ by a temperature-independent factor, preserving the relation between forward and reverse rates. The resulting values for $\partial \ln Y_a/\partial \ln \vev{\sigma v}_i$ are given in \tab \ref{tab:importantReactions}.
As usual in BBN simulations, the slow $\beta$-decays of tritium and \bese\ are accounted for by adding on the T and \bese\ abundances to \heth\ and \lise\ respectively at the end of the run, when other nuclear reactions have frozen out.
\begin{table}
\centering
\begin{tabular}{|l|c|c|c|c|c|c|}
\hline
Reaction  &    $Q$ value [MeV] & D & \heth & \hefo & \lisi & \lise \\
\hline\hline
$n\leftrightarrow p$ 	& 1.29 & -0.80 & -0.30 & -1.48 & -2.76 & -0.92 \\ \hline
$p(n,\gamma)d$	 	& 2.22 & -0.2 &  0.1 &  0 & -0.2 &  1.3 \\ \hline
$d(p,\gamma)$\heth 	& 5.49 & -0.3 &  0.4 &  0 & -0.3 &  0.6 \\ \hline
$d(d,n)$\heth 		& 3.27 & -0.5 &  0.2 &  0 & -0.5 &  0.7 \\ \hline
$d(d,p)t$ 		& 4.03 & -0.5 & -0.3 &  0 & -0.5 &  0.1 \\ \hline
$d(\alpha,\gamma)$\lisi & 1.47 &  0   &  0   &  0 &  1.0 &  0   \\ \hline
\heth$(n,p)t$		& 0.76 &  0   & -0.2 &  0 &  0   & -0.3 \\ \hline
\heth$(d,p)$\hefo	& 18.35&  0   & -0.8 &  0 &  0   & -0.7 \\ \hline
\heth$(\alpha,\gamma)$\bese & 1.59& 0 &  0   &  0 &  0   &  1.0 \\ \hline   
\lisi$(p,\alpha$)\heth 	& 4.02 &  0   &  0   &  0 & -1.0 &  0   \\ \hline
\bese$(n,p)$\lise  	& 1.64 &  0   &  0   &  0 &  0   & -0.7 \\ \hline
\hline
\bese(n,a)\hefo      & 18.99&  0 &  0 &  0 &  0 & -0.01 \\ \hline
T(p,g)\hefo          & 19.81&  0 &  0 &  0 &  0 &  0.02 \\ \hline   
\lise(p,a)\hefo      & 17.35&  0 &  0 &  0 &  0 & -0.06 \\ \hline     
T(a,g)\lise          & 2.47&  0 &  0 &  0 &  0 &  0.03 \\ \hline                  
T(d,n)\hefo          & 17.59&  0 & -0.01 &  0 &  0 & -0.02 \\ \hline  
\bese(d,pa)\hefo     & 16.77&  0 &  0 &  0 &  0 & -0.01 \\ \hline
\end{tabular}
\caption[Leading dependence of abundances on thermal averaged cross-sections]{Leading dependence of abundances on thermal averaged cross-sections $\partial \ln Y_a/\partial \ln \vev{\sigma v}_i$ for important reactions (1st part) and less important reactions (2nd part)} \label{tab:importantReactions}
\end{table}

For all reactions in Tab.~\ref{tab:importantReactions}, the comments on $Q$-value (and hence binding energy) dependence of the preceding sections do apply. In the following sections we will give details on those reactions for which a more detailed dependence on fundamental parameters is known.

\subsection{The $n \leftrightarrow p$ reaction rate}

The $n\leftrightarrow p$ weak interactions influence every abundance nontrivially. Opposed to most of the reaction rates which can only be determined experimentally, a closed formula is known for the $n \leftrightarrow p$ weak reaction \cite{Scherrer83, Lopez97, WeinbergGRT}.
\beq
\lambda_{n\rightarrow p} = \lambda_{en\rightarrow \nu p} + \lambda_{\nu n\rightarrow e p} + \lambda_{n\rightarrow p e \nu} 
\eeq
\beq
\lambda_{p\rightarrow n} = \lambda_{ep\rightarrow \nu n} + \lambda_{\nu p\rightarrow e n} + \lambda_{p e \nu \rightarrow n} 
\eeq
\beq
\label{eqn:nDecayWithTemp}
\lambda_{n\rightarrow p e \nu} = K \int_1^q d \epsilon \frac{ (\epsilon - q)^2 (\epsilon^2 - 1)^{1/2} \epsilon}{[1 + \exp(-\epsilon z)][1 + \exp( (\epsilon - q) z_{\nu} - \xi_e])} 
\eeq
\beq
\lambda_{n \nu \rightarrow p e } = K \int_q^{\infty} d \epsilon \frac{ (\epsilon - q)^2 (\epsilon^2 - 1)^{1/2} \epsilon}{[1 + \exp(-\epsilon z)][1 + \exp( (\epsilon - q) z_{\nu} - \xi_e])} 
\eeq
\beq
\lambda_{n e \rightarrow p \nu } = K \int_1^{\infty} d \epsilon \frac{ (\epsilon + q)^2 (\epsilon^2 - 1)^{1/2} \epsilon}{[1 + \exp(\epsilon z)][1 + \exp( -(\epsilon + q) z_{\nu} - \xi_e])} \, .
\eeq
Here $q \define (\mneutron - \mproton)/m_e$, $z = m_e/T_{\gamma} $, $z_{\nu} = m_e/T_{\nu} $ and $\xi_e$ is the electron neutrino degeneracy parameter which we always set to zero. The constant $K$ is obtained by demanding $\lambda_{n \rightarrow p}(T \rightarrow 0) \equiv \tau_n^{-1}$. The corresponding $p \rightarrow n$ rates are obtained via $\lambda_{p \rightarrow n} = \lambda_{n \rightarrow p} (-q, - \xi_e)$. 

The equations given above derive from first-order electroweak theory and contain some approximations to allow an easy numerical evaluation. However, higher precision of measurements demands a higher accuracy for theoretical simulations. Hence, modern simulations of BBN apply corrections to the equations given above. Those are radiative corrections in the form of zero-temperature and thermal Coulomb corrections \cite{Lopez98}, as well as finite nucleon mass corrections given in \cite{Lopez97}. For example, thermal Coulomb corrections are applied by multiplying the integrand of the rates for $n \leftrightarrow pe \nu$ and $ep \leftrightarrow n \nu$ with the Fermi factor
\beq
\frac{2\pi \alpha / \beta}{1 - \exp( - 2\pi \alpha/\beta)} \; , 
\eeq
where $\beta = \sqrt{1-\epsilon^{-2}}$. Similar prescriptions exist for the other corrections \cite{Lopez97,Lopez98}.

\newpage

\subsection{The $n +p \rightarrow D + \gamma$ reaction rate}
\label{sec:npdgamma}
Chen \etal derived an effective theory for the strong $n+p \rightarrow D + \gamma$ cross section which is now widely used \cite{Chen99}. They use an effective theory which contains direct nucleon-nucleon interactions and photon exchange, but neglects pion exchange\footnote{In EFT treatments of processes where the external momenta are much smaller than the pion mass (as is the case at BBN with nucleon energies $E_N \lesssim 1$ MeV, $m_{\pi} \approx 135$ MeV), a pion-less EFT turns out to describe the process sufficiently well \cite{Gegelia98,ChenRupak99}.}. They estimate the error in the energy regime of BBN to be $\lesssim 4\%$. Using nucleon-nucleon phase-shift data and the cross section for cold neutron capture as input data, they obtain the folloing equations

\beq
\sigma_{np \rightarrow d \gamma} = \frac{4\pi \alpha (\gamma^2 + P^2)^3}{\gamma^3 \mnucleon^4 P} \left( \tilde{X}_{M1}^2 + \tilde{X}_{E1}^2 \right) \; ,
\eeq
where $P = \sqrt{\mnucleon E}$, $\gamma = \sqrt{B_D \mnucleon}$, $B_D$ is the deuteron binding energy and $E$ is the cross section energy in the center of mass system. $\tilde{X}_{M1}^2$ and $\tilde{X}_{E1}^2$ are given by
\beq
\tilde{X}_{E1}^2 = \frac{P^2 \mnucleon^2 \gamma^4}{(\gamma^2 + P^2)^4} \left[ 1 + \gamma \rho_d + (\gamma \rho_d)^2 + (\gamma \rho_d)^3 + \frac{\mnucleon \gamma}{6\pi} \left( \frac{\gamma^2}{3} + P^2 \right) C_1 \right] \; ,
\eeq
\beq
\tilde{X}_{M1}^2 = \frac{ \kappa_1^2 \gamma^4 \left(\frac{1}{a_1} - \gamma \right)^2}{\left( \frac{1}{a_1^2} + P^2 \right) \left( \gamma^2 + P^2 \right)^2} \left[ 1 + \gamma \rho_d -r_0 \frac{ \left( \frac{\gamma}{a_1} + P^2 \right) P^2}{\left( \frac{1}{a_1^2} + P^2 \right) \left( \frac{1}{a_1} - \gamma \right)} - \frac{L_{np} \mnucleon }{2\pi \kappa_1} \frac{ \gamma^2 + P^2}{\frac{1}{a_1} - \gamma} \right] \; ,
\eeq
with the EFT fitting constants \cite{Chen98, Chen99}\\

\begin{tabular}{cccl}
$\rho_d $&=& $1.764 \mbox{ fm}$ & (effective range parameter)\\
$C_1 $&=& $-1.49 \mbox{ fm}^4$ & (P-wave interaction constant) \\
$\kappa_1$ &=& $2.352945$ & (isovector nucleon magnetic moment) \\
$a_1$ &=& $-23.714 \mbox{ fm}$ & (scattering length in the $^1S_0$ channel) \\
$r_0$ &=& $2.73 \mbox{ fm}$ & (effective range in the $^1S_0$ channel) \\
$L_{np}$ &=& $-4.513 \mbox{ fm}^2$ & ($M1$ capture constant) . \\
\end{tabular}

\subsection{Charged particle reaction rates}
\label{sec:Sfactor}
Theoretical models describing the rough shape of charged particle reaction cross sections have long been available \cite{Fowler67}. In absence of resonances, the cross sections are a product of the Gamow factor and an ``S-factor'',
\beq
\sigma (E) = S(E) \frac{e^{-2\pi\tilde{\eta}}}{E} \; , 
\eeq
where 
\beq
\tilde{\eta} \equiv \alpha Z_1Z_2\sqrt{\mu/2E} \, , 
\eeq
$Z_{1,2}$ the atomic numbers in the initial state and $\mu$ is the reduced mass. The S-factor may be expanded in a Maclaurin series to quadratic order in energy, which is usually sufficient to account for any smoothly-varying dependence,
\beq
S(E) = S_0 + S_1 E + S_2 E^2 \, . 
\eeq
To obtain the Maxwell-Boltzmann averaged cross section (\eqn \eqref{eqn:MaxwellBoltzmannCrossSection}) one has to perform an integration over energy, resulting in a sum of terms for $\vev{\sigma v}$ with specific dependence on $\alpha$ and the reduced mass $\mu$. A generic cross section has the form \cite{Bergstrom99}\footnote{\cite{Bergstrom99} only gives the dependence on $\alpha$, we have recomputed the expansion including the dependence on the reduced mass $\mu$.}
\begin{multline}
\vev{\sigma v} = a_1 T_9^{-\frac{2}{3}} \left(\frac{\alpha}{\alpha_0}\right)^{\frac{1}{3}} \left(\frac{\mu}{\mu_0}\right)^{-\frac{1}{3}}  e^{-3 \kappa^{1/3}}\\
 \times \left[ 1 + \frac{5}{36} \kappa^{-\frac{1}{3}} + a_2 T_9^{\frac{2}{3}} \left(\frac{\alpha}{\alpha_0}\right)^{\frac{2}{3}} \left(\frac{\mu}{\mu_0}\right)^{\frac{1}{3}} \cdot \left(1 + \frac{35}{36} \kappa^{-\frac{1}{3}} \right) \right.\\
\left. + a_3 T_9^{\frac{4}{3}} \left(\frac{\alpha}{\alpha_0}\right)^{\frac{4}{3}} \left(\frac{\mu}{\mu_0}\right)^{\frac{2}{3}} \cdot \left(1 + \frac{89}{36} \kappa^{-\frac{1}{3}} \right) \right] + \mbox{[resonance terms]}, 
\label{eqn:StermExpansion}
\end{multline}
where
\beq
\kappa \define \pi^2 \alpha^2 Z_1^2 Z_2^2 \frac{\mu}{2 k_B T} 
\eeq
and the $a_i$ are fitting constants which are fit to the measured cross-sections. Some cross-sections are fit with an additional exponential term $\tilde{S}(0)e^{-\beta E}$ \cite{Fowler75}. In addition, non-resonant terms may be multiplied by a cutoff factor $f_{\rm cut}=e^{-(T/T_{\rm cut})^2}$, where $T_{\rm cut}$ has been argued to be proportional to $\alpha^{-1}$ \cite{Bergstrom99,Fowler75}. 

\subsubsection{Resonances}
Where the cross-section as a function of energy shows one or more resonances, they contribute to the thermal averaged rate as
\beq
\vev{\sigma v}_{\rm res} = g(T)e^{-\bar{E}/T} \; ,
\eeq
where $g(T)$ and $\bar{E}$ are fitting parameters corresponding to the shape and position of the resonance. Usually a power-law is taken for $g(T)$, thus $g(T)=c T^p$. In principle one should consider the variation of the resonance parameters if this term is significant. But since the major contribution to the resonance energy $\bar{E}$ probably arises from $\lqcd$ which we take as our (non-varying) unit, it seems a reasonable first guess to keep the resonance parameters fixed.

For the code, the NACRE formulae \cite{NACRE99} fitted at the level of the thermal averaged cross-sections $\vev{\sigma v}$ are not suitable as they do not allow to incorporate the dependence on physical parameters. We use the functional forms of rates from \cite{Smith92, Bergstrom99} which have been described above in \eqn \eqref{eqn:StermExpansion} and fit the free parameters of these rates to reproduce the NETGEN rates \cite{NETGEN} as closely as possible. We also checked that the resulting cross-sections are consistent with experimental data. In the case of $d(\alpha,\gamma)$\lisi\ we found a set of parameters which seems to fit the experimental cross-section at low energies \cite{Kiener6Li} better than the NACRE fit. But note that this cross-section is not measured directly, rather it is derived from experimental data under various assumptions.

\newpage

Replacing the NETGEN rates in the code with our fitted rates, the obtained abundances change as follows:
\begin{itemize}
\item $Y_{\rm D}$ differs by $-0.3\%$,
\item $Y_{3\rm He}$ by $+0.9\%$,
\item $Y_{4\rm He}$ by less than $0.1\%$, and
\item $Y_{7\rm Li}$ by $+3\%$. 
\end{itemize}
Hence we do not consider this refitting as significant, except in the case of the $d(\alpha,\gamma)$\lisi\ reaction. Depending on whether this reaction is fit to NETGEN, or to the cross-section values of \cite{Kiener6Li}, we found a \lisi\ abundance larger by a factor of 1.02, or 3.3, respectively. However, given the unclear observational status of \lisi\ this discrepancy is not currently worth pursuing (see \sect \ref{sec:ObsLithium6}).

\section{The simulation of the BBN process}

The BBN process has been described and simulated more than 40 years ago \cite{Wagoner66}. One starts with the initial conditions given in \sect \ref{sec:InitialConditions} and evolves the element synthesis processes for a set of nuclei as described by the differential equations in \sect \ref{sec:ElementSynthesisProcess} in the background of an expanding universe. As we are working at a time when the Universe was radiation-dominated, the expansion of the Universe as given by the Friedmann equation \eqref{eqn:Friedmann1} is dominated by the photon, neutrino and electron contributions\footnote{The computer code also includes the evolution of baryonic and dark matter. However, this does not make any impact on the obtained abundances.}.
Hence the energy density $\rho$ entering in \eqn \eqref{eqn:Friedmann1} is 
\beq
\rho = \rho_{\gamma} + \rho_{\nu} + \rho_e 
\eeq
and the corresponding pressure
\beq
p = \frac{1}{3} \rho_{\gamma} c^2 + \frac{1}{3} \rho_{\nu} c^2 + p_e \, . 
\eeq

Whilst photons and neutrinos are ultra relativistic throughout BBN, the electrons move from the relativistic to the non relativistic domain in the course of the BBN simulation ($10 \mev \gtrsim T \gtrsim 0.001 \mev$). Hence the electron density and pressure have to be described very carefully.

For the electron energy and pressure density one introduces the electron chemical potential
\beq
\phi_e \sim \frac{\pi^2 (\hbar c)^3 n_B Y_p}{2 (k_B T z)^3} \left[ \frac{1}{\sum_n (-)^{n+1} n L(nz)} \right] \; , 
\eeq
where $z=\frac{m_e c^2}{k_B T}$. The electron and positron energy and pressure densities are then \cite{FowlerHoyle64}
\beq
\rho_e \equiv \rho_{e^-} + \rho_{e^+} = \frac{2 (m_e c^2)^4}{\pi^2 (\hbar c)^3} \sum_{n=1}^{\infty} (-)^{n+1} \cosh(n \phi_e) M(nz) 
\eeq
\beq
\frac{p_{e^-} + p_{e^+}}{c^2} = \frac{2 (m_e c^2)^4}{\pi^2 (\hbar c)^3} \sum_{n=1}^{\infty} \frac{(-)^{n+1}}{nz} \cosh(n \phi_e) L(nz) \; . 
\eeq
The functions $L$ and $M$ are related to the modified Bessel functions $K_i$ via
\beq
L(z) \define K_2(z)/z 
\eeq
\beq
M(z) \define \frac{1}{z} \left[ \frac{3}{4} K_3(z) + \frac{1}{4} K_1(z) \right] \; . 
\eeq
\hfill\\
In our BBN simulation, we evolve the following set of parameters with time
\begin{itemize}
\item temperatures $T_{\gamma}$, $T_{\nu}$
\item electron chemical potential $\phi_e$
\item baryon density $\rho_B$ (or equivalently the scale factor $a$)
\item the abundances $Y_i$
\end{itemize}
according to the differential equations given in this chapter, and the Friedmann equation \eqref{eqn:Friedmann1} is used to evolve the baryon density and temperatures according to the expansion of the Universe.

\subsection{Numerical aspects of the BBN simulation}
\label{sec:NumericalAspectsofBBN}
The main task of a BBN code is to numerically integrate the set of coupled differential equations which have been given in the preceding sections. In our code, the differential equations are solved using an adaptive second-order Runge-Kutta method, which is important to correctly account for the nuclear reactions\footnote{Some nuclear reactions can have both high forward and backward reaction rates. A naive solution of the differential equations for the nuclear reactions via $y(t_o + \Delta t) = y(t_0) + \Delta t \cdot y'(t_0)$ would not take into account that produced nuclei might be immediately destroyed after creation via the back reaction process.}. Given a differential equation
\beq
y'(t) = f(t,y) 
\eeq
and a starting value at time $t_0$, $y(t_0) = y_0$, the task is to derive the value of $y$ at time $t_0 + \Delta t$. The second order Runge-Kutta method does this in a two-step approach.
First one derives the value in linear approximation, \ie
\beq
\tilde{y}(t_o + \Delta t) = y(t_0) + \Delta t \cdot y'(t_0) \; . 
\eeq
The ``average'' derivative between $t_0$ and $t_0 + \Delta t$ is then defined as
\beq
\tilde{y}' \define \frac{f(t_0, y_0) + f(t_0 + \Delta t, \tilde{y}(t_0 + \Delta t))}{2} 
\eeq
and the final value
\beq
y(t_0 + \Delta t) = y_0 + \tilde{y}' \cdot \Delta t \;. 
\eeq
The step width $\Delta t$ is determined adaptively by demanding that the changes in abundances per time step do not exceed a certain value.

We have adapted the code to the capabilities of modern computer technologies, which means that we have increased the internal numerical precision, implemented more precise integration routines and added several further numerical improvements. These corrections were necessary to remove numerical inaccuracies present in the available BBN codes. 

In general, the uncertainties of the input parameters for BBN result in much higher uncertainties of the resulting abundances than do the numerical inaccuracies of the code. However, in the next chapter we will derive linearized dependences of the final abundances $Y$ on input parameters $X$, $\partial \ln Y / \partial \ln X$. This will be achieved by slightly varying the input parameters $X$ and observing the resulting changes in the abundances $Y$, where the variations in $X$ are much below their uncertainty in order to get the correct linear order approximation. Hence, we have to ensure that the changes in the final results of the BBN simulations under small changes of the input parameters are not affected by numerical effects but solely by changes in the physical processes.

\section{Observational situation and uncertainties}
\label{sec:ObsTheoBBN}
One of the biggest successes of standard BBN is the matching of theoretically predicted and observed primordial abundances for major elements. For a review of the theoretical and observational status and obstacles see for instance \cite{Steigman05}. 

\subsection{\hefo}
\begin{figure}
\begin{center}
\includegraphics[width=8cm]{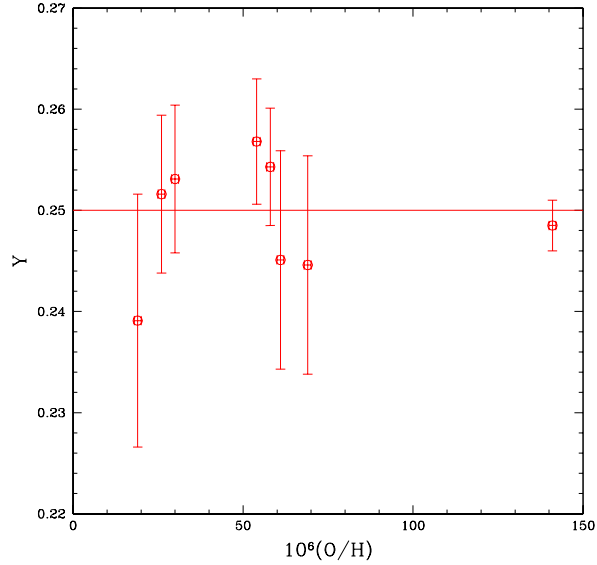}
\end{center}
\caption[\hefo\ abundances versus oxygen abundance]{\hefo\ abundances versus oxygen abundance from \cite{Steigman05}. The solid line is the weighted mean of for the 8 systems. Data from \cite{OliveSkillman04}} \label{fig:HefovsOxygen}
\end{figure}

The highest precision observation is that of the \hefo\ abundance (conventionally written $Y_P$). The post-BBN evolution of \hefo\ is simple. In stars, hydrogen is burned to \hefo\, which increases the abundance of \hefo\ above its primordial value. Hence one expects the \hefo\ abundance to decrease once one goes to stars with lower metallicity\footnote{Elements with higher mass number than Helium (which are in astronomy called ``metals'') are only produced in stars. Hence a low metallicity of a system is a strong indication of element abundances which are assumed to be close to the primordial ones.}, and a \hefo\ ``plateau'' is expected for sufficiently low metallicity. Olive \etal \cite{OliveSkillman04} analyze 8 systems with very low oxygen abundance, which are displayed in \fig \ref{fig:HefovsOxygen}.
They argue that the observational data indicated a primordial abundance of
\beq
Y_P = 0.249 \pm 0.009 \; , 
\eeq
which we take to be a 1$\sigma$ range. However, given the probable dominance of systematic effects, instead of using 2-$\sigma$ bounds to later determine the range of allowed variations\footnote{See \sect \ref{sec:BoundsOnSeparateVariations}.}, we rather use the ``conservative allowable range'' of $Y_P$ given in \cite{OliveSkillman04} as
\beq
0.232 \le Y_P \le 0.258.
\eeq

\subsection{Deuterium}
For deuterium, the post-BBN evolution is as clear as it is for \hefo. Deuterium which was formed during primordial nucleosynthesis is burned in stars to \heth\ and higher elements. Also, any deuterium which is newly produced in stars is immediately burned into \heth\ as deuterium is the most weakly bound light nucleus. Hence, the deuterium abundance should increase when going back in time (opposed to \hefo, which decreases). However, the almost identical absorption spectra of D and ``normal'' hydrogen ($^1$H) complicate spectroscopic determinations of deuterium significantly. 

\begin{figure} 
\begin{center}
\includegraphics[width=8cm]{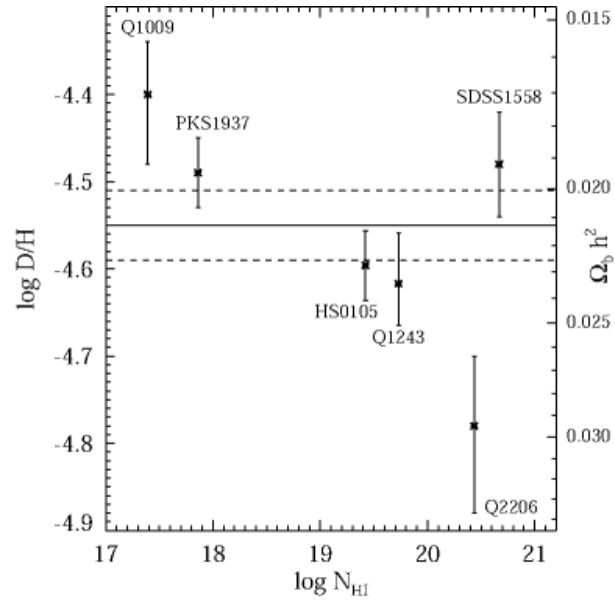}
\end{center}
\caption[Deuterium versus neutral column density $N_{HI}$]{Deuterium versus neutral column density $N_{HI}$ for a set of low-metallicity absorption spectra along QSO sightlines. From \cite{O'Meara06}. ($N_{HI}$ in cm$^{-2}$, its value is of no relevance here). The horizontal line represents the weighted mean, the right axis gives the derived SBBN values for $\OmegaBaryon h^2$.} \label{fig:DeuteriumvsneutralColumnDensity}
\end{figure}

Due to this complication, the determination of the primordial deuterium abundance follows from a small number of observed systems. One basically observes absorption spectra of interstellar deuterium from sightlines of Quasars which have to fulfill certain quality conditions (see \cite{Kirkman03} for details). As D/H is of the order $10^{-5}$, the hydrogen column density $N_{HI}$ must be large enough in order to be able to observe deuterium with modern high-resolution spectrographs\footnote{The column density $N_{HI}$ gives a two-dimensional measure of the density (here number density) of a cloud of material, measured in cm$^{-2}$.}. We use the determinations of \cite{O'Meara06,Kirkman03} which give a value of 
\beq
\mbox{D/H} = (2.8 \pm 0.4) \times 10^{-5} \, .
\eeq
The analysis of \cite{O'Meara06} is displayed in Fig.~\ref{fig:DeuteriumvsneutralColumnDensity}, where the large scatter between values determined from different systems should be noted.
After our studies have been performed, a new analysis appeared \cite{Pettini08} which reduces the error of $D/H$ by a factor of 2.

\subsection{\heth}
\label{Sec:BBNHe3}
\begin{figure} 
\begin{center}
\includegraphics[width=8cm]{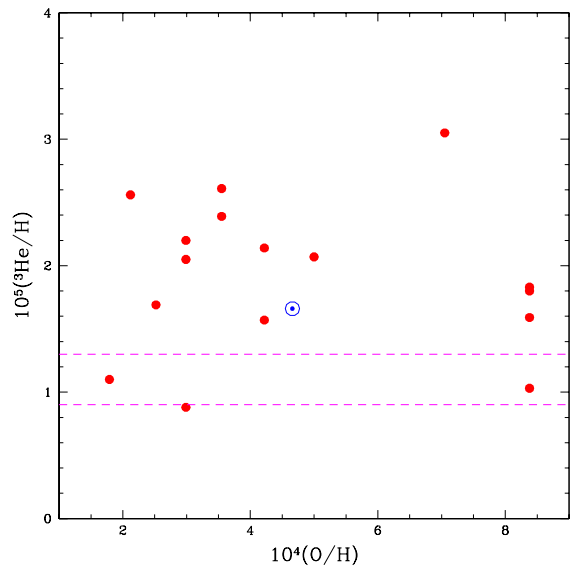}
\end{center}
\caption[\heth\ abundances versus oxygen]{\heth\ abundances versus oxygen from \cite{Steigman05}. Data from \cite{Bania02}. The dashed lines show the $1 \sigma$ band given by \cite{Bania02}, the blue spot indicates the \heth\ abundance for the pre-solar nebula.} \label{fig:HeThDeterminations}
\end{figure}

The post-BBN development of \heth\ is quite complex, as it is both produced and destroyed in stars. As quantitative analyses of these competing processes are quite model dependent, the primordial \heth\ abundance cannot be extracted easily from spectra of old stars. This complexity is revealed by determinations of the \heth\ abundance, which typically have a large scatter as can be seen in Fig.~\ref{fig:HeThDeterminations}. Hence the common understanding is that \heth\ abundance determinations cannot be considered as good tracers for the primordial abundance \cite{Vangioni-Flam02}.

\subsection{\lise}
\begin{figure} 
\begin{center}
\includegraphics[width=12cm]{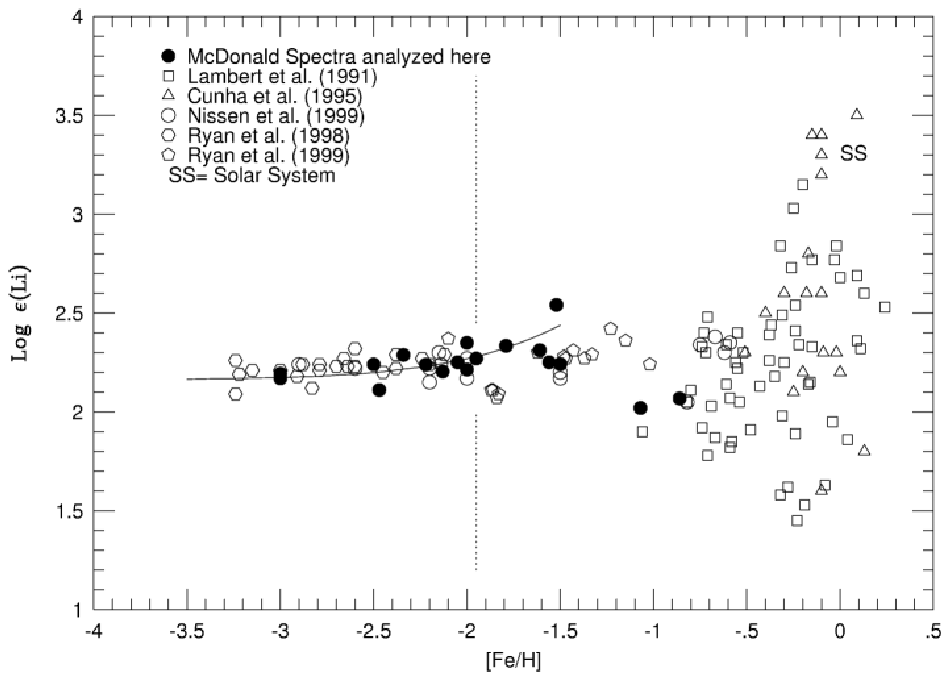}
\end{center}
\caption[\lise\ abundances versus metallicity]{\lise\ abundances versus metallicity from \cite{Steigman05}. $\log \epsilon({\rm Li}) \equiv 12 + \log({\rm Li/H})$} \label{fig:LiseAbundance}
\end{figure}

The post-BBN \lise\ abundance increases due to cosmic ray spallation/fusion reactions and \lise\ production in stars. Even though \lise\ is also easily destroyed inside stars, some stars observationally appear to be ``super-lithium rich'' \cite{Steigman05}, supporting the assumption that (most) stars are in fact net producers of lithium. Hence, one expects the lithium abundance to decrease when going back in time, which is supported by observations of the \lise\ abundance which show a plateau at old, metal-poor halo stars (see Fig.~\ref{fig:LiseAbundance}). One assumes that the plateau value is closely related to the primordial one. 

The most recent determinations of the abundance are quite small: \lise/H = $(1.3 \pm 0.3) \cdot 10^{-10}$ \cite{Bonifacio06} (see also \cite{Asplund05}). It has been suggested that there are unresolved systematic errors relating to the effective temperature of the stars \cite{Charbonnel05} which may imply a value as large as \lise/H = $(1.64 \pm 0.3) \cdot 10^{-10}$.
To account for this possible systematic, we adopt a value
\beq
\mbox{\lise/H} = (1.5 \pm 0.5) \times 10^{-10}.
\eeq
It turns out that the observed \lise\ abundance is about a factor of 3 smaller than the standard theoretical prediction, which is called the ``lithium problem''.

\newpage

\subsection{\lisi}
\label{sec:ObsLithium6}
A possible detection of \lisi\ was discussed in \cite{Asplund05}, who claim to have found signals of \lisi\ in nine stars at the $\ge 2 \sigma$ significance level. Their observation suggests a \lisi\ plateau at the level of
\beq
\mbox{\lisi/H} \approx 6.2 \times 10^{-12} \; . 
\eeq
If this detection is correct, the \lisi\ abundance is about a factor 100 larger than the SBBN prediction. However, given the unclear observational status and post-BBN history of this isotope, we do not include \lisi\ in the final analysis.

\subsection{Theoretical predictions}
Theoretically predicted primordial abundances also come with an error, mainly due to cross-section uncertainties. Our numerical procedures do not provide error estimates, so we adopt the $1\sigma$ ranges from \cite{Serpico04}, using a baryon density $\OmegaBaryon h^2=0.0224$ \cite{ManganoTalk07, HinshawWMAP5}:
\bea
\mbox{D/H}     &=& (2.61 \pm 0.04) \times 10^{-5} \nonumber \\
\mbox{\heth/H} &=& (1.03 \pm 0.03) \times 10^{-5} \nonumber \\
Y_P            &=& 0.2478 \pm 0.0002 \nonumber \\
\mbox{\lise/H} &=& (4.5  \pm 0.4 ) \times 10^{-10}.
\eea
A compilation of the recent observational and theoretical primordial abundances which we will use for BBN is given in Tab.~\ref{tab:ObsTheoAbundances}.
\begin{table}
\center
\begin{tabular}{|c|c|c|}
\hline
Abundance & Observational & Theoretical\\
\hline \hline
D/H & $(2.8 \pm 0.4) \times 10^{-5}$ & $(2.61 \pm 0.04) \times 10^{-5}$\\
\hline
$Y_P$ & $0.249 \pm 0.009$ & $0.2478 \pm 0.0002$\\
\hline
\lise / H & $(1.5 \pm 0.5) \times 10^{-10}$ & $(4.5 \pm 0.4) \times 10^{-10}$\\
\hline
\end{tabular}
\caption{Current observational and theoretical primordial abundances}
\label{tab:ObsTheoAbundances}
\end{table}

%%%%%%%%%%%%%%%%%%%%%%%%%%%%%%%%%%%%%%%%%%%%%%%%%%%%%%%%%%%%%%%%%%%%%%%%%
%%%%%%%%%%%%%%%%%     BBN with varying constants     %%%%%%%%%%%%%%%%%%%%
%%%%%%%%%%%%%%%%%%%%%%%%%%%%%%%%%%%%%%%%%%%%%%%%%%%%%%%%%%%%%%%%%%%%%%%%%

\chapter{BBN with varying constants}
\label{chap:BBNvaryingNucl}

\section{Nuclear and fundamental parameters}

Big Bang Nucleosynthesis happens at energies in the keV regime, much below the energy scale of the Standard Model of particle physics where the parameters which are nowadays considered as fundamental (Tab.~\ref{Tab:CstOfNature}) are relevant. In particular, the theory of QCD cannot be applied at BBN as it happens well below the QCD scale $\lqcd \approx 200 \mev$. Quarks have combined into nucleons and the strong coupling constant $\alphastrong$ is in the non-perturbative regime. Hence, rather than the parameters of the SM, a set of effective low-energy parameters (masses of nucleons, binding energies, cross-sections and lifetimes) enters in the BBN simulations. We call these parameters ``nuclear parameters'' which stresses that those are the actual relevant (effective) parameters which enter in the nuclear physics processes.

\section{Nuclear parameters relevant for BBN}

The following set of effective low-energy (``nuclear'') parameters enters the BBN simulations:
\begin{itemize}
\item Neutron and proton mass $\mneutron$, $\mproton$, combined to
\item Average nucleon mass $\mnucleon \define (\mneutron + \mproton)/2$
\item Neutron proton mass difference $Q_N \define \mneutron - \mproton$
\item Neutron lifetime $\tau_n$
\item Binding energies of D, T, \heth, \hefo, \lisi, \lise, \bese.
\end{itemize}
Additionally, several fundamental parameters are also used in the BBN simulation:
\begin{itemize}
\item Electron mass $m_e$
\item Gravitational constant $\gnewton$
\item Fine structure constant $\alpha$.
\end{itemize}
The baryon to photon ratio $\eta$ enters as a further important parameter which is of cosmological origin. If one had a closed picture for the origin of our Universe, this ratio would derive from ``fundamental'' parameters, but lacking any theory this value only enters as an observed cosmological quantity. We combine all parameters which enter in the BBN simulation into a set $X_i$ which we will call ``set of nuclear parameters''.\footnote{Note that for instance variations with respect to the nuclear parameter $\alpha$ are different from variations with respect to the fundamental parameter $\alpha$. This is due to the fact that variations w.r.t.~the fundamental parameter incorporate \textit{all} dependences, whereas variations w.r.t.~the nuclear parameter only include \textit{some} dependences. For instance the dependence of $\tau_n$ on $\alpha$ will only be included in the fundamental parameter dependence.}

\section[Nuclear parameter dependence]{Nuclear parameter dependence and the response matrix}

We consider the set of primordial abundances $Y_a$ with $a=($D, \heth, \hefo, \lisi, \lise) and study its dependence on the variation of our set of nuclear physics parameters $X_i$. Here the index $i$ denotes the parameters which enter the calculation of nuclear abundances, listed in the preceding section. Our central quantity is the response matrix $C$ with matrix elements \cite{MSW04}
\beq \label{eq:Cdef}
 c_{ai} = \frac{\partial \ln Y_a}{\partial \ln X_i}.
\eeq
It indicates the leading linear dependence for small deviations of the abundances about the values obtained given the nuclear parameters inferred from present laboratory experiments. The matrix $C$ is extracted by varying the quantities $X_i$ independently in the BBN code, a procedure which includes variation of the reaction cross-sections and rates that have a physical dependence on $X_i$. If all variations in parameters are taken to be small, all necessary information can indeed be extracted from the response matrix.

Our results for the nuclear response matrix are shown in \tab \ref{tab:dlnYdlnX}. The first thirteen rows constitute the transposed nuclear response matrix $C^T$. We also quote the dependence of the abundances on $\eta$ in the last row. Values are quoted to 2 d.\,p.\ or to 2 sig.\ fig.\ when the magnitude exceeds 1. 

\begin{table}
\centering
\begin{tabular}{|c|c|c|c|c|c|}
\hline
$\partial\ln Y_a/\partial\ln X_i$  & D & \heth & \hefo & \lisi & \lise \\
\hline \hline
$\tau_n$      &  0.41 &  0.15 &  0.73 &  1.4  &  0.43 \\ \hline
$Q_N$         &  0.83 &  0.31 &  1.55 &  2.9  &  1.00 \\ \hline
$m_N$         &  3.5  &  0.11 & -0.07 &  2.0  &-12    \\ \hline
$B_{\rm D}$   & -2.8  & -2.1  &  0.68 & -6.8  &  8.8  \\ \hline
$B_{\rm T}$   & -0.22 & -1.4  &  0    & -0.20 & -2.5  \\ \hline
$B_{\rm 3He}$ & -2.1  &  3.0  &  0    & -3.1  & -9.5  \\ \hline
$B_{\rm 4He}$ & -0.01 & -0.57 &  0    &-59    &-57    \\ \hline
$B_{\rm 6Li}$ &  0    &  0    &  0    & 69    &  0    \\ \hline
$B_{\rm 7Li}$ &  0    &  0    &  0    &  0    & -6.9  \\ \hline
$B_{\rm 7Be}$ &  0    &  0    &  0    &  0    & 81    \\ \hline
\hline
$\gnewton$         &  0.94 &  0.33 &  0.36 &  1.4  & -0.72 \\ \hline
$\alpha$      &  2.3  &  0.79 &  0.00 &  4.6  & -8.1  \\ \hline
$m_e$         & -0.16 & -0.02 & -0.71 & -1.1  & -0.82 \\ \hline
\hline
$\eta$        & -1.6  & -0.57 &  0.04 & -1.5  &  2.1  \\ \hline
\end{tabular}
\caption{Response matrix $C$, dependence of abundances on nuclear parameters.} \label{tab:dlnYdlnX}
\end{table}

%%%%%%%%%%%%%%%%%%%%%%%%%%%%%%%%%%%%%%%%%%%%%%%%%%%%%%%%%%%%%%%%%%%%%%%%%
%%%%%%%%%%     FROM NUCLEAR TO FUNDAMENTAL PARAMETERS     %%%%%%%%%%%%%%%
%%%%%%%%%%%%%%%%%%%%%%%%%%%%%%%%%%%%%%%%%%%%%%%%%%%%%%%%%%%%%%%%%%%%%%%%%

\chapter{From nuclear to fundamental parameters}
\label{chap:BBNvaryingFund}

\section{From nuclear to fundamental parameters}
\label{sec:BBNFromNucToFunParams}

Looking at the Standard Model of particle physics, it is clear that the nuclear parameters given in the preceding section (Tab.~\ref{tab:dlnYdlnX}) are degenerate in a sense that the set of 10 non-fundamental parameters only depends on about 5 to 6 fundamental parameters. Hence, in a next step, we will derive relations between a set of Standard Model parameters $G_k$ and the nuclear physics parameters $X_i$. This is encoded in a second response matrix $F$ with entries
\beq \label{eq:Fdef}
 f_{ik} = \frac{\partial \ln X_i}{\partial \ln G_k} \, .
\eeq
The variation of abundances with respect to the fundamental parameters $G_k$ is then expressed by the ``fundamental response matrix'' $R$ with elements $r_{ak}$,
\beq \label{eqn:Rdef}
 \frac{\Delta Y_a}{Y_a} = r_{ak} \frac{\Delta G_k}{G_k} \, .
\eeq
The matrix $R$ is obtained from $C$ and $F$ by simple matrix multiplication,
\beq \label{eq:RequalsCF}
R = CF \, .
\eeq
We consider the following six fundamental parameters $G_k$:
\begin{itemize}
\item Gravitational constant $\gnewton$
\item Fine structure constant $\alpha$
\item Electron mass $m_e$
\item Light quark mass difference $\delta_q \equiv m_d-m_u$
\item Averaged light quark mass $\hat{m}\equiv (m_d+m_u)/2 \propto m_\pi^2$
\item Higgs v.e.v.\ $\vev{\phi}$.
\end{itemize}

An additional parameter is the strange quark mass $m_s$. We have omitted it from our treatment of BBN, as it enters the parameters (binding energies, neutron and proton mass) with high uncertainty. We have computed that the final dependence of abundances on the strange quark mass due to the known proton and neutron dependence is less than $3\%$ of that of the light quark mass dependence (given in Tab.~\ref{tab:dlnYdlnG}), hence the dependence on $m_s$ is much lower than the model uncertainty\footnote{However, it might be that binding energies get a substantial contribution from strange effects which we so far cannot quantify \cite{Flambaum02}.} (\eg for nuclear binding energies). However, in our study of variations in \chap \ref{chap:VariationsFromBBNtoTodayinGUT} we will include the strange quark mass contributions, as the parameters studied there are more directly influenced by strange effects. This is why we will also include the strange contribution in our following derivation of fundamental parameter dependences.

In the next sections, we give details on how the nuclear parameters depend on our set of fundamental parameters. In effective theories of nuclear forces, the pion appears as an effective mediator of the strong force. Hence, we will also introduce the pion mass $m_{\pi}$ as an intermediate parameter.

\subsection{Pion mass}

The pion as the lightest of all mesons is the dominant mediator of the strong interaction in effective low energy theories of QCD. Its light mass is due to the fact that the pion is the pseudo Goldstone boson of the only slightly broken chiral symmetry of QCD. This fact implies that one can make several predictions about pion properties in chiral perturbation theory. In first order chiral perturbation theory, one obtains the famous Gell-Mann-Oakes-Renner relation \cite{Gell-Mann68}, which relates the pion mass to fundamental parameters\footnote{Note that there are in fact three pions, $\pi^+$, $\pi^-$ and $\pi^0$ with roughly the same mass ($m_{\pi^{\pm}} = 139.6 \mev$, $m_{\pi^{0}} = 135.0 \mev$). We will only work with the first order dependences on fundamental parameters which are the same for all three pions. Further details on specific mass contributions can be found in \cite{Gasser82}.}\cite{Gasser82},
\beq
\label{eq:GMOR_relation}
m_{\pi}^2 \simeq - \frac{1}{f_{\pi}^2} (m_u + m_d) \vev{0|\bar{u}u + \bar{d}d|0} \; .  
\eeq
The nonvanishing v.e.v.~of $\bar{u}u$ and $\bar{d}d$ is a measure of the chiral asymmetry of the vacuum, signaling the spontaneous breakdown of chiral symmetry \cite{Gasser82}. As can be seen from \eqn \eqref{eq:GMOR_relation}, the values of $\vev{\bar{q}q}$ are negative. 
$f_{\pi}$ is the decay constant of the pion, which does not depend on quark masses at first order and which has a value of \cite{PDG08}
\beq
f_{\pi} \approx 130 \mev \;. 
\eeq

The formulas mentioned above are pure chiral perturbation theory equations neglecting any electromagnetic contributions. In fact, the pion mass difference of $m_{\pi^{\pm}} - m_{\pi^0} = 4.6 \mev$ is dominantly due to electromagnetic effects \cite{Gasser82}: up to a small uncertainty of $0.1 \mev$ the virtual photon cloud surrounding the \pipm accounts for the \piZero -\pipm mass difference,
$$m^{\gamma}_{\pi^{\pm}} - m^{\gamma}_{\pi^{0}} = 4.6 \pm 0.1 \mev $$ 
with $m^{\gamma}_{\pi^{0}} \approx 0$. We can safely neglect this contribution and only work with the first order dependence on the light quark mass, which can be read off from \eqn \eqref{eq:GMOR_relation},
\beq
m_{\pi}^{2} \propto \hat{m} 
\eeq
or
\beq
\Delta \ln m_{\pi} = \frac{1}{2} \Delta \ln \hat{m} \; . 
\eeq

\subsection{Neutron and proton mass}
The different contributions to the neutron and proton mass and the origin of the proton neutron mass difference have been studied by Gasser and Leutwyler more than 25 years ago \cite{Gasser75, Gasser82}. Using ``improved chiral perturbation theory'', they estimate the contributions coming from electromagnetic and (effective) strong self energy as well as from different quark masses. Knowing the properties of electron proton scattering they could estimate the electromagnetic contributions to the masses at Born level\footnote{Born level contributions are proportional to $\alpha$, so a rescaling of $\alpha$ by a certain factor scales the values given in \eqn \eqref{eq_nuclElmContributions} by the same factor.} from the electromagnetic form factors as
\beq
\label{eq_nuclElmContributions}
\begin{aligned}
\mproton^{\gamma} &=& 0.63 \mev \\
\mneutron^{\gamma} &=& -0.13 \mev  
\end{aligned}
\eeq
to a very high accuracy. A further electromagnetic correction might come from possible intermediate states, a contribution which could not be calculated but has been estimated to 
\beq
\Delta m_{res} = \pm 0.2 \mev \;.
\eeq
Thus, the total electromagnetic contributions to the nucleon masses are the values given in \eqref{eq_nuclElmContributions} with an error of $\pm 0.2 \mev$. For the neutron-proton electromagnetic mass difference one obtains
\beq
\label{eqn:ProtonNeutronElmContribution}
(\mneutron - \mproton)^{\gamma} = - 0.76 \pm 0.30 \mev \;.
\eeq
The bare QCD masses of the nucleons are hence
\beq
\label{eqn:nucleonQCDmasses}
\begin{aligned}
\mproton^{QCD} &=& 937.64 \pm 0.20 \mev \;, \\
\mneutron^{QCD} &=& 939.70 \pm 0.20 \mev \;. 
\end{aligned}
\eeq
The difference in neutron and proton mass of 
\beq
\label{eqn:ProtonNeutronQuarkContribution1}
(\mneutron - \mproton)^{QCD} = 2.05 \pm 0.30 \mev 
\eeq
can be explained in terms of the different quark content, $p=(uud)$, $n=(udd)$, which turns out to be in lowest order quark mass expansion \cite{Gasser82}
\beq
\label{eqn:ProtonNeutronQuarkContribution2}
(\mneutron - \mproton)^{QCD} = (m_d - m_u)\frac{1}{2 \mnucleon} <p|\bar{u}u - \bar{d}d|p> = \delta_q \frac{1}{2m_N} <p|\bar{u}u - \bar{d}d|p> \;. 
\eeq
Equations \eqref{eqn:ProtonNeutronElmContribution}, \eqref{eqn:ProtonNeutronQuarkContribution1} and \eqref{eqn:ProtonNeutronQuarkContribution2} can be combined into
$$
\Delta Q_N \simeq (-0.76 \Delta \ln \alpha + 2.05 \Delta \ln \delta_q) \mev 
$$
\beq
\label{eqn:QNDependence}
\Rightarrow \Delta \ln Q_N \simeq (-0.59 \Delta \ln \alpha + 1.59 \Delta \ln \delta_q) \, . 
\eeq

Subtracting the electromagnetic contributions, one is left with the pure QCD masses as given in Eqn~\eqref{eqn:nucleonQCDmasses}, which is far from only being the sum of the masses of the three valence quarks of the nucleons. A recent study using ``heavy baryon chiral perturbation theory'' \cite{Borasoy96} yields that the baryon mass in the chiral limit is
\beq
\stackrel{o}{m} = 767  \pm 110 \mev \, . 
\eeq
This quantity is purely due to QCD effects which only depend on $\lqcd$, and since we are holding this quantity fixed, $\stackrel{o}{m}$ remains constant. The contributions to the final nucleon mass come from the u, d, and s quarks, whose contributions can be estimated using the two quantities \cite{Borasoy96}
\beq
\sigma_{\pi N} (0) \define \hat{m} <p| \bar{u}u + \bar{d}d |p> = 45 \pm 10 \mev 
\eeq
\beq
y \define \frac{ 2<p|\bar{s}s|p>}{<p|\bar{u}u + \bar{d}d|p>} = 0.21 \pm 0.20 \; . 
\eeq
They are connected to the quark mass dependence of the nucleons via the Feynmann-Hellmann theorem \cite{Feynman39, Hellmann37},
\beq
\sigma_{\pi N}(0) = \hat{m} \frac{\partial m_N}{\partial \hat{m}} 
\eeq
and equivalently
\beq
<p|\bar{s}s|p> = \frac{ \partial{m_N}}{\partial m_s} \;. 
\eeq
This can be translated into
\beq
\label{eqn:NucleonDependence1}
\frac{ \partial \ln m_N}{\partial \ln \hat{m}} = m_N^{-1} \sigma_{\pi N}(0) = 0.048 \pm 0.011 
\eeq
\beq
\label{eqn:NucleonDependence2}
\frac{ \partial \ln m_N}{\partial \ln m_s} = \frac{m_s}{\hat{m}} \frac{y}{2} m_N^{-1} \sigma_{\pi N}(0) = 0.12 \pm 0.12  
\eeq
using $\frac{m_s}{\hat{m}} \approx 25$ \cite{PDG08}.

\subsection{Neutron lifetime}
\label{sec:NeutronLifetime}
In the electroweak model the neutron lifetime can be evaluated analytically. One derives \cite{WeinbergGRT} (see also \eqn \eqref{eqn:nDecayWithTemp} with zero temperature),
\beq
\tau_n^{-1} = \lambda_{n->p+e^- + \bar{\nu}_e} = \frac{1 + 3g_A^2}{2\pi^3} G_F^2 \int_{m_e}^{Q_N} x^2 \left( Q_N -x \right)^2 \sqrt{ 1 - \frac{m_e^2}{x^2}} d x \;. 
\eeq
Using $q \define \frac{Q_N}{m_e}$, this integral evaluates to
\beq
\lambda_{n->p+e^- + \bar{\nu}_e} = \frac{1 + 3g_A^2}{120\pi^3} G_F^2 m_e^5 \left[ \sqrt{q^2-1} \left( 2 q^4 - 9q^2 - 8 \right) + 15 q \ln \left( q + \sqrt{q^2 - 1} \right) \right] \;. 
\eeq
Here, $G_F$ is the Fermi coupling constant, which is related to the Higgs v.e.v.~via \eqn \eqref{eqn:RelationFermiConstHiggs} and $g_A$ is the nucleon axial vector coupling constant, $g_A \approx 1.27$ \cite{PDG08}. Recent developments in lattice QCD, combined with chiral perturbation theory, allow to compute $g_A$, yielding \cite{Hemmert03}
\beq
g_A = (1.2 \pm 0.1) - 6.9 \left(\frac{m_{\pi}}{1 \gev}\right)^2 \ln \frac{m_{\pi}}{1 \gev} - (10.4 \pm 4.8) m_{\pi}^2 \gev^{-2} + {\cal O}(m_{\pi}^3) \; . 
\eeq
Obviously the chiral value $\stackrel{o}{g_A} = 1.2 \pm 0.1$ dominates the physical value of $g_A$, and for the derivative we obtain
\beq
\frac{ \partial \ln \tau_n}{ \partial \ln m_{\pi}} = 0.006 \quad \Longrightarrow \quad \frac{ \partial \ln \tau_n}{ \partial \ln \hat{m}} = 0.003 \, . 
\eeq
Thus $g_A$ can be assumed constant in our treatment. Using the known dependences of $Q_N$ from the preceding section, we arrive at a fundamental parameter dependence of the neutron lifetime of
\beq
\Delta \ln \tau_n = 3.86 \Delta \ln \alpha + 4 \Delta \ln \vev{\phi} + 1.52 \Delta \ln m_e - 10.4 \Delta \ln \delta_q \; .
\eeq

\subsection{Binding energies}

The dependence of nuclear binding energies on the pion mass and $\alpha$ have been estimated in \cite{Pudliner97} and \cite{Pieper01} using quantum Monte Carlo calculations with realistic models of nuclear forces (similar values for the $\alpha$ dependence appear in \cite{Nollett02}). We use the Pudliner and Pieper values which give for the dependence of binding energies on $\alpha$:
\begin{multline}
 \Delta \ln (B_{\rm D}, B_{\rm T}, B_{{3}\rm He}, B_{{4}\rm He}, B_{{6}\rm Li}, B_{{7}\rm Li}, B_{{7}\rm Be}) = \\
 (-0.0081, -0.0047, -0.093, -0.030, -0.054, -0.046, -0.088) \Delta \ln \alpha.
\end{multline}

The pion mass determines the range of attractive nuclear forces, and the quantum Monte Carlo calculations of \cite{Pudliner97,Pieper01} which include pion exchange accurately reproduce many experimental properties. One-pion exchange and two-pion exchange are dominant contributions within the expectation values of the two- and three-nucleon potentials respectively. Currently such studies have not been extended to determine the functional dependence of binding energies on the pion mass in general \cite{Flambaum07,Damour07}. This dependence would in any case have uncertainties due to subleading effects of pion mass (or equivalently light quark masses) on other terms in the nucleon-nucleon potential \cite{BeaneSavage02}.

However, the dependence of the deuteron binding energy on the pion mass has been extensively studied within low-energy effective theory \cite{Epelbaum02,BeaneSavage02}: the result may be expressed as 
\beq
 \Delta \ln B_{\rm D} = r \Delta \ln m_\pi = \frac{r}{2} \Delta \ln \hat{m}
\eeq
for small variations about the current value \cite{YooScherrer02}, with $-10\leq r \leq -6$.\footnote{Our definition of $r$ differs by a sign from \cite{YooScherrer02}.} We will also take this dependence as a guide for the likely pion mass dependence of other binding energies. Although the size of the deuteron binding appears due to an accidental cancellation between attractive and repulsive forces, its derivative with respect to $m_\pi$ (which is just $B_{\rm D}/m_\pi$ times $r$) is not expected to be subject to any cancellation.
We also expect that the pion contribution to the total binding energy should increase with the number of nucleons; a proportionality to $(A-1)$ seems reasonable to obtain correct scaling at both small and large $A$\footnote{Flambaum \etal \cite{Flambaum07} obtain roughly a scaling of type $(A-1)$ for nuclei with $3 \le A \le 7$, but their results have a large uncertainty.}. Hence to estimate the effect of pion mass on the binding energy of a nucleus $B_i$ we set
\beq \label{eqn:dBdmpi}
 \frac{\partial B_i}{\partial m_\pi} = f_i (A_i-1) \frac{B_{\rm D}}{m_\pi} r \simeq -0.13 f_i (A_i-1) \; ,
\eeq
taking $r\simeq-8$. The numerical constants $f_i$ are expected to be of order unity, but will differ between light nuclei due to peculiarities of the shell structure, {\em etc}. Our normalization corresponds to $f_{\rm D}=1$. 
Then the nontrivial dependences of nuclear parameters on $\hat{m}$ are
\begin{multline}
 \Delta \ln (B_{\rm D}, B_{\rm T}, B_{3\rm He}, B_{4\rm He}, B_{6\rm Li}, B_{7\rm Li}, B_{7\rm Be}, m_N) 
 \simeq \\
 (0.5r, 0.26f_{\rm T}r, 0.29f_{3\rm He}r, 0.12f_{4\rm He}r, 0.17f_{6\rm Li}r, 
 0.17f_{7\rm Li}r, 0.18f_{7\rm Be}r, 0.048) \Delta \ln \hat{m} \; ,
\end{multline}
where the dependence of $m_N$ is taken from \eqn \eqref{eqn:NucleonDependence1}. For the $\hat{m}$ dependence of abundances due to the variation of binding energies we then have
\beq
 \left.\frac{\partial \ln Y_a}{\partial \ln \hat{m}}\right|_{B} = \frac{r}{2}
 \sum_i f_i \frac{(A_i-1)B_{\rm D}}{B_i}
 \frac{\partial \ln Y_a}{\partial \ln B_i} \; ,
\eeq
where the dependence $\frac{\partial \ln Y_a}{\partial \ln B_i}$ is obtained from the BBN code by varying the binding energies $B_i$.
Taking account also of the small effect of $\hat{m}$ on the nucleon mass $m_N$, the resulting dependence of abundances on $\hat{m}$ is 
\bea
 \frac{\partial \ln Y_{\rm D}}{\partial \ln \hat{m}} &\simeq& 
 11  
 + 0.5 f_T + 5 f_{3\rm He} \nonumber \\
 \frac{\partial \ln Y_{3\rm He}}{\partial \ln \hat{m}} &\simeq& 
 8 %f_D 
 + 3 f_T - 7 f_{3\rm He} \nonumber \\
 \frac{\partial \ln Y_{4\rm He}}{\partial \ln \hat{m}} &\simeq& 
 -2.7 %f_D 
 \nonumber \\
 \frac{\partial \ln Y_{6\rm Li}}{\partial \ln \hat{m}} &\simeq& 
 27 %f_D
 + 0.4 f_{\rm T} + 7 f_{\rm 3He} + 55 f_{4\rm He} - 96 f_{6\rm Li} \nonumber \\
 \frac{\partial \ln Y_{7\rm Li}}{\partial \ln \hat{m}} &\simeq& 
 - 36 %f_D
 + 5 f_{\rm T} + 22 f_{3\rm He} + 54 f_{4\rm He} + 9 f_{7\rm Li} -115 f_{7\rm Be} \; ,
\eea
where we have neglected subleading terms. Even if we consider that some contributions could cancel against one another due to the values of the $f_i$, the magnitude of these variations is striking, particularly concerning the lithium abundances. To get an idea of the possible effect of cancellations, we may set all $f_i$ to unity and find the dependences
\beq
\Delta \ln (Y_{\rm D}, Y_{3\rm He}, Y_{4\rm He}, Y_{6\rm Li}, Y_{7\rm Li}) \simeq 
(17, 5, -2.7, -6, -61)
\Delta \ln \hat{m}.
\eeq
One may also consider to what extent varying $\hat{m}$ or the pion mass may affect reaction cross-sections beyond the $npd\gamma$ reaction. It seems very likely that matrix elements would acquire nontrivial dependence on $m_\pi$; however, since the dependence of abundances on reaction cross-sections is relatively mild (see Table~\ref{tab:importantReactions}),
the dependence via reaction matrix elements is unlikely to compete with the very large effects arising through the variation of binding energies.

\newpage

\section{The response matrices}

The results of the previous section can be summarized in the response matrix $F$ defined in \eqn \eqref{eq:Fdef} which relates variations in fundamental parameters $G_k$ to variations in nuclear parameters $X_i$. Its values are shown in \tab \ref{tab:dlnXdlnG}.
\begin{table}
\centering
\begin{tabular}{|c||c|c|c|c|c|c|}
\hline 
$\partial \ln X_i/\partial\ln G_k$ & $\gnewton$ & $\alpha$ & $\vev{\phi}$ & $m_e$ & $\delta_q$ & $\hat{m}$ \\
\hline \hline
$\gnewton$ & 1 & 0 & 0 & 0 & 0 & 0 \\
\hline 
$\alpha$ & 0 & 1 & 0 & 0 & 0 & 0 \\
\hline
$\tau_n$ & 0 & 3.86 & 4 & 1.52 & -10.4 & 0 \\
\hline
$m_e$ & 0 & 0 & 0 & 1 & 0 & 0 \\
\hline
$Q_N$ & 0 & -0.59 & 0 & 0 & 1.59 & 0 \\
\hline
$m_N$ & 0 & 0 & 0 & 0 & 0 & 0.048 \\
\hline
$B_{\rm D}$ & 0 & -0.0081 & 0 & 0 & 0 & $-4$ \\
\hline
$B_{\rm T}$ & 0 & -0.0047 & 0 & 0 & 0 & $-2.1 f_{\rm T}$ \\
\hline
$B_{3\rm He}$ & 0 & -0.093 & 0 & 0 & 0 & $-2.3 f_{3\rm He}$ \\
\hline
$B_{4\rm He}$ & 0 & -0.030 & 0 & 0 & 0 & $-0.94 f_{4\rm He}$ \\
\hline
$B_{6\rm Li}$ & 0 & -0.054 & 0 & 0 & 0 & $-1.4 f_{6\rm Li}$ \\
\hline
$B_{7\rm Li}$ & 0 & -0.046 & 0 & 0 & 0 & $-1.4 f_{7\rm Li}$ \\
\hline
$B_{7\rm Be}$ & 0 & -0.088 & 0 & 0 & 0 & $-1.4 f_{7\rm Be}$ \\
\hline
\end{tabular}
\caption[Response matrix $F$, dependence of nuclear on fundamental parameters]{Response matrix $F$, dependence of nuclear parameters $X_i$ on fundamental parameters $G_k$}\label{tab:dlnXdlnG}
\end{table}
These results can be combined with the nuclear response matrix, \tab \ref{tab:dlnYdlnX}, to the ``fundamental response matrix'' $R$ defined in \eqn \eqref{eqn:Rdef} according to \eqn \eqref{eq:RequalsCF}. The matrix $R$ relates variations in fundamental parameters to variations in final abundances and is shown in \tab \ref{tab:dlnYdlnG}. This table is the central result of this part of the thesis.
\begin{table}
\centering
\begin{tabular}{|c||c|c|c|c|c|}
\hline 
$\partial \ln Y_a/\partial\ln G_k$ &  D   & \heth & \hefo & \lisi  & \lise \\ \hline \hline
$\gnewton$                        & 0.94 & 0.33  & 0.36  &  1.4   & -0.72 \\ \hline 
$\alpha$                           & 3.6  & 0.95  & 1.9   &  6.6   &-11    \\ \hline
$\vev{\phi}$                       & 1.6  & 0.60  & 2.9   &  5.5   &  1.7  \\ \hline
$m_e$                              & 0.46 & 0.21  & 0.40  &  0.97  & -0.17 \\ \hline
$\delta_q$                         &-2.9  &-1.1   &-5.1   & -9.7   & -2.9  \\ \hline
$\hat{m}$                          &17    & 5.0   &-2.7   & -6     &-61    \\ \hline
\hline
$\eta$                             &-1.6  &-0.57  & 0.04  & -1.5   &  2.1  \\ \hline
\end{tabular}
\caption[Response matrix $R$, dependence of abundances on fundamental parameters]{Response matrix $R$, dependence of abundances $Y_i$ on fundamental parameters $G_k$. All $f_i$ are set to unity.}\label{tab:dlnYdlnG}
\end{table}
In treating the $\hat{m}$-dependences, which arise from the nuclear binding energies with their uncertain values of $f_i$, we have given the values which arise when setting all $f_i$ to unity. In our further treatment of BBN we will neglect the model uncertainty\footnote{Given the current status of low-energy QCD, it seems hard to quantify possible ranges of uncertainty for the $f_i$.} in the binding energies and always assume $f_i \equiv 1$.

\section{Comparison to other studies}

The \hefo\ dependence was previously calculated in \cite{MSW04} by semi-analytic methods \cite{Esmailzadeh91}. Their findings are displayed in Tab.~\ref{tab:MSWResults}, our results for the dependence on fundamental parameters are similar.
\begin{table}
\centering
\begin{tabular}{|c||c|c|c|c|c|c|}
\hline 
$\partial \ln Y_a/\partial\ln G_k$ &  $\gnewton$   & $\alpha$ & $\vev{\phi}$  & $m_e$ & $\delta_q$ & $\hat{m}$\\ \hline 
\hefo                     & 0.41 & 1.94  & 3.36  &  0.389  & -5.358 & -1.59  \\ \hline
\end{tabular}
\caption[Dependence of abundances on fundamental parameters found by \cite{MSW04}]{Dependence of abundances $Y_i$ on fundamental parameters $G_k$ found by \cite{MSW04}}\label{tab:MSWResults}
\end{table}
The dependence on $\gnewton$ can be compared with the results of \cite{Scherrer03} and \cite{Chamoun05,Landau04}, once one translates from units where $\gnewton$ is constant to ours where $\lqcd$ is constant. The latter ones also give values for the dependence on $\vev{\phi}$ (variation of $G_{\rm F}\propto \vev{\phi}^{-2}$) and on $\alpha$. Their results are shown in Tab.~\ref{tab:Landau04Results}. They roughly match with our estimates most times except for a few numbers, \eg the much smaller variations of D and \heth\, under variations of the Higgs v.e.v. We assume that these deviations are due to the applied semi-analytical method of \cite{Esmailzadeh91} which might not give appropriate results in the case of abundances other than \hefo.
\begin{table}
\centering
\begin{tabular}{|c||c|c|c|c|c|}
\hline 
$\partial \ln Y_a/\partial\ln G_k$ &  D   & \heth & \hefo & \lisi  & \lise \\ \hline \hline
$\gnewton$                        & 0.4 & 0.6  & 0.3  &  0.3   & -1.0 \\ \hline 
$\alpha$                           & 2.3  & 1.0  & 2.3   &  7.4   &-9.5    \\ \hline
$\vev{\phi}$                       & 0.07  & 0.1  & 3.1   &  4.1   &  $\approx 0.5$  \\ \hline
\end{tabular}
\caption[Dependence of abundances on fundamental parameters from \cite{Chamoun05} and \cite{Landau04}]{Dependence of abundances $Y_i$ on fundamental parameters from \cite{Chamoun05} and \cite{Landau04}}\label{tab:Landau04Results}
\end{table}
\begin{table}
\centering
\begin{tabular}{|c||c|c|c|c|c|}
\hline 
$\partial \ln Y_a/\partial\ln G_k$ &  D   & \heth & \hefo & \lisi  & \lise \\ \hline \hline
$\alpha$                           & $\approx 3.5$  & $\approx 0.9$  & $\approx 2.0$   &  -   & $\approx -7$    \\ \hline
\end{tabular}
\caption[Dependence of abundances on $\alpha$ from \cite{Bergstrom99} and \cite{Nollett02}]{Dependence of abundances $Y_i$ on $\alpha$ from \cite{Bergstrom99} and \cite{Nollett02}}\label{tab:BergstromNollettResults}
\end{table}
Bergstr\"om and Nollett \cite{Bergstrom99, Nollett02} use the full BBN code \cite{Kawano92} to constrain the variation of $\alpha$, incorporating $\alpha$-dependent reaction cross-sections as described in \sect \ref{sec:Sfactor}. From their graphical results we extract the dependences given in Tab.~\ref{tab:BergstromNollettResults} which again match with our findings.

\chapter[Constraints on variations]{Constraints on variations of fundamental parameters}
\label{chap:BBNobsVStheo}

\section{Bounds on separate variations of fundamental couplings}
\label{sec:BoundsOnSeparateVariations}
The first application of our results from the preceding chapter is in setting bounds on the variation of each fundamental parameter considered separately, under the assumption that only one parameter varies at once. We may consider three observational determinations of primordial abundances (see \sect \ref{sec:ObsTheoBBN}): deuterium, \hefo\ and \lise. However, the observed \lise\ abundance deviates by a factor two to three from the value predicted by standard BBN theory (SBBN), and systematic uncertainties related to stellar evolution exist \cite{Korn06}. Thus, we use the former two, D and \hefo, to constrain the allowed variations of the fundamental constants individually.
For deuterium we take $2 \sigma$ limits;
for \hefo\ we consider instead the ``conservative allowable range'' of \cite{OliveSkillman04}. The resulting constraints are given in \tab \ref{tab:AllowedVariations}.
\begin{table}
\centering
\begin{tabular}{|clcc|}
\hline
$-19 \% $&$\le\ \Delta \ln \gnewton        $&$\le$&$ +10 \%  $\\ \hline
$-3.6\% $&$\le\ \Delta \ln \alpha     $&$\le$&$ +1.9\%  $\\ \hline
$-2.3\% $&$\le\ \Delta \ln \vev{\phi} $&$\le$&$ +1.2\%  $\\ \hline
$-17 \% $&$\le\ \Delta \ln m_e        $&$\le$&$ +9.0\%  $\\ \hline
$-0.7\% $&$\le\ \Delta \ln \delta_q   $&$\le$&$ +1.3\%  $\\ \hline
$-1.3\% $&$\le\ \Delta \ln \hat{m}    $&$\le$&$ +1.7\%  $\\
\hline
\end{tabular}
\caption[Allowed individual variations of fundamental couplings]{Allowed individual variations ($2\sigma$ or ``conservative allowable range'', see \sect \ref{sec:ObsTheoBBN}) of fundamental couplings}\label{tab:AllowedVariations}
\end{table}
 
\section{Variations of abundances in unified models}
\label{sec:VariationsOfAbundancesInUnifiedModels}
A major problem of using BBN to constrain variations of fundamental parameters is degeneracy. There are only three observational abundances (D, \hefo\ and \lise) but six fundamental parameters which can vary. Hence, the three observed values cannot be used to constrain a set of six parameters.
In Chapter~\ref{chap:SUSYandGUT} we have introduced unified models where the variations of fundamental couplings satisfy relations that reduce the number of free parameters. Here we apply GUT models in order to be able to visualize our findings for BBN. 
In the simplest case every variation of a parameter $G_k$ is determined by a single underlying degree of freedom, for instance a variation of the unified coupling $\Delta \ln \alphagut$. We may also eliminate $\Delta \ln \alphagut$ in favor of some observable parameter, which we will choose to be the fine structure constant $\alpha$.

For grand unified theories, it makes sense to change from a system of units with constant $\lqcd$ to a system where $\mgut$, the mass scale of the GUT, is set to constant. Furthermore, for the scenarios we will use in this section, we will take the Planck mass fixed relative to the unification scale, thus $\Delta (\mplanck/\mgut)=0$. We set all $f_i$ to unity and, for simplicity, we take the Yukawa couplings to be constant\footnote{See also the comments on variations of the Yukawa couplings in \sect \ref{sec:GUTrels}.}, thus the electron and quark masses are proportional to $\vev{\phi}$,
\beq
 \Delta \ln \frac{m_e}{\mgut} = \Delta \ln \frac{\delta_q}{\mgut} = \Delta \ln \frac{\hat{m}}{\mgut} = \Delta \ln \frac{\vev{\phi}}{\mgut} \, .
\eeq
Finally we define an exponent $\gamma$ which relates the variation of $\vev{\phi}$ with respect to $\mgut$ to the variation of $\lqcd/\mgut$ as
\beq
 \frac{\vev{\phi}}{\mgut} = \mbox{const.}\left(\frac{\lqcd}{\mgut}\right)^{\gamma}.
\eeq

We study three non-supersymmetric GUT scenarios, $\gamma = 0$, $\gamma = 1$ and $\gamma = 1.5$, and use the fine structure constant $\alpha$ to parametrize the variations. As has been shown in \sect \ref{sec:VariationsInGutFramework}, the variation of $\alpha$ can be related to variations of particle masses, the GUT coupling $\alphagut$ and the QCD invariant scale $\lqcd$, see equation \eqref{eqn:dlnAlpha}. 

\subsubsection{Scenario $\gamma = 0$}
In the first scenario, $\gamma = 0$, the Higgs v.e.v.\ and elementary fermion masses are all proportional to the unification scale, $\Delta(\mgut/\mplanck,\vev{\phi}/\mplanck,m_{e,q}/\mplanck)=0$. Then the variations of fundamental couplings are related as
\beq 
 \Delta \ln \frac{\lqcd}{\mgut} = \Delta \ln \frac{\lqcd}{\vev{\phi}}
 = \frac{3\pi}{40\alpha} \Delta \ln \alpha \simeq 32.3 \Delta \ln \alpha \, .
\eeq
In a system with constant $\lqcd$ we get
\beq \label{dGGUT2}
 \Delta \ln(\gnewton,\alpha,\vev{\phi},m_e,\delta_q,\hat{m}) \simeq (64.5,1,-32.3,-32.3,-32.3,-32.3) \Delta \ln \alpha. 
\eeq
We then obtain variations of abundances
\beq
 \Delta \ln (Y_{\rm D}, Y_{3\rm He}, Y_{4\rm He}, Y_{6\rm Li}, Y_{7\rm Li}) \simeq 
 (-450, -130, 170, 380, 1960) \Delta \ln \alpha \; .
\eeq

\subsubsection{Scenario $\gamma = 1$}
In the second scenario with $\gamma = 1$, all low-energy mass scales of particle physics are proportional to $\lqcd$. Setting $\lqcd$ constant, we have
\beq \label{dGGUT1}
 \Delta \ln(\gnewton,\alpha,\vev{\phi},m_e,\delta_q,\hat{m}) \simeq (78,1,0,0,0,0) \Delta \ln \alpha.
\eeq
In this case the variations of abundances are not subject to the theoretical uncertainty of varying $m_s/\lqcd$. 
We obtain 
\beq
 \Delta \ln (Y_{\rm D}, Y_{3\rm He}, Y_{4\rm He}, Y_{6\rm Li}, Y_{7\rm Li}) \simeq 
 (77, 27, 30, 120, -68) \Delta \ln \alpha \, .
\eeq

\subsubsection{Scenario $\gamma = 1.5$}
In the third unified scenario with $\gamma = 1.5$ we consider the case when the Higgs v.e.v.\ and fermion masses vary {\em more}\/ rapidly (with respect to the unification scale) than the QCD scale $\lqcd$ does. Converting back to a system where $\lqcd$ is constant, we find that the variations of fundamental couplings are related as
\beq \label{dGGUT3}
 \Delta \ln(\gnewton,\alpha,\vev{\phi},m_e,\delta_q,\hat{m}) \simeq (87,1,21.5,21.5,21.5,21.5) \Delta \ln \alpha. 
\eeq
The variations of abundances are then
\beq
 \Delta \ln (Y_{\rm D}, Y_{3\rm He}, Y_{4\rm He}, Y_{6\rm Li}, Y_{7\rm Li}) \simeq 
 (430,130,-65,-60,-1420) \Delta \ln \alpha.
\eeq

\subsection{Linear results}
\begin{figure} 
\begin{center}
\includegraphics[width=7.5cm]{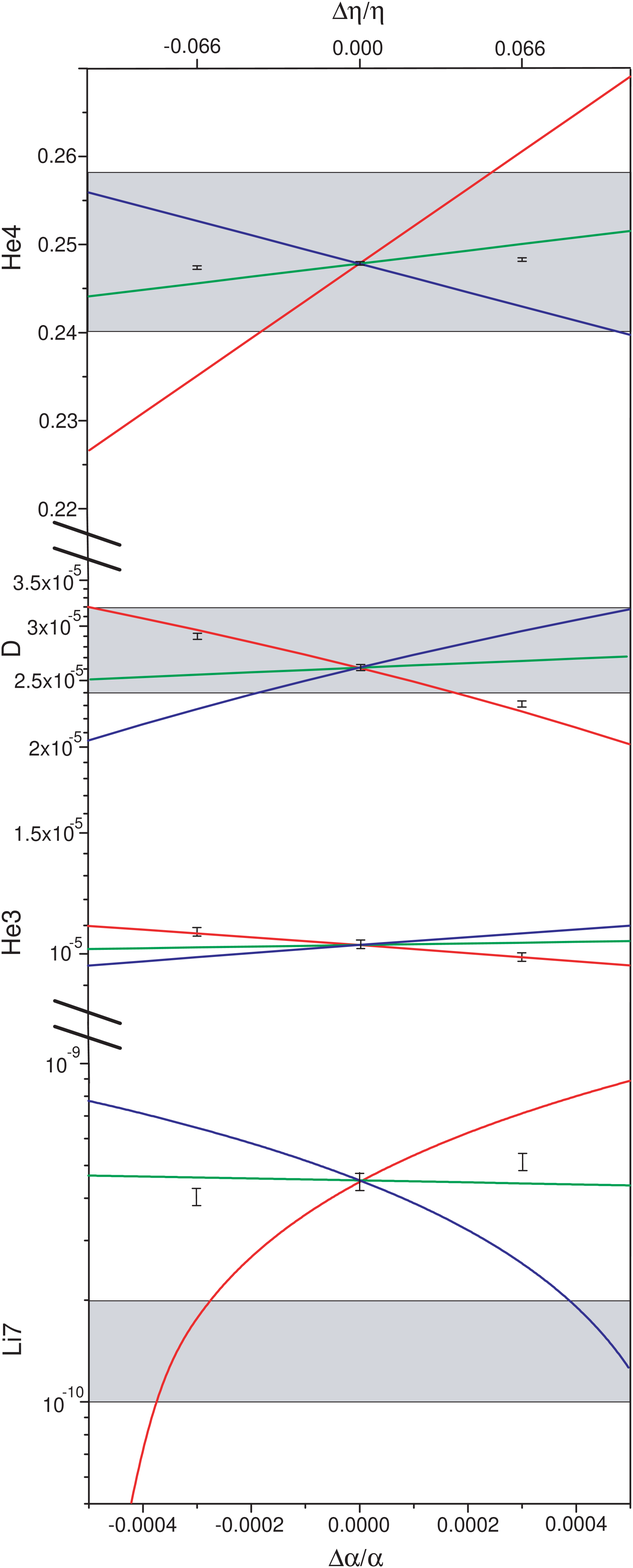}
\end{center}
\caption[Variation of primordial abundances with $\alpha$ in three GUT scenarios]{Variation of primordial abundances with $\alpha$ in three GUT scenarios. Red lines show the first $\gamma=0$ scenario; green lines the second $\gamma = 1$ and blue lines the third $\gamma=1.5$ scenario. Highlighted regions give the observational 1$\sigma$ limits (as explained in \sect \ref{Sec:BBNHe3}, no observational limits can be given for \heth). Error bars indicate the standard BBN abundances with theoretical 1$\sigma$ error \cite{Serpico04}, for three different values of $\eta$ about the WMAP central value, as indicated on the upper horizontal axis.}\label{fig:VaryGUT}
\end{figure}

In Fig.~\ref{fig:VaryGUT} we show the abundance variations given by the three GUT models, as a function of the variation of $\alpha$. First we plot only the linear dependence of abundances on $\alpha$ as given by the fundamental response matrix $R$ (Tab.~\ref{tab:dlnYdlnG}). We also show the 1$\sigma$ observational bounds as highlighted regions. Also included in the plot is the effect on the standard BBN predictions of varying the baryon-to-photon ratio $\eta$ over the 2$\sigma$ range allowed by WMAP 3 year data\footnote{Updating to WMAP5 values does not lead to any significant change, there the $2\sigma$ range is $5.9 \leq 10^{10} \eta \leq 6.5$.}, $5.7\leq 10^{10}\eta \leq 6.5$.

It can be seen that in the $\gamma=0$ scenario a reduction of $\alpha$ by about $0.025 \%$ (\ie a fractional variation of $-2.5\times 10^{-4}$) would bring theory and observation into agreement within 2$\sigma$ bounds, while remaining in the linear regime. Conversely, in the $\gamma=1.5$ model an increase of $\alpha$ by about $0.04 \%$, \ie $\Delta \ln \alpha = 4\times 10^{-4}$, brings theory and observation into agreement within 1$\sigma$ bounds. 
Considering the variations of fundamental parameters in the three scenarios Eqns.~\eqref{dGGUT2}, \eqref{dGGUT1} and \eqref{dGGUT3}, the behavior of the weak scale $\vev{\phi}$ and fermion masses is decisive for the variation of abundances. However, if $\Delta \ln Y_a$ becomes larger than 1 (as in the case of \lise) the results are affected by higher order terms and the linear approximation can no longer be applied.

\subsection{Nonlinear results} 
\label{sec:nonlinear}
The linear analysis in the previous section suggests that it is possible, and may even be natural, to obtain a large negative variation in the \lise\ abundance, and considerably smaller variations in other measurable abundances: positive in the case of deuterium and negative for \hefo. Agreement between theory and data in all three abundances could then be possible for a narrow range of values in the variation of fundamental parameters. The required fractional variation in \lise\ is so large (a factor two or more in $Y_{7\rm Li}$) that a linear analysis using matrix multiplication is inaccurate.

We improve the analysis by including the relations between nuclear and fundamental parameters and the three GUT relations for the fundamental parameters in our BBN simulation code. Then we run the code for a set of parameters $\Delta \ln \alpha$ and obtain the full nonlinear parameter dependence of abundances on variations of fundamental parameters. Note that this method is impractical to investigating the full parameter space: fundamental parameters span a six-dimensional and nuclear parameters a 13-dimensional parameter space which cannot be analyzed numerically due to time reasons. It is only practicable if the dimensionality of the parameter space is reduced by applying unification relations. 

For most nuclear parameters $X_i$ the dependence on fundamental parameters $G_k$ is only known to linear order or the nonlinear dependence involves a high uncertainty. However, we tested that for the unified models considered here, the fractional variations in the nuclear parameters $X_i$ remain small, well below $0.1$. A linear approximation for the relation between nuclear and fundamental parameters is therefore appropriate. The main nonlinear effects enter at the level of nuclear reactions, in equations where nuclear parameters as binding energies enter in a known power law dependence.

The nuclear parameters affecting most the large variation in \lise\ abundance are mainly the deuterium and \bese\ binding energies, with the \heth\ and \hefo\ binding energies playing a smaller role. A decrease of $B_{\rm D}$ causes BBN to happen later, which means that the nucleon density is lower and reaction rates smaller. The abundances of $A>4$ elements are rate-limited and thus decrease with decreasing $B_{\rm D}$. This accounts for about two-thirds of the change in $Y_{7\rm Li}$. In addition, the cross-section of the \heth$(\alpha,\gamma)$\bese\ reaction depends strongly on the $Q$-value, hence on the \bese\ binding energy. Both these effects are computationally under control, therefore we believe that the specific nonlinear dependence in the scenarios we consider is well estimated within our code.
\begin{figure} 
\begin{center}
\includegraphics[width=7.5cm]{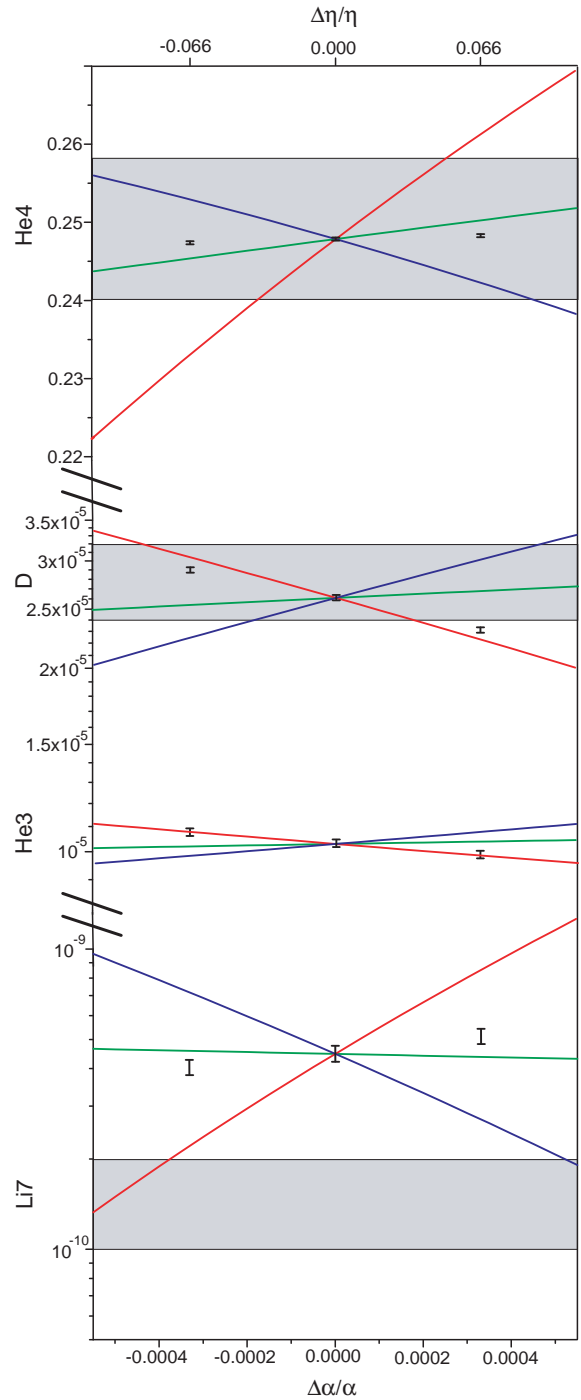}
\end{center}
\caption[Variation of primordial abundances with $\alpha$ in three GUT scenarios including nonlinear effects]{Variation of primordial abundances with $\alpha$ in three GUT scenarios including nonlinear effects. Labels as in Fig.~\ref{fig:VaryGUT}.}\label{fig:VaryGUTnonlin}
\end{figure} 
We show in \fig \ref{fig:VaryGUTnonlin} the primordial abundances including nonlinear effects, \ie without using a linear approximation for the relation between $Y_a$ and $X_i$. For our three GUT models we find a slightly different behavior of the \lise\ abundance, which now has an approximately power-law dependence on variation of $\alpha$. It is only slightly more difficult to bring the present observational abundances into agreement with standard BBN and the WMAP determination of $\eta$ ; still, if we allow a variation of $0.00045 \lesssim \Delta \ln \alpha \lesssim 0.0005$ in the $\gamma=1.5$ model, the predicted abundances are all very close to the 1$\sigma$ allowed regions.

%%%%%%%%%%%%%%%%%%%%%%%%%%%%%%%%%%%%%%%%%%%%%%%%%%%%%%%%%%%%%%%%%%%%%%%%%%%%%
%%%%%%%%%%%%%%%%%%%%%%%%%%%%%    Part 3     %%%%%%%%%%%%%%%%%%%%%%%%%%%
%%%%%%%%%%%%%%%%%%%%%%%%%%%%%%%%%%%%%%%%%%%%%%%%%%%%%%%%%%%%%%%%%%%%%%%%%%%%%
\part{Unifying cosmological and late-time variations}

%%%%%%%%%%%%%%%%%%%%%%%%%%%%%%%%%%%%%%%%%%%%%%%%%%%%%%%%%%%%%%%%%%%%%%%%%%%%%
%%%%%%%%%%%%%%%%%%%    Experimental tests of variations     %%%%%%%%%%%%%%%%%
%%%%%%%%%%%%%%%%%%%%%%%%%%%%%%%%%%%%%%%%%%%%%%%%%%%%%%%%%%%%%%%%%%%%%%%%%%%%%

\chapter{Experimental tests of variations}
\label{chap:ExpTestofVariations}
In recent years possible variations in the constants of nature have been tested in various ways. Whilst direct laboratory measurements do not point towards any variation, some astrophysical tests yield slight variations. Still, those non-zero results neither constitute compelling evidence for variations since the applied fitting and analysis methods are still under debate, nor do the different non-zero results seem to be consistent with each other\footnote{For a thorough comparison of measurements of variations, a more profound theoretical background is needed, as we will lay out in the rest of this thesis.}. Hence the question if there are and have been variations remains open from the experimental side. 

Here we review and discuss the observational data that we will consider in our effort to obtain a unified picture of time variation of couplings. We summarize the results that are most relevant for our analysis in \tab \ref{tab:CollectionOfMeasurements}.

\section{BBN}
The earliest processes for which Standard Model physics can be tested is BBN ($z \sim 10^{10}$). The influence of varying constants on BBN has been laid out in chapters \ref{chap:BBNvaryingNucl} to \ref{chap:BBNobsVStheo}. In an extension of our previous treatment, we include in \sect \ref{sec:CMBeta} the possible effect of varying constants at last scattering (formation of the CMB) on the input parameter $\eta$ of our BBN procedure.

The uncertainty in the $\eta$ determination, $\eta = (6.20 \pm 0.16) \times 10^{-10}$ (WMAP5 plus BAO and SN, \cite{WMAP5} ) yields a further correlated error for the abundances, which can be treated using the method of \cite{Fiorentini98}. For any given set of fundamental variations we can define
\beq
 \chi^2 \equiv \sum_{i,j} (Y_i - Y_i^{obs}) w_{ij} (Y_j - Y_j^{obs}),
\eeq
with the inverse weight matrix
\beq
 w_{ij} = \left[ \sigma^{2, \eta}_{ij} + \delta_{ij} (\sigma^2_{obs,i} +
 \sigma^2_{th,i}) \right]^{-1},
\eeq
where
\beq
 \sigma^{2, \eta}_{ij} \equiv Y_i Y_j \frac{ \partial \ln Y_i}{\partial \ln \eta}
 \frac{ \partial \ln Y_j}{\partial \ln \eta} 
 \left( \frac{ \Delta \eta}{\eta} \right)^2.
\eeq
We take, as in Tab.~\ref{tab:dlnYdlnX},
\beq
 \frac{\partial \ln ({\rm D/H}, Y_{\rm p}, \mbox{\lise/H})}{\partial \ln \eta} =
 (-1.6,\ 0.04,\ 2.1).
\eeq
The $1(2) \sigma$ error contour is given by $\chi^2/\nu \le 1(4)$ where $\nu$ is the number of degrees of freedom. As the final abundances depend on variations of all fundamental constants, we have to evaluate the variations allowed by BBN for every model separately.

In the light of complex astrophysics which may affect the extraction of the primordial \lise\ fraction, we also consider bounding the variations using deuterium and \hefo\ alone (see also \sect \ref{sec:BoundsOnSeparateVariations}). This yields a value consistent with zero for variations at BBN, since these abundances are consistent with standard BBN.

\section{CMB}
A further far-reaching test of varying constants is the cosmic microwave background (CMB) ($z \sim 10^3$). In principle, $\alpha$ and $\gnewton$ are bounded by CMB observations. A variation of $\alpha$ affects the formation of the CMB through Thomson scattering and the recombination history. However, bounds on variations from CMB are typically very weak as there are significant degeneracies with other cosmological parameters \cite{Martins03,Rocha03}, see also the discussion in \sect \ref{sec:CMBeta}. Current bounds on $\alpha$ are \cite{Rocha03}
\beq
 0.95<\frac{\alpha_{\rm CMB}}{\alpha_0}<1.02 \qquad (2\sigma). 
\eeq
The CMB anisotropies may also be used to constrain the variation of Newton's constant $\gnewton$. The resulting bound depends on the form of the variation of $\gnewton$
from the time of CMB to now. Using a step function one finds \cite{Chan07,Zahn02}
\beq 
 0.95 \le \frac{\gnewton}{{\gnewton}_{,0}} \le 1.05 \qquad (2 \sigma), 
\eeq
where the 
instantaneous change in $\gnewton$ may happen at any time between now and CMB decoupling. 
Using instead a linear function of the scale factor $a$, the bound is
\beq 
 0.89 \le \frac{\gnewton}{{\gnewton}_{,0}} \le 1.13 \qquad (2 \sigma) .
\eeq
Note that here, as in most studies of time-dependent $\gnewton$, units are implicitly defined such that the elementary particle masses (and thus the mass of gravitating bodies, if gravitational self-energy is neglected) are constant. The relevant bound on {\em dimensionless}\/ parameters concerns $\gnewton m_N^2 \equiv (\mnucleon/\mplanck)^2 (8 \pi)^{-1}$. 

\newpage

\subsection{Effect of ``varying constants'' at CMB and $\eta$} 
\label{sec:CMBeta}

In our study on BBN we used the WMAP determination of the baryon number density parameter $\eta\equiv n_B/n_\gamma$ directly to reduce by one the number of unknown parameters. However, we should also consider the effect of possible variations of fundamental parameters at the epoch of CMB decoupling, as CMB measurements are used to derive the value of $\eta$. Hence, it seems important to study how possible variations at the epoch of CMB decoupling affect the determination of $\eta$ and hence also the outcomes of BBN simulations.  

Fundamental parameters affecting the CMB are the proton and electron masses, the gravitational constant and the fine structure constant, as well as the mass of any dark matter particle present. In Planck units, these reduce to the particle masses and $\alpha$. The relevant cosmological parameters are the amplitude, spectral index (and possible running, \etc) of primordial perturbations; the baryon, dark matter, dark energy (cosmological constant, \etc) and curvature densities normalized to the critical density; the Hubble constant; and the reionization optical depth. Of these, the baryon density $\OmegaBaryon h^2$ will vary linearly with the proton mass in Planck units, for a fixed baryon-to-photon ratio $\eta$. Conversely, given a measurement of $\OmegaBaryon h^2$, the correct value of $\eta$ varies inversely with the proton mass. The conversion factor between $\OmegaBaryon h^2$ and $\eta_{10}\equiv 10^{10} \eta$ is then (see \eqn \eqref{eqn:EtaOmegaB})
\beq
273.9 (\mproton\sqrt{\gnewton})_{|0} (\mproton\sqrt{\gnewton})^{-1} \simeq 273.9 (1-\Delta\ln(\mnucleon/\mplanck)_{|\rm CMB} ) \, ,
\eeq
where we approximate the proton and neutron masses by their average $\mnucleon$.

If, therefore, we allow the proton mass (or the gravitational constant, in QCD units) to vary arbitrarily at the CMB epoch, $\eta$ is undetermined by WMAP and we must consider it as an extra free parameter or try to impose independent cosmological bounds. However, we impose that the size of variations away from the present value of $\mproton/\mplanck$ is a monotonically decreasing function of time: thus $\Delta\ln(m_N/M_{\rm P})_{|\rm CMB}\leq \Delta\ln(m_N/M_{\rm P})_{|\rm BBN}$. Hence we would have a self-consistent treatment of this parameter if the secondary discrepancies in primordial abundances due to an incorrectly estimated $\eta$ were smaller than the primary effect of varying $m_N/M_{\rm P}$ at BBN. The maximum effect due to rescaling of $\eta$ would occur when $\Delta \ln (m_N/M_{\rm P})_{|\rm CMB}$ = $\Delta \ln (m_N/M_{\rm P})_{|\rm BBN}$, adding the nucleon mass induced variation in $\eta$,
\beq
 \Delta \ln \eta = \Delta \ln (\mnucleon/\mplanck)_{|\rm BBN} \; . 
\eeq
In our treatment of BBN in the next chapter, we will consider this effect by studying two limiting cases. First, when $\Delta (\mnucleon/\mplanck)_{|\rm CMB} \ll \Delta (\mnucleon/\mplanck)_{|\rm BBN}$, then our previous results hold. In the second case, with $\Delta (\mnucleon/\mplanck)_{|\rm CMB} \simeq \Delta (\mnucleon/\mplanck)_{|\rm BBN}$, the value of $\eta$ may be significantly rescaled.

\section{Quasar absorption spectra}

The observation of absorption spectra of distant interstellar clouds allows to probe atomic physics over large time scales. Comparing observed spectra with the spectra observed in the laboratory, together with the in general well known dependence of the spectra on fundamental constants, gives bounds on the possible variation of couplings. Different kinds of spectra (atomic, molecular, \ldots) are sensitive to different parameters, which we will list in the following paragraphs.

\paragraph{Tests of $\alpha$}
\ \\
Atomic spectra are primarily sensitive to $\alpha$. Several groups using various methods of modeling and numerical analysis have published results; we quote here only the latest bounds.
Murphy and collaborators \cite{Murphy03.2} studied the spectra of 143 quasar absorption systems over the redshift range $0.2<z_{abs}<4.2$. Their most robust estimate is a weighted mean 
\beq
  \frac{\Delta \alpha}{\alpha} = (-0.57 \pm 0.11) \times 10^{-5}.
\eeq
Dividing the data into low ($ z < 1.8$) and high ($z > 1.8$) redshift subsamples, they obtain
\begin{align}
z &< 1.8, & N_{sys} &= 77, & \vev{z_{abs}} &= 1.07, & \frac{\Delta\alpha}{\alpha} &= (-0.54 \pm 0.12) \times 10^{-5} \nonumber\\
z &> 1.8, & N_{sys} &= 66, & \vev{z_{abs}} &= 2.55, & \frac{\Delta\alpha}{\alpha} &= (-0.74 \pm 0.17 ) \times 10^{-5},
\end{align}
where $N_{sys}$ is the number of absorption systems in the sample and $\vev{z_{abs}}$ is the averaged sample redshift.

In discussing unified models in \sect \ref{sec:Epochs_and_evolutionFactors}, we will define various ``epochs'' for the purpose of collating data and comparing them with models over certain ranges of redshift. The 143 data points are then assigned to different epochs: we choose to put boundaries at $z=0.81$ and $z=2.4$, thus we obtain three sub-samples 
\begin{align} 
	z &< 0.81, & N_{sys} &=18, & \vev{z}&= 0.65, & \frac{\Delta\alpha}{\alpha} 
	&= (-0.29 \pm 0.31)\times 10^{-5} \nonumber \\
	0.81 < z &< 2.4 & N_{sys} &=85, & \vev{z}&=1.47, &\frac{\Delta\alpha}{\alpha} 
	&= (-0.58 \pm 0.13)\times 10^{-5} \nonumber \\
	z &> 2.4, & N_{sys} &=40, & \vev{z}&= 2.84, & \frac{\Delta\alpha}{\alpha} 
	&= (-0.87 \pm 0.37)\times 10^{-5}. \label{eq:Murphysplit2}
\end{align}
Here we have used the ``fiducial sample'' of \cite{Murphy03.1}, the weighted average has been taken, and we have included \cite{Murphyprivate} the 15 additional samples used in \cite{Murphy03.2}. For convenience we will refer to these results as ``M$\alpha$''.

Further results have been obtained by Levshakov \etal \cite{Levshakov07.1}, and reported in \cite{Fujii07}:
\begin{align}
 \frac{\Delta \alpha}{\alpha} &= (-0.01 \pm 0.18)\times 10^{-5}, & z_{abs} &= 1.15 \nonumber \\
 \frac{\Delta \alpha}{\alpha} &= (0.57 \pm 0.27)\times 10^{-5}, & z_{abs} &= 1.84.
\end{align}
We note that the value for $z=1.84$ has an opposite sign of variation to the M$\alpha$ result, though the variation does not have high statistical significance. The observational situation is clearly unsatisfactory.

\paragraph{Tests of $\mu$}
\ \\
Vibro-rotational transitions of molecular hydrogen $H_2$ are sensitive to $\mu \equiv \mproton/m_e$. From H$_2$ lines of two quasar absorption systems (at $z = 2.59$ and $z=3.02$) a variation is found \cite{Reinhold06} of
\beq \label{Reinholdmu}
 \frac{\Delta \mu}{\mu} = (2.4 \pm 0.6) \times 10^{-5},
\eeq
taking a weighted average. We will refer to this result as ``R$\mu$'' after Reinhold {\it et al}. The individual systems yield \cite{Reinhold06}
\begin{align} \label{eq:Reinholdsystems}
 \frac{\Delta \mu}{\mu} &= (2.78 \pm 0.88) \times 10^{-5}, & z_{abs} &= 2.59 \nonumber \\
 \frac{\Delta \mu}{\mu} &= (2.06 \pm 0.79) \times 10^{-5}, & z_{abs} &= 3.02. 
\end{align}
Recently the $z=3.02$ system has been reanalyzed \cite{Wendt08}, with the result that the claimed significance of \eqn \eqref{eq:Reinholdsystems} was not reproduced, and the absolute magnitude of the variation is bounded by $|\Delta \mu / \mu| \le 4.9 \times 10^{-5}$ at $2 \sigma$, or
\beq
 |\Delta \mu / \mu| \le 2.5 \times 10^{-5}, \qquad z_{abs} = 3.02 \quad (1 \sigma). 
\eeq

Very recently, a new determination of the variation of $\mu$ appeared \cite{King08} reporting a reanalysis of spectra from the same two H$_2$ absorption systems as \cite{Reinhold06}, and adding one additional system at $z\simeq 2.8$. The results of the new analysis are not consistent with the previous claim indicating a nonzero variation, either considering all three systems or the two previously considered. Instead, \cite{King08} obtain a null bound, $\Delta \mu/\mu = (2.6 \pm 3.0) \times 10^{-6}$. As these results were published after this study has been finished, they are not considered any further. 

The inversion spectrum of ammonia has been used to bound $\mu$ precisely at lower redshift \cite{FlambaumNH3}. Recently the single known NH$_3$ absorber system at cosmological redshift has been analyzed \cite{MurphyNH3}, yielding
\beq
 \frac{\Delta \mu}{\mu} = (0.74\pm 0.89) \times 10^{-6}, \qquad z = 0.68.
\eeq

\paragraph{Tests of $y$}
\ \\
The 21cm HI line and molecular rotation spectra are sensitive to $y \equiv \alpha^2 g_p$, where $g_p$ is the proton g-factor. Bounds on this quantity from \cite{Murphy01} are
\begin{align}
\frac{\Delta y}{y} &= (-0.20 \pm 0.44) \times 10^{-5}, &z &= 0.247 \nonumber\\
\frac{\Delta y}{y} &= (-0.16 \pm 0.54) \times 10^{-5}, &z &= 0.685.
\end{align}

\paragraph{Tests of $x$}
\ \\
Further, the comparison of UV heavy element transitions with HI line probes for variations of $x \equiv \alpha^2 g_p \mu^{-1}$  \cite{Tzanavaris06}: the weighted mean of nine analyzed systems yields
\beq
  \frac{\Delta x}{x} = (0.63 \pm 0.99) \times 10^{-5},\quad 0.23<z_{abs}<2.35.
\eeq
However, we note that i) the systems lie in two widely-separated low-redshift ($0.23<z<0.53$) and high-redshift ($1.7<z<2.35$) ranges; and ii) these two sub-samples have completely different scatter, $\chi^2/\nu$ about the mean for the low- and high-redshift systems being $0.33$, and $2.1$, respectively. Hence we consider two samples, with average redshift $z=0.40$ (5 systems) and $z=2.03$ (4 systems). With expanded error bars in the high-redshift sample 
(after ``method 3'' of \cite{Tzanavaris06}) we find
\begin{align} 
 \frac{\Delta x}{x} &= (1.02 \pm 1.68) \times 10^{-5}, & \vev{z} &= 0.40 \nonumber \\
 \frac{\Delta x}{x} &= (0.58 \pm 1.94) \times 10^{-5}, & \vev{z} &= 2.03. 
\end{align}

\paragraph{Tests of $F$}
\ \\
The comparison of HI and OH lines is sensitive to changes in $F \equiv g_p \left[ \alpha^2 \mu \right]^{1.57}$ \cite{Kanekar05} and yields
\beq
 \frac{\Delta F}{F} = (0.44 \pm 0.36^{stat} \pm 1.0^{sys}) \times 10^{-5},\quad z = 0.765. 
\eeq

\paragraph{Tests of $F'$}
\ \\
A similar method comparing CII and CO lines has very recently been proposed at high redshift \cite{Levshakov07.2} yielding the best bound at redshifts $> 4.5$. The following bounds on $F'\equiv \alpha^2/\mu$ 
are obtained for two systems:
\begin{align} 
 \frac{\Delta F'}{F'} &= (0.1 \pm 1.0)\times 10^{-4}, & z & = 6.42 \nonumber \\
 \frac{\Delta F'}{F'} &= (1.4\pm 1.5)\times 10^{-4},  & z & = 4.69. 
\end{align}

\section{The Oklo natural reactor}
\label{sec:Oklo}
In Oklo/Gabon, a natural fission reactor formed by naturally enriched uranium in a rock formation with a water moderator was operating about 2 billion years ago ($\Delta t\simeq 1.8\times 10^9\,$y, $z \sim 0.14$ with WMAP5 best fit cosmology). The resulting isotopic ratios in this rock nowadays differ radically from any other terrestrial material. By modeling the nuclear fission process, one can in principle bound the variation of $\alpha$ over this period. The determination of $\Delta \ln \alpha$ at the time of the reactions results from considering the possible shift, due to variation of electromagnetic self-energy, in the position of a very low-lying neutron capture resonance of $^{149}$Sm. The analysis of \cite{Petrov05} gives the bound (taken as 1$\sigma$) 
\beq \label{eqn:Okloalpha}
 -5.6\times 10^{-8} < \Delta\alpha/\alpha < 6.6\times 10^{-8}.
\eeq
For a linear time dependence this results in the bound
\beq
 |\dot{\alpha}/\alpha| \leq 3 \times 10^{-17} {\rm y}^{-1}. 
\eeq

Note that these results concern varying $\alpha$ only. If other parameters affecting nuclear forces, in particular light quark masses, are allowed to vary, the interpretation of this bound becomes unclear \cite{Olive02, Flambaum02} since it depends on a nuclear resonance of ${^{150}}$Sm whose properties are very difficult to investigate from first principles. In the absence of a resolution to this problem we consider Oklo as applying only to the $\alpha$ variation in each model. In scenarios where several couplings vary simultaneously we do not consider strong cancellations. Nevertheless, we allow for a certain degree of accidental cancellation and therefore multiply the error on the bound \eqn \eqref{eqn:Okloalpha} by a factor three.

\newpage

\section{Meteorite dating}
Meteorites which have formed at about the same time as the solar system, $t_{\rm Met}\simeq 4.6\times 10^{9}\,$y ago ($z \simeq 0.44$) contain long-lived $\alpha$- or $\beta$-decay isotopes. The decay rates of those isotopes may be sensitive probes of cosmological variation \cite{Olive02,Olive03,Sisterna90}. Their (generally) small $Q$-values result from accidental cancellations between different contributions to nuclear binding energy, depending on fundamental couplings in different ways, thus the sensitivity of the decay rate may be enhanced by orders of magnitude.

The best bound concerns the ${^{187}}$Re $\beta$-decay to osmium with $Q_\beta=2.66\,$keV. The decay rate $\lambda_{187}$ is measured at present in the laboratory, and also deduced by isotopic analysis of meteorites formed about the same time as the solar system, $4.6\times 10^{9}$ years ago. More precisely, the ratio $\lambda_{187}/\lambda_{\rm U}$, averaged over the time between formation and the present, is measurable \cite{Olive03,Fujii03}, where $\lambda_{\rm U}$ is the rate of some other decay (for example uranium) used to calibrate meteorite ages.

The experimental values of $\lambda_{187}$ imply (setting $\lambda_{U}$ to a constant value)
\beq \label{meteoriteintegral}
	t_{\rm Met}^{-1} \int_{-t_{\rm Met}}^{0} \frac{\Delta \lambda_{187}(t)} {\lambda_{187}} 
	\, dt = 0.016\pm 0.016 .
\eeq
Since the redshift back to $t_{\rm Met}$ is relatively small, we obtain bounds on recent time variation by assuming a linear evolution up to the present, for which the left hand side is $-(t_{\rm Met}/2) \dot{\lambda}_{187}/ \lambda_{187}$ and the fractional rate of change is bounded by 
\beq
	\frac{\dot{\lambda}_{187}}{\lambda_{187}} \simeq (-7.2\pm6.9) \times10^{-12}\,{\rm y}^{-1}.
\eeq
Projected back to $t_{\rm Met}$ this gives the bound 
\beq
\Delta \ln \lambda_{187} \simeq 0.033\pm0.032 \qquad (z\simeq 0.44).
\eeq
This is a conservative bound unless the time variation has recently accelerated, or there are significant oscillatory variations over time.

Since the possible dependence of ``control'' decay rates $\lambda_{\rm U}/m_N$ on nuclear or fundamental parameters is much weaker than that of $\lambda_{187}/m_N$, we use this result for the variation of $\lambda_{187}$ in units where $\lambda_{\rm U}$ is constant, \ie $\Delta \ln (\lambda_{187}/\lambda_{\rm U})\simeq \Delta \ln (\lambda_{187}/m_N)$. We find the decay rate dependence to be \cite{DSW08.1}
\beq
	\Delta \ln \frac{\lambda_{187}}{m_N} \simeq 
	-2.2\times 10^4 \Delta \ln \alpha - 1.9\times 10^4 \Delta \ln \frac{\hat{m}} 
	{\lqcd} + 2300\, \Delta \ln \frac{\delta_q}{\lqcd}
	-580\, \Delta \ln \frac{m_e}{\lqcd} .
\eeq

\section{Bounds on the variation of $\gnewton$}
Variations of Newton's constant have been studied in the solar system and in astrophysical effects. Whilst all references give bounds exclusively on a potential variation of $\gnewton$, one should note that besides $\gnewton$ also nuclear parameters (neutron / proton masses and parameters of nuclear forces) can vary, which would in general add degeneracies and make the results less stringent. It has generally been assumed that particle masses are constant, thus the resulting bounds actually constrain variation of $\gnewton \mnucleon^2 \propto (\mnucleon/\mplanck)^2$.

In the solar system, changes of $\gnewton$ induce changes in the orbits of planets. Range measurements to Mars from 1976 to 1982 can be used to obtain \cite{Hellings83}
\beq
 \dotgnewton/\gnewton = 2 \pm 4 \times 10^{-12} \mbox{y}^{-1}. 
\eeq
Lunar laser ranging from 
1970 to 2004 yields \cite{Williams04}
\beq 
 \dotgnewton/\gnewton = (4 \pm 9) \times 10^{-13} \mbox{y}^{-1}. 
\eeq
The stability of the orbital period of the binary pulsar PSR 1913+16 \cite{Damour88} may be used to deduce 
\beq
 \dotgnewton/\gnewton = (1.0 \pm 2.3) \times 10^{-11} \mbox{y}^{-1}. 
\eeq
All these results apply at the present epoch $z=0$.

A bound on the behavior of $\gnewton$ over the lifetime of the Sun (approximately $4.5\times 10^{9}$y, $z = 0.43$) was found by Guenther \etal \cite{Guenther98} by considering the effect of the resulting discrepancy in the helium/hydrogen fraction on p-mode oscillation spectra. The claimed constraint is
\begin{align}
 |\dotgnewton/\gnewton| &\leq 1.6 \times 10^{-12}\,{\rm y}^{-1} \nonumber \\
 |\Delta \ln \gnewton| &\leq  7.2 \times 10^{-3} \qquad z=0.43,
\end{align}
where the assumed form of variation is a power law in time since the Big Bang, which may be approximated over the last few billion years as a linear dependence. For models with significantly nonlinear time dependence the bound may be reevaluated: since the bound arises from the accumulated effect of hydrogen burning since the birth of the Sun, it may be expressed as an integral of the variation over the Sun's lifetime analogous to \eqn \eqref{meteoriteintegral}.

The mass of neutron stars is determined by the Chandrasekhar mass 
\beq
 M_{\rm Ch} \simeq \frac{ 1
}{\gnewton^{3/2} m_n^2} 
\eeq
where $m_n$ is the neutron mass. This may be reexpressed in terms of the baryon number of the star $n_B \propto M_{\rm Ch}/m_n \propto (\gnewton m_n^2)^{-3/2}$, which is expected to be constant up to small corrections from matter accreting onto it. Thus the relative masses of neutron stars measured at the same epoch probes the fractional variation of $\gnewton m_n^2$ between their epochs of formation. From the comparison of masses of young and old neutron stars in binary systems (where the oldest neutron stars are up to 12 Gy old, $z \sim 3.3$), it is found \cite{Thorsett96} that the variation of the average neutron star mass $\mu_n$ is 
$ \dot{\mu}_n = -1.2 \pm 4.0 (8.5) \times 10^{-12} M_{\odot}\, \mbox{y}^{-1} $
at 60\% (95\%) confidence level. In units where particle masses are constant, we have
\beq
 \dotgnewton/\gnewton = -0.6 \pm 2.0\, (4.2) \times 10^{-12} \mbox{y}^{-1}, 
\eeq
where the averaging is performed over the last $12\times 10^{9}$y, and the bound should be reinterpreted for variations which are not linear in time. The absolute variation over this period is then bounded at $1\sigma$ as
\beq
 \Delta \ln \gnewton = (-0.7 \pm 2.4) \times 10^{-2}, \quad z=3.3.
\eeq

\section{Atomic clocks} 
\label{sec:clocks}
Atomic transitions can be measured in the laboratory to very high precision over periods of years. As each atomic transition depends differently on fundamental constants (\eg $\alpha$, $\mu$), comparisons of different atomic transitions over long periods of time give very sensitive results. 

Recently, stringent bounds on the present time variation of the fine structure constant and the electron-proton mass ratio have been obtained by \cite{Blatt08},
\begin{align}
 d \ln \alpha / dt &= (-0.31 \pm 0.3)\times 10^{-15}\, \mbox{y}^{-1} \nonumber \\
 d \ln \mu / dt &= (1.5 \pm 1.7) \times 10^{-15}\, \mbox{y}^{-1}.
\end{align}

Fortier \etal \cite{Fortier07} obtain stronger bounds, $|\dot{\alpha}/\alpha| < 1.3\times 10^{-16}\,$y$^{-1}$, if other relevant parameters are assumed not to vary. If other atomic physics parameters are allowed to vary, this bound becomes considerably weaker, depending on a possible relative variation of the Cs magnetic moment and the Bohr magneton.
Direct comparison of optical frequencies may yield bounds at the level of $10^{-17}$ per year; limits on variation of $\alpha$ from this method are reported with uncertainty $2.3 \times 10^{-17}$y$^{-1}$ \cite{Rosenband08} but designated as preliminary. If these bounds are used then our limits from atomic clocks via $\alpha$ variation should be tightened by about an order of magnitude.

Extrapolating the results of \cite{Blatt08} to the time of Oklo ($z=0.14$, $t = 1.8 \times 10^9\,$y) gives 
\begin{align}
  \Delta \ln \alpha &= (-0.56 \pm 0.54) \times 10^{-6}, \nonumber \\
  \Delta \ln \mu &= (-0.27 \pm 0.31) \times 10^{-5}.
\end{align}

\newpage
%--------------------

\begin{sidewaystable}
\hspace{-1.5cm}
\begin{tabular}{|c||c|c|c|c|c|c|c|c|c|}
\hline
Method & redshift & $\Delta \ln \alpha $ & $\Delta \ln \mu$ & $\Delta \ln \gnewton \mnucleon^2$ & $\Delta \ln x$ & $\Delta \ln y$ & $\Delta \ln F$ & $\Delta \ln F'$ & $\Delta \ln \lambda_{187}$\\ 
& & $[10^{-6}]$&$[10^{-5}]$&$[10^{-2}]$&$[10^{-5}]$&$[10^{-5}]$&$[10^{-5}]$&$[10^{-4}]$ & $[10^{-2}]$ \\
\hline
\hline
Oklo $\alpha$ \cite{Petrov05} & 0.14 & $0.00 \pm 0.06$ & & & & & & &\\
\hline
21cm \cite{Murphy01}       & 0.247 & & & & & $-0.20 \pm 0.44$ & & &\\
Sun \cite{Guenther98}          & 0.43 & & & $ 0 \pm 0.72$ & & & & &\\
Heavy/HI, low-z \cite{Tzanavaris06} & 0.40 & & & & $1.0 \pm 1.7$ & & & &\\
Meteorite \cite{Olive03}  & 0.44 & & & & & & & & $3.3\pm 3.2$ \\
M$\alpha$ epoch 2 \cite{Murphy03.2} & 0.65 & $-2.9 \pm 3.1$ & & & & & & &\\
Ammonia \cite{FlambaumNH3}      & 0.68 & & $0.06 \pm 0.19$ &  & & & & &\\
21cm \cite{Murphy01}       & 0.685 & & & & & $-0.16 \pm 0.54$ & & &\\
HI / OH \cite{Kanekar05}   & 0.765 & & & & & & $0.4 \pm 1.1$ & &\\
\hline
Absorption \cite{Fujii07}  & 1.15 & $-0.1 \pm 1.8$ & &  & & & & &\\
$M\alpha$ epoch 3 \cite{Murphy03.2} & 1.47 & $-5.8 \pm 1.3$ & & & & & & &\\
Absorption \cite{Levshakov07.1} & 1.84 & $5.7 \pm 2.7$ & & & & & & &\\
Heavy/HI, high-z \cite{Tzanavaris06} & 2.03 & & & & $0.6 \pm 1.9$ & & & &\\
\hline
$H_2$ \cite{Reinhold06}    & 2.59 & & $2.78 \pm 0.88$ & & & & & &\\
$M\alpha$ epoch 4 \cite{Murphy03.2} & 2.84 & $-8.7 \pm 3.7$ & & & & & & &\\
$H_2$ \cite{Reinhold06}    & 3.02 & & $2.06 \pm 0.79$ &  & & & & &\\
Neutron stars \cite{Thorsett96} & 3.3 & & & $-0.7 \pm 2.4$  & & & & &\\
CII / CO \cite{Levshakov07.2} & 4.69 & & & & & & & $1.4 \pm 1.5$ &\\
CII / CO \cite{Levshakov07.2} & 6.42 & & & & & & & $0.1 \pm 1.0$ &\\
\hline
CMB \cite{Martins03}, \cite{Chan07} & $10^3$ & $0^{+1\times 10^4}_{-3\times 10^4} $ & & $0^{+7}_{-6}$ & & & & &\\
\hline
\end{tabular}
\caption[Observational $1 \sigma$ bounds on variations]{Observational $1 \sigma$ bounds on variations. Observables are defined as $\mu\equiv \mproton/m_e$, $x\equiv \alpha^2 g_p\mu^{-1}$, $y\equiv \alpha^2g_p$, $F\equiv g_p[\alpha^2 \mu]^{1.57}$, $F'\equiv \alpha^2/\mu$. The given redshift may denote a single measurement, or an averaged value over a certain range: see main text. The two CMB bounds are independent of each other. Our BBN bounds cannot be displayed in this form.} \label{tab:CollectionOfMeasurements}
\end{sidewaystable}

%%%%%%%%%%%%%%%%%%%%%%%%%%%%%%%%%%%%%%%%%%%%%%%%%%%%%%%%%%%%%%%%%%%%%%%%%
%%%%%%%%%%%%%
%%%%%%%%%%%%%%%%%%%%%%%%%%%%%%%%%%%%%%%%%%%%%%%%%%%%%%%%%%%%%%%%%%%
\chapter[Variations from BBN to today in GUTs]{Variations from BBN to today in unified scenarios}
\label{chap:VariationsFromBBNtoTodayinGUT}

As has been laid out in \sect \ref{sec:GUT}, we will use the concept of grand unification to reduce the number of potentially varying parameters. The main idea is that the GUT relations interrelate variations of the ``fundamental'' parameters $G_k$ which we defined in \sect \ref{sec:BBNFromNucToFunParams}.

In this chapter, we consider the hypothesis that, for all redshifts, all fractional variations in the ``fundamental'' parameters $G_k$ are proportional to one nontrivial variation with fixed constants of proportionality. If the variation of the unified gauge coupling $\Delta \ln \alphagut$ is nonvanishing, we may write
\beq \label{eq:dkdef}
	\Delta \ln G_k = d_k\Delta \ln\alphagut
\eeq
for some constants $d_k$, assuming small variations. Different unification scenarios correspond to different sets of values for the ``unification coefficients'' $d_k$.
Considering the values of $\Delta\ln G_k$ as coordinates in an $N_k$-dimensional space, this assumption restricts variations to a single line passing through zero. The variation then constitutes exactly one degree of freedom.
We will go beyond this hypothesis in the next chapter where we also consider models with growing neutrinos and oscillating variations (see \sect \ref{sec:GrowingNeutrinoModels}) for which a fixed linear relation (\ref{eq:dkdef}) is not realized for all $z$.

\section{GUT relations}
\label{sec:GUTrels}
GUT relations have the property that variations of the Standard Model gauge couplings and mass ratios can be determined in terms of a smaller set of parameters describing the unified theory and its symmetry breaking. Hence, if nonzero variations in different observables are measured at similar redshifts, models of unification may be tested without referring to any specific hypothesis for the overall cosmological history of the variation. We need only assume that for a given range of $z$ the time variation is slow and approximately homogeneous in space, hence $\Delta \ln\alphagut$ depends only on redshift $z$ to a good approximation.
The relevant unified parameters are the unification mass $\mgut$ (relative to the Planck mass), the GUT coupling $\alphagut$ defined at the scale $\mgut$, the Higgs v.e.v.\ $\vev{\phi}$ and, for supersymmetric theories, the soft supersymmetry breaking masses $\tilde{m}$, which enter in the renormalization group (RG) equations for the running couplings. Then, for the variations at any given $z$ we can write 
\beq \label{eq:unifdefs}
\Delta \ln \frac{\mgut}{\mplanck} = d_M l, \quad \Delta \ln \alphagut = d_X l, \quad \Delta \ln \frac{\vev{\phi}}{\mgut} = d_H l, \quad \Delta \ln \frac{\tilde{m}}{\mgut} = d_S l,
\eeq
where $l(z)$ is the ``evolution factor'' introduced for later convenience. If $\alphagut$ varies nontrivially we may normalize $l$ via $d_X=1$. In supersymmetric theories we set $\alphagut = 1/24$, in nonsupersymmetric theories we set $d_S \equiv 0$ and $\alphagut = 1/40$ (see \sect \ref{sec:GUT}).

We make the simplifying assumption that the masses of Standard Model fermions all vary as the Higgs v.e.v., \ie Yukawa couplings are constant at the unification scale: 
\beq \label{eq:fermionMX}
 \Delta \ln \frac{m_e}{\mgut} = \Delta \ln \frac{\delta_q}{\mgut} = \Delta \ln \frac{\hat{m}}{\mgut} = \Delta \ln \frac{m_s}{\mgut}
 = \Delta \ln \frac{\vev{\phi}}{\mgut}.
\eeq
Like the gauge couplings, also the fermion masses vary under variations of the unified coupling $\alphagut$ due to the renormalization group running of fermion masses.
However, we have explicitly calculated the effect of varying couplings and found that it is at the order of a 1\% correction\footnote{For low-energy observables such as $m_q(Q^2)/\lqcd$ we consider an RG scale $Q^2$ that is fixed relative to $\lqcd$, $Q^2=\,$const$\cdot\lqcd^2$. Thus the variation of $m_q(Q^2)/m_q(\mgut^2)$ is entirely due to the dependence on $\alpha_3(\mgut)$, which is suppressed by a loop factor $\alphagut/\pi$ compared to the nonperturbative dependence of $\lqcd/\mgut$ on $\alphagut$. We find $\Delta \ln (\bar{m}_q(Q^2)/\bar{m}_q(\mgut^2)) = 2/7\Delta\ln \alphagut \simeq (9\alphagut/7\pi)\Delta \ln (\lqcd/\mgut)$ under variation of $\alphagut$, where $\bar{m}_q$ is the running quark mass.}, which is already smaller than our uncertainties in hadronic and nuclear physics\footnote{Langacker \etal \cite{Langacker01} arrive at the same conclusion.}.
Hence we can apply the assumption (\ref{eq:fermionMX}). Using the relations \eqref{eqn:dlnAlpha} and \eqref{eqn:dlnLambdadlnMgut}, one finds for the QCD scale
\beq
	\frac{\Delta \ln (\lqcd/\mgut)}{l} = \frac{2\pi}{9\alphagut}d_X + \frac{2}{9}d_H + \frac{4}{9}d_S
\eeq
and for the fine structure constant,
\beq
	\frac{\Delta \ln \alpha} {l} = \frac{80 \alpha}{27 \alphagut} d_X + \frac{43}{27}\frac{\alpha}{2\pi} d_H + \frac{257}{27} \frac{\alpha}{2\pi} d_S.
\eeq

For the nucleon mass we include possible strange quark contributions. In our treatment of BBN, we neglected strange quark contributions, as the final dependence on $m_s$ was much below the model uncertainty. Here we include the roughly known strange contribution to the nucleon mass. The uncertainty in the strangeness content is an indicator of the overall uncertainty that may arise due to $m_s$ variation. We found (\eqns \eqref{eqn:NucleonDependence1},  \eqref{eqn:NucleonDependence2} and \eqref{eqn:QNDependence})
\begin{align}
 	\Delta \ln \frac{\mnucleon}{\lqcd} &= 0.048 \Delta \ln \frac{\hat{m}}{\lqcd} + (0.12 \pm 0.12) \Delta \ln \frac{m_s}{\lqcd}, \label{eq:strangeness}\\
 	\Delta \ln \frac{Q_N}{\lqcd} &= -0.59 \Delta \ln \alpha + 1.59 \Delta \ln \frac{\delta_q}{\lqcd},
\end{align}
and thus
\begin{align} 
	\frac{\Delta \ln \mu}{l} &= (0.58 \mp 0.08) \frac{d_X}{\alphagut} + (0.37 \mp 0.05) d_S + (-0.65 \pm 0.09) d_H, \\
	\frac{\Delta \ln (\gnewton \mnucleon^2)}{l} &= 2 d_M + (1.16 \mp 0.17)	\frac{d_X}{\alphagut} + (0.74 \mp 0.11) d_S + (0.71 \pm 0.19) d_H, \label{DelGmNsquared}
\end{align}
where the upper or lower signs correspond to the positive or negative signs in \eqn \eqref{eq:strangeness} respectively. 

The largest contribution to variations of the proton g-factor $g_p$ has been argued to arise from the pion loop \cite{Murphy03.2}, yielding at first order a dependence on the light quark mass of
\begin{align}
  \Delta \ln g_p &\simeq -0.087 \Delta \ln \hat{m}/\lqcd, \nonumber \\ 
  \frac{\Delta \ln g_p}{l} &\simeq 0.06 \frac{d_X}{\alphagut} - 0.07 d_H + 0.04 d_S.
\end{align}
Hence the variations of observables including $g_p$ are
\begin{align}
 \frac{\Delta \ln x}{l} &= (-0.48 \pm 0.08) \frac{d_X}{\alphagut} + (0.59 \mp 0.09) d_H + (-0.31 \pm 0.05) d_S \nonumber \\
 \frac{\Delta \ln y}{l} &= 0.10 \frac{d_X}{\alphagut} -0.06 d_H + 0.06 d_S \nonumber \\
 \frac{\Delta \ln F}{l} &= (1.04 \mp 0.13) \frac{d_X}{\alphagut} + (-1.08 \pm 0.14) d_H + (0.65 \mp 0.08) d_S \nonumber \\
 \frac{\Delta \ln F'}{l} &= (-0.54 \pm 0.08) \frac{d_X}{\alphagut} + (0.65 \mp 0.09) d_H + (-0.35 \pm 0.05) d_S.
\end{align}
We have now expressed the variations accessible to observation in terms of three (four) variables: $l$, $d_X$, $d_H$ (and $d_S$), where one parameter may be eliminated by normalization. Different unified scenarios will be characterized by different relations among these parameters. 

\section{Variations in six different unified scenarios}
\label{sec:SixScenarios}
We will now investigate six different scenarios for the variation of the grand unified parameters $\alpha_X$, $\mgut/\mplanck$, $\vev{\phi}/\mgut$ and $\tilde{m}/\mgut$. These will fix the unification coefficients $d_k$. For each unified scenario we display the $z$-dependence of the fractional variation. Each figure shows the available information from observations of different couplings, interpreted as constraints on the variation of a single parameter. Since we have only one free variable we can plot all observations simultaneously as a function of redshift. Inspection ``by eye'' permits to judge if a smooth and monotonic evolution of the varying parameter is consistent or not. Most data points are upper bounds on a possible variation, and for the non-zero variations we can study two immediate questions in each scenario:

First, whether claimed nonzero variations of $\alpha$ \cite{Murphy03.2} and $\mu$ \cite{Reinhold06} at redshift $2$--$3$ are compatible with one another, since the ratio of their fractional variations is predicted in each scenario.

Second, we consider whether there is an indication of nonzero variation at BBN. For no variation at BBN we obtain $\chi^2 = 17.9$ for 3 measured abundances (\hefo, D, \lise). This discrepancy between theory and observation is exclusively due to \lise. (Considering only \hefo\ and D, the value of $\chi^2$ is $0.24$.) If we wish to solve or ameliorate the ``lithium problem'' by a nonzero variation, we will require $\chi^2 / \nu$ to be not much larger than unity, taking $\nu = 2$ as appropriate for one adjustable parameter. If there is no significant range where the three abundances have a $2\sigma$ fit ($\chi^2/\nu\leq 4$) then we give up the hypothesis that the \lise\ problem is solved by coupling variations and instead assume that the observed depletion %of \lise\ 
is due to some astrophysical effect. In this case we consider only D and \hefo\ abundances as observational bounds on the size of variations at BBN.

\subsection{Varying $\alpha$ alone}
\label{sec:VaryingAlphaAlone}
\begin{figure}[p]
\begin{center}
\includegraphics[width=13.5cm]{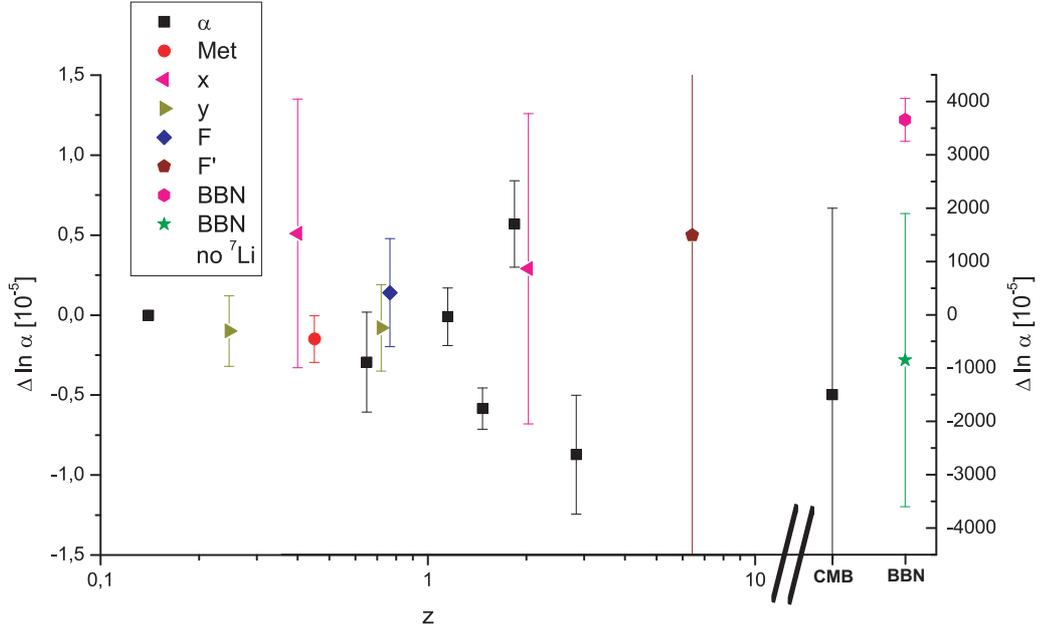}
\end{center}
\caption[Variations for varying $\alpha$ alone]{Variations for varying $\alpha$ alone. Only observations constraining $\alpha$ variation are shown; the BBN fit including \lise\ is poor ($\chi^2/\nu\ge 7.8/2$) hence we also display a conservative bound from \hefo\ and D abundances neglecting \lise.}
\label{fig:alphaonly}
\end{figure}

Before describing the six different grand unified scenarios, we consider a variation of the fine structure constant $\alpha$ alone. Clearly here we are unable to account for any nonzero variation in $\mu$ or other quantities independent of $\alpha$. The cosmological history is dominated by the nonzero variation of the M$\alpha$ values at redshifts $z\simeq 1$ to $4$.
We find that there is almost no $2 \sigma$ match of the BBN values ($\chi^2 / \nu \ge 3.9$): the 2-sigma range is
\beq
	3.25 \% \ge \Delta \ln \alpha_{BBN} \ge 4.06 \%.
\eeq
Hence it seems unlikely that the ``lithium problem'' can be solved by a variation of $\alpha$ alone. If we regard the \lise\ discrepancy as due to systematic or astrophysical effects 
we can set a conservative bound on $\alpha$ variation from \hefo\ and D abundances
\beq
	-3.6\% \ge \Delta \ln \alpha_{BBN} \ge 1.9\%,
\eeq
where we imposed that neither the D nor \hefo\ abundance should deviate by more than $2\sigma$ from observational values. See Fig.~\ref{fig:alphaonly} for a summary of the bounds in this case.
\begin{figure}[p] 
\begin{center}
\includegraphics[width=13.5cm]{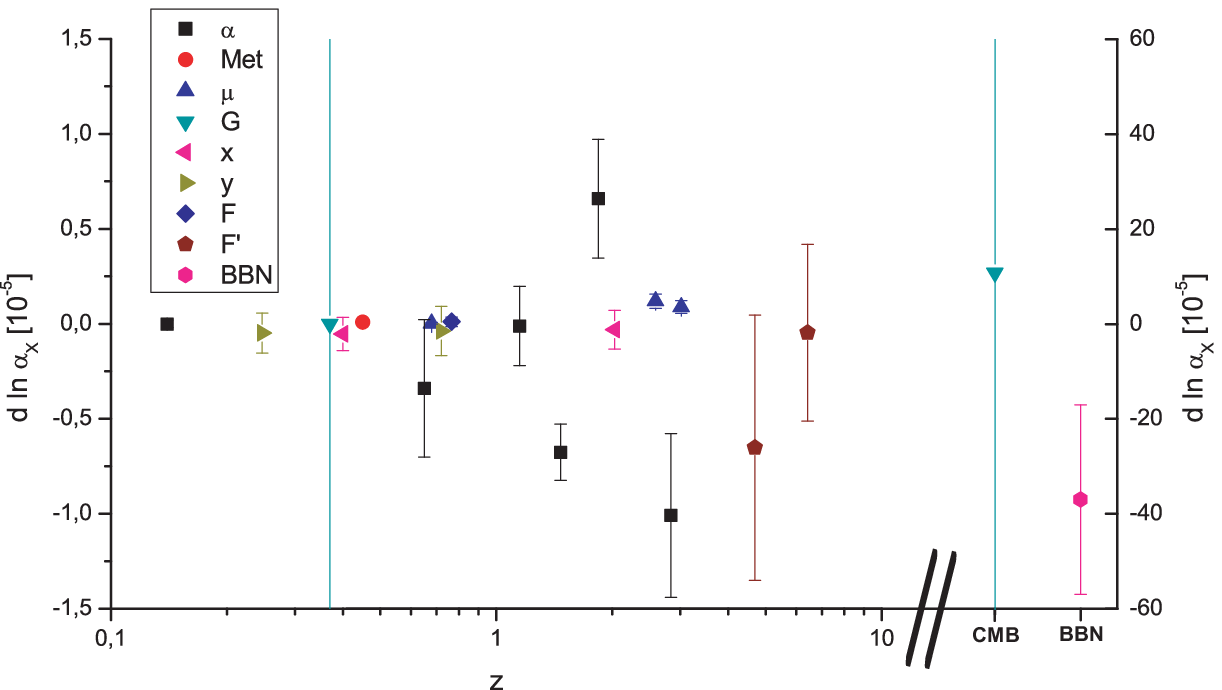}
\end{center}
\caption[Variations for scenario 2]{Variations for scenario 2; BBN bounds are $2 \sigma$ bounds.}
\label{fig:scen2}
\end{figure} 

\subsection{Scenario 1: Varying gravitational coupling}
In this scenario we have only $d_M$ nonvanishing, 
\beq
	d_H = d_S = d_X = 0,
\eeq
therefore
\beq
	\Delta \ln \frac{\mgut}{\mplanck} = \frac{1}{2} \Delta \ln \gnewton \lqcd^2.
\eeq
We find that there is no value of $\Delta \ln \gnewton \lqcd^2$ for which BBN is consistent with the three observed abundances within $2 \sigma$. The best fit values are $\chi^2 / \nu \ge 7.7$ for no variation of $\mnucleon/\mplanck$ at CMB and  $\chi^2 / \nu \ge 5.9$ if the variation of $\mnucleon/\mplanck$ has the same size at BBN and CMB.
Assuming that the discrepancy in the \lise\ abundance is due to some other effect, we find the allowed region of variation of $\gnewton$ at BBN under which primordial D and \hefo\ abundance lie within the observed range at $1 \sigma$ ($2 \sigma$),
\beq
	-5\% \; (-13\%) \le \Delta \ln \gnewton \lqcd^2 \le 12\% \; (22\%)
\eeq
If the variation of $\mnucleon/\mplanck$ has the same size at BBN and CMB one finds
\beq
	-4\% \; (-11\%) \le \Delta \ln \gnewton \lqcd^2 \le 10\% \; (16\%).
\eeq

The bounds on time variation of $\gnewton \lqcd^2$ are much weaker than for many other varying couplings.
This scenario also predicts a vanishing value of $\eta$ in E{\" o}tv{\" o}s experiments (see \sect \ref{sec:WEP} for details). Thus, to any one of the following scenarios we may add an additional nonzero $d_M$ of similar size to $d_X$, $d_H$ or $d_S$ without changing the results significantly.

\subsection{Scenario 2: Varying unified coupling}
In the first GUT scenario without SUSY we consider the case when only $d_X$ is nonvanishing,
\beq
	d_H = d_S = d_M = 0, \qquad \alphagut = 1/40.
\eeq
Within a supersymmetric theory the same relations will apply except that $\alphagut=1/24$ and the variations of observables are scaled by a factor $24/40$ relative to $\Delta \ln \alphagut$: we designate this as Scenario 2S. In both cases we find here
\beq
	\frac{\Delta \ln \mu}{\Delta \ln \alpha} = 27. 
\eeq
It is then highly unlikely for the nonzero M$\alpha$ result for variation of $\alpha$ 
to coexist with the determination of $\mu$ at redshift around $3$ \cite{Reinhold06}, even if the latter is interpreted as an upper bound on the absolute size of variation \cite{Wendt08}.

For the BBN fit, we find without SUSY (excluding modifications of the baryon fraction $\eta$ due to varying $\mnucleon$) 
no range of values fitting at $1\sigma$ level ($\chi^2 / \nu \ge 2.3$). At $2\sigma$ the abundances, including \lise, become consistent for the range
\beq
	-5.7 \times 10^{-4} \le \Delta \ln \alphagut \le -1.7 \times 10^{-4} \qquad (2 \sigma).
\eeq
If one includes a variation of $m_N$ at the time of CMB with the same magnitude as at BBN the result remains unchanged ($\chi^2 / \nu \ge 2.45$), with the same $2 \sigma$ range. For this scenario we may consider a nonzero variation at BBN, but more recent probes must all be viewed as increasingly tight null bounds.
\begin{figure}[p]
\begin{center}
\includegraphics[width=13.5cm]{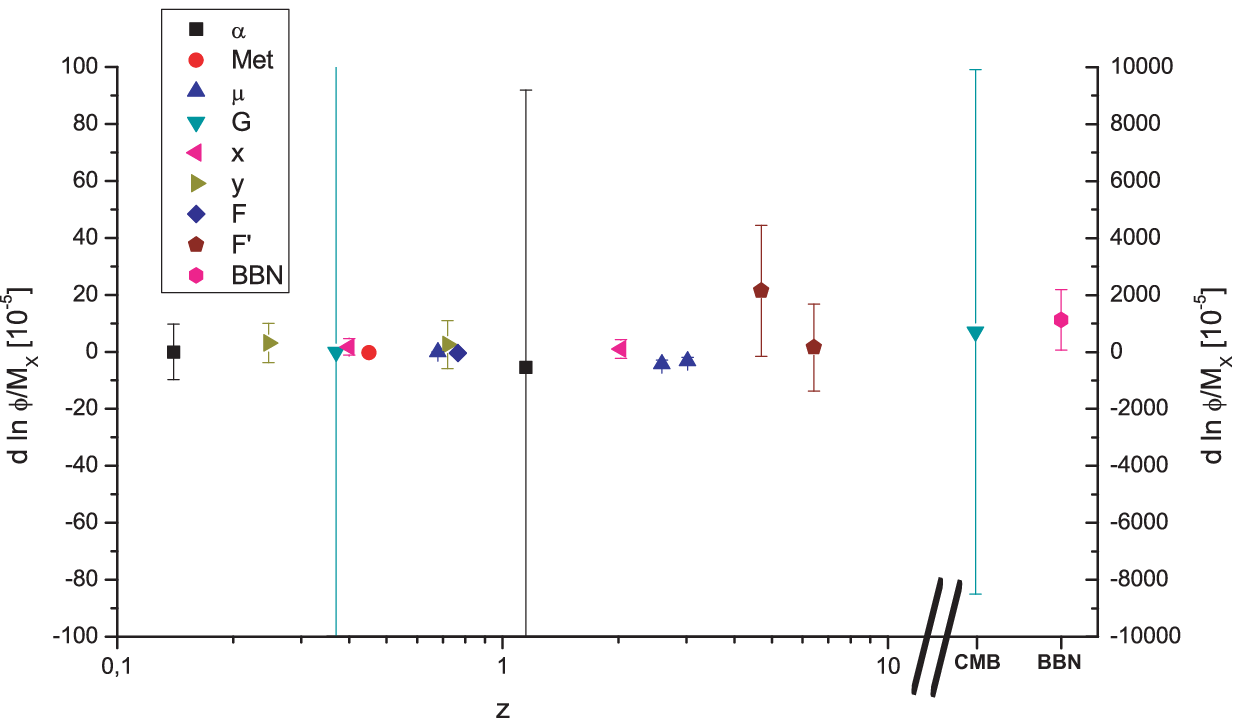}
\end{center}
\caption[Variations for scenario 3]{Variations for scenario 3; BBN bounds are $2 \sigma$. Note that due to the very large ratio $\Delta \ln \mu/ \Delta \ln \alpha$ in this scenario, points indicating any nonzero variation of $\alpha$ fall well outside the range of the graph.}
\label{fig:scen3_small}
\end{figure} 

\subsection{Scenario 3: Varying Fermi scale}
In this scenario we consider the case when the variation arises solely from a change in the Higgs expectation value relative to the unified scale, thus only $d_H$ is nonzero:
\beq
 d_S = d_M = d_X = 0, \qquad \alphagut = 1/40.
\eeq
This scenario implies
\beq
 \frac{\Delta \ln \mu}{\Delta \ln \alpha} = -325.
\eeq
Whether we interpret the determination of $\mu$ \cite{Reinhold06} as a detection or an upper bound, any variation in $\alpha$ at large redshift should be orders of magnitude smaller than current observational sensitivity.

We find for BBN including \lise\ ($\nu=2$) no $1\sigma$ range ($\chi^2 / \nu \ge 1.95$) but 
\beq
6 \times 10^{-3} \le \Delta \ln \vev{\phi}/\mgut \le 22 \times 10^{-3} \qquad (2 \sigma).
\eeq
A variation of $m_N$ at the time of CMB with the same magnitude as at BBN does not change this result.

\subsection{Scenario 4: Varying Fermi scale and SUSY-breaking scale}
\begin{figure}[p]
\begin{center}
\includegraphics[width=13.5cm]{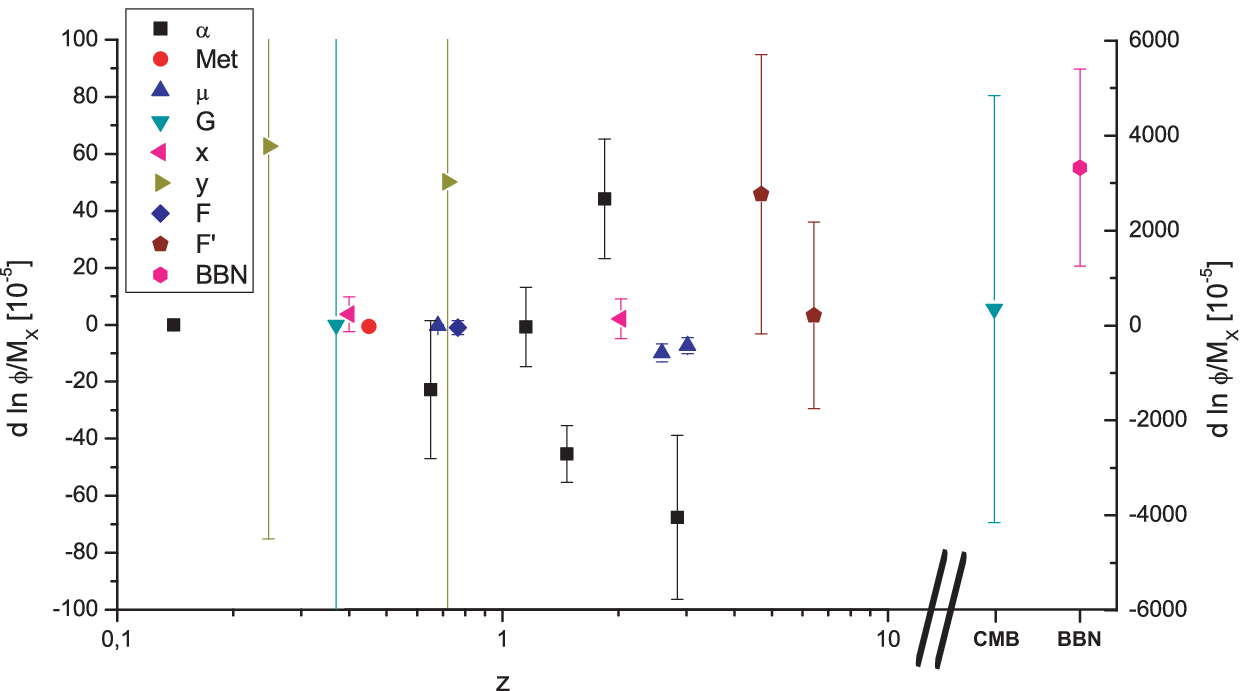}
\end{center}
\caption[Variations for scenario 4]{Variations for scenario 4; BBN bounds are $2 \sigma$}\label{fig:scen4}
\end{figure}

This scenario corresponds to scenario 3, but includes supersymmetry and assumes that the mass-generating mechanism for SM particles and their superpartners gives rise to the same variation:
\beq
 d_M = d_X = 0, \qquad d_S = d_H, \qquad \alphagut = 1/24.
\eeq
We find here
\beq
 \frac{\Delta \ln \mu}{\Delta \ln \alpha} = -21.5, 
\eeq
such that again the claimed nonzero variations in $\alpha$ and $\mu$ cannot be compatible and the variation in $\alpha$ at redshift $3$ must be below current sensitivities. We demonstrate this in Fig.~\ref{fig:scen4}, where we show for this scenario the bounds on the variable $d_Hl = \Delta \ln (\vev{\phi}/\mgut)$ that arise from various observations. 

We find for BBN including \lise ($\nu=2$) no $1\sigma$ fit ($\chi^2 / \nu \ge 1.60$), while at $2\sigma$ 
\beq
1.25 \times 10^{-2} \le \Delta \ln \vev{\phi}/\mgut \le 5.4 \times 10^{-2} \qquad (2 \sigma).
\eeq
If one includes a variation of $m_N$ at the time of CMB with the same magnitude as at BBN the allowed range becomes slightly restricted ($\chi^2 / \nu \ge 1.72$),
\beq
1.20 \times 10^{-2} \le \Delta \ln \vev{\phi}/\mgut \le 4.9 \times 10^{-2} \qquad (2 \sigma).
\eeq

\subsection{Scenario 5: Varying unified coupling and Fermi scale}
In this scenario we study a combined variation of the unified coupling and the Higgs expectation value:
\beq
 d_M = d_S = 0, \qquad d_H = \tilde{\gamma} d_X, \qquad \alphagut = 1/40.
\eeq
The parameter $\tilde{\gamma}$ can be related to the parameter $\gamma \equiv \frac{ \Delta \ln \vev{\phi} / \mgut}{\Delta \ln \lqcd / \mgut}$ which was introduced in \sect \ref{sec:VariationsOfAbundancesInUnifiedModels} via
\beq
	\gamma = \tilde{\gamma} \left( \frac{2\pi}{9 \alphagut} + 
	\frac{2}{9} \tilde{\gamma} \right)^{-1}.
\eeq
There we examined the cases $\gamma = (0, 1, 1.5)$ which correspond to $\tilde{\gamma} = (0, 36, 63)$. Here we find that the best BBN fit is reached for $\tilde{\gamma} \approx 50$ with $\chi^2/\nu = 1.45$. Note that we have the freedom to adjust $\tilde{\gamma}$ such that nonzero variations of $\alpha$ and $\mu$ at redshift $\simeq 3$ are consistent with each other. 
We have
\beq
	\frac{\Delta \ln \mu}{\Delta \ln \alpha} = \frac{23.2 -0.65\tilde{\gamma}}
	{0.865 + 0.002\tilde{\gamma}}\,.
\eeq
We choose for illustration $\tilde{\gamma} = 42$, for which
\beq
  \Delta \ln \mu = -5.6\, \Delta \ln \alpha 
\eeq
and the $2 \sigma$ contour for BBN is 
\beq
  7.5 \times 10^{-4} \le \Delta \ln \alphagut \le 28 \times 10^{-4}.
\eeq
For a variation of $m_N$ at the time of CMB with the same magnitude as at BBN the fit becomes worse ($\chi^2 / \nu \ge 1.68$). However, a $2 \sigma$ fit to BBN is obtained over a wide range of $0 \le \tilde{\gamma} \le 26 $ (negative $\Delta \ln \alphagut$) and $40 \le \tilde{\gamma} < \infty$ (positive $\Delta \ln \alphagut$). 

\begin{figure}[p]
\begin{center}
 \includegraphics[width=13.5cm]{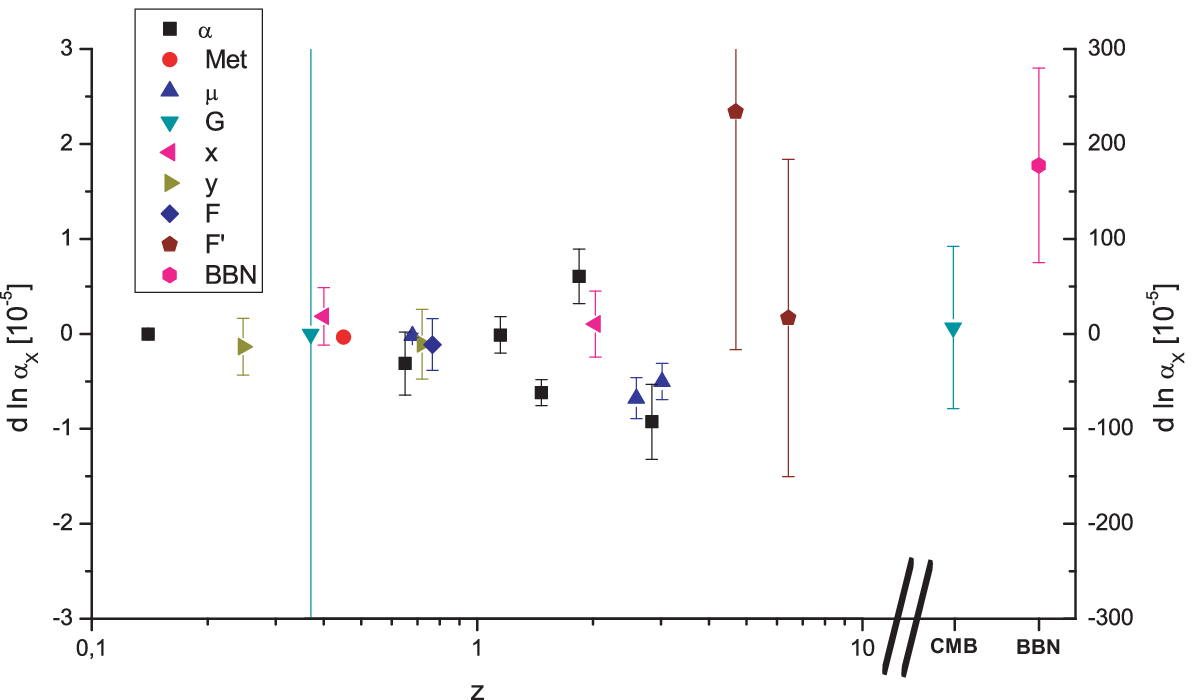}
\end{center}
\caption[Variations for scenario 5, $\tilde{\gamma} = 42$]{Variations for scenario 5, $\tilde{\gamma} = 42$; BBN bounds are $2 \sigma$} \label{fig:scen542}
\end{figure}

Assuming that the apparent \lise\ mismatch at BBN is due to systematic astrophysical effects, we may bound $\alphagut$ with only D and \hefo\ abundances. Here we find at $1 \sigma$ 
\beq
  -5.5 \times 10^{-4} \le \Delta \ln \alphagut \le 1.44 \times 10^{-3} .
\eeq
In Fig.~\ref{fig:scen542} we again plot simultaneously all observations for this scenario. This shows that the bound from BBN including \lise\ is not consistent with the claimed nonzero variations of $\alpha$ and $\mu$ for a monotonic evolution over $z$. 
\begin{figure}[p]
\begin{center}
 \includegraphics[width=13.5cm]{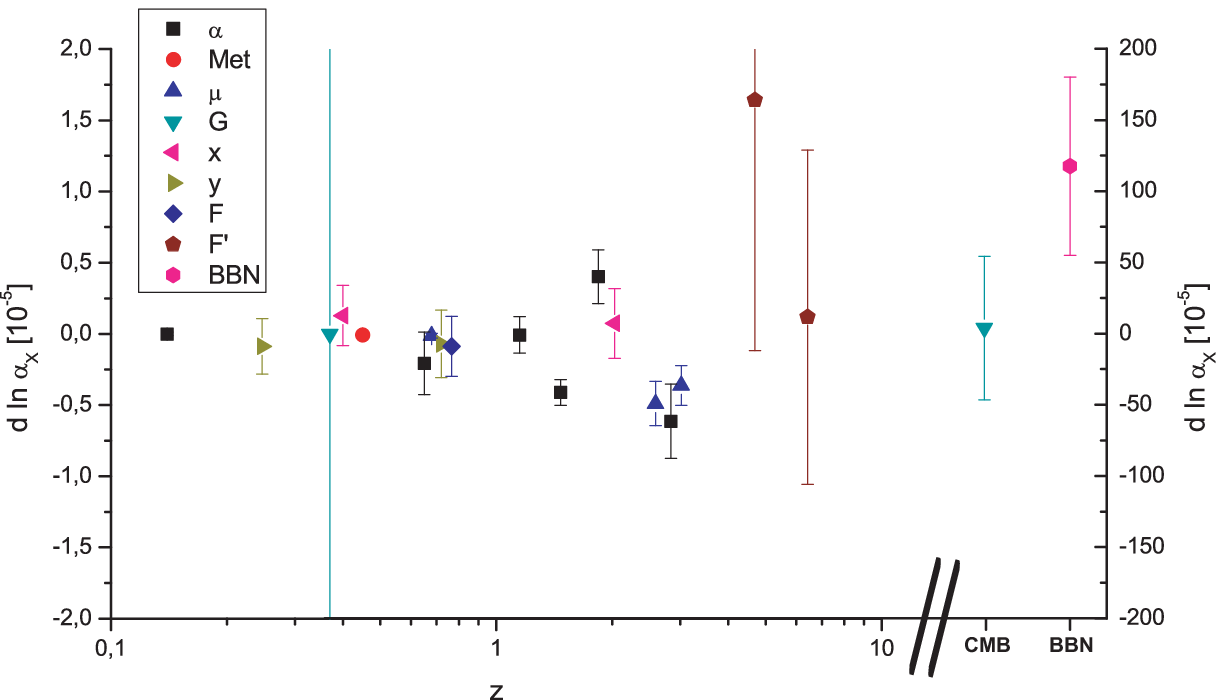} 
\end{center}
\caption[Variations for scenario 6, $\tilde{\gamma} = 70$]{Variations for scenario 6, $\tilde{\gamma} = 70$; BBN bounds are $2 \sigma$} \label{fig:scen670}
\end{figure} 
\begin{figure}
\begin{center}
 \includegraphics[width=13.5cm]{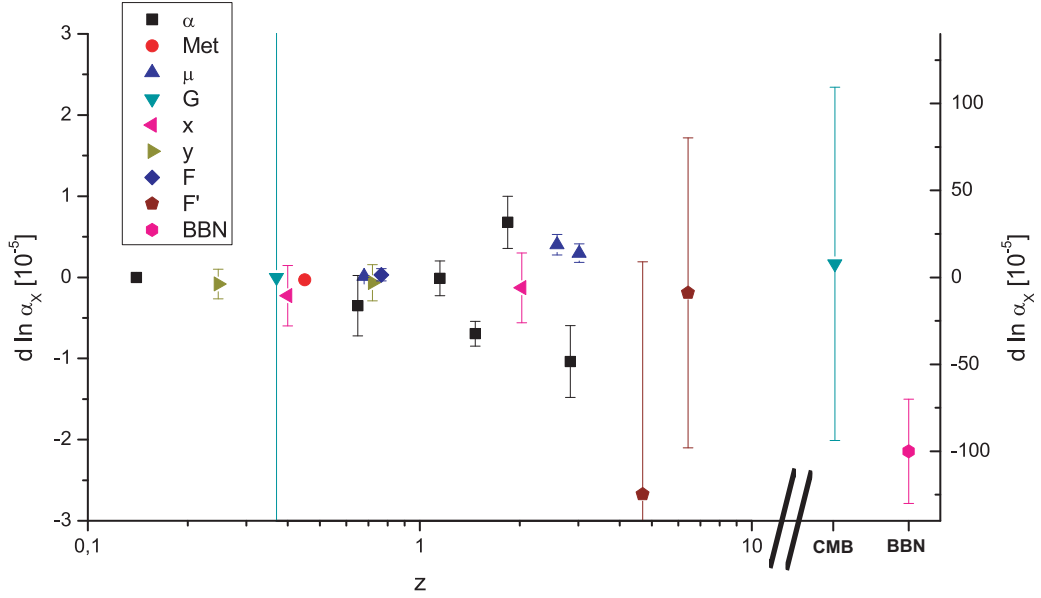}
\end{center}
\caption[Variations for scenario 6, $\tilde{\gamma} = 25$]{Variations for scenario 6, $\tilde{\gamma} = 25$; BBN bounds are $2 \sigma$} \label{fig:scen625}
\end{figure} 

\subsection{Scenario 6: Varying unified coupling and Fermi scale with SUSY}
In this scenario we study a combined variation of the unified coupling and the Higgs v.e.v.\ including SUSY, where as in Scenario 4 we tie the variations of the superpartner masses and Fermi scale together:
\beq
 d_M = 0, \qquad d_S \simeq d_H = \tilde{\gamma} d_X, \qquad \alphagut = 1/24.
\eeq
Now the relation to $\gamma$ is modified as
\beq
  \gamma = \tilde{\gamma} \left( \frac{2\pi}{9 \alphagut} + \frac{2}{3} \tilde{\gamma} \right)^{-1} \; .
\eeq
One may again adjust $\tilde{\gamma}$ to make nonzero variations in $\alpha$ and $\mu$ self-consistent, where now
\beq
	\frac{\Delta \ln \mu}{\Delta \ln \alpha} = \frac{14 - 0.28\tilde{\gamma}}
	{0.52 +  0.013\tilde{\gamma}} \,.
\eeq
We find that a good fit to BBN is obtained over a large range of $\tilde{\gamma}$, ranging from $\tilde{\gamma} = 100$ to infinity with minimal $\chi^2 / \nu = 1.45$. This shows that the main effect in the SUSY model comes from the variation of the Higgs v.e.v. Including a variation of $m_N$ at the time of CMB with the same magnitude as at BBN the fits gets worse ($\chi^2 / \nu \ge 1.8$). A $2 \sigma$ fit can be obtained for $0 \le \tilde{\gamma} \le 28$ (for negative $\Delta \ln \alphagut$ at BBN) and for $58 \le \tilde{\gamma} < \infty$ (positive $\Delta \ln \alphagut$). 

First, we study the case $\tilde{\gamma} = 70$ for which 
\beq
  \Delta \ln \mu = -3.9 \Delta \ln \alpha \qquad (\tilde{\gamma} = 70)
\eeq
and BBN is fit with a $2 \sigma$ range
\beq
  5.5 \times 10^{-4} \le \Delta \ln \alphagut \le 18 \times 10^{-4}.
\eeq
Neglecting \lise, we obtain a $1 \sigma$ bound from BBN
\beq
  -3.5 \times 10^{-4} \le \Delta \ln \alphagut \le 9.3 \times 10^{-4}.
\eeq
Secondly, we study the case $\tilde{\gamma} = 25$ where 
\beq
  \Delta \ln \mu = 8.3 \Delta \ln \alpha \qquad (\tilde{\gamma} = 25),
\eeq
and where the $2 \sigma$ contour for BBN is
\beq
  -13 \times 10^{-4} \le \Delta \ln \alphagut \le -7 \times 10^{-4}.
\eeq
In this second case the Murphy $\alpha$ measurement and BBN point into the same direction. The difference between the two values of $\tilde{\gamma}$ can be seen from a comparison of Figs.~\ref{fig:scen670} and \ref{fig:scen625}.

\section{Epochs and evolution factors}
\label{sec:Epochs_and_evolutionFactors}

\subsection{Epochs}
As a next step, we group the information on experimental bounds on variations of couplings in the different unified scenarios (displayed in Figs.~\ref{fig:alphaonly} to \ref{fig:scen625}) into different cosmological epochs. This produces a first quantitative estimate of the possible time evolution for the various unified scenarios. The choice of epochs is somewhat arbitrary. Two epochs are singled out by events in early cosmology, namely the last scattering surface of CMB, and BBN. The very recent epoch comprises present day laboratory experiments and the Oklo natural reactor, for which a linear interpolation to the present rate of varying couplings seems reasonable. We further divide the observations at intermediate redshift into three epochs.
\begin{itemize}
\item \textbf{Epoch 1:} Today until Oklo\\
 Contains Oklo and laboratory measurements. For the laboratory measurements, we extrapolate the rate of change of the couplings to finite changes at the redshift $z=0.14$ ($t = 1.8 \times 10^9\,$y) of the Oklo event.
\item \textbf{Epoch 2:} $0.2 \le z \le 0.8$\\
 Contains absorption spectra and isotopic abundance measurements in meteorites. We chose a boundary $z=0.8$ since the Murphy dataset \cite{Murphy03.2} has relatively few systems around this redshift,
 making a natural division.
\item \textbf{Epoch 3:} $0.8 \le z \le 2.4$\\
 Contains several absorption spectra measurements. The end of the Tzana\-varis dataset \cite{Tzanavaris06} sets the cut at $z=2.4$.
\item \textbf{Epoch 4:} $2.4 \le z \le 10$\\
 Contains absorption spectra measurements and bounds on $\gnewton$ from neutron stars.
\item \textbf{Epoch 5:} CMB, $z \approx 1100$
\item \textbf{Epoch 6:} BBN, $z \approx 10^{10}$
\end{itemize}

\subsection{Evolution factors}
We define ``evolution factors'' $l_n$ for epochs $n=1,\ldots,6$ by
\beq \label{eq:defln1}
	\Delta \ln G_{k,n} = d_{k} l_n .
\eeq
For each unification scenario we proceed to a quantitative estimate of $l_n$, shown in \tab \ref{tab:multiplierFactors}. The usefulness of considering the evolution factors $l_n$ is that the unknown (and possibly not monotonic) behavior of the mechanism driving the coupling variations is rolled into a finite number of parameters. For a monotonic behavior they satisfy $l_n<l_p$ whenever $z_n<z_p$. The basic assumption remains the proportionality $\Delta \ln G_k(z_n) = d_k l(z_n) = d_k l_n$, with constant unification coefficients $d_k$ independent of the epoch.
The normalization of $l_n$ is arbitrary, and we take for scenarios 2, 5 and 6
\beq
\label{eq:defEvolutionCoefficient}
l_n = \Delta \ln \alpha_{X,n},
\eeq
while for scenarios 3 and 4 we take
\beq
l_n = \Delta \ln (\vev{\phi}/\mgut)_{,n}.
\eeq
For each epoch and scenario, we compute the evolution coefficients $l_n$ as a weighted average over the measurements in the epoch. The representative redshift $z_n$ is the average over the redshifts of observations inside the corresponding epoch. It is shown together with the resulting values for $l_n$ in Tab.~\ref{tab:multiplierFactors}. This table summarizes our results under the assumption of proportionality.

\paragraph{Rates of time variation in the present epoch} 
\ \\
For epoch 1 we incorporate the laboratory measurements for rates of varying couplings by linear extrapolation in time to the Oklo redshift $z_1=0.14$. The logarithmic time derivatives may be approximated by linear interpolation
\beq
	\frac{\dot{G}_k}{G_k} = \partial_t \ln G_k \simeq - \frac{d_k l_1}{t_0-t_1},
\eeq
where $t_1 = 1.8 \times 10^9$y is the time 
corresponding to the redshift $z_1 = 0.14$. 

\paragraph{Method of averaging}
\ \\
We evaluate the weighted average using all values listed in Tab.~\ref{tab:CollectionOfMeasurements}. This procedure may be quite problematic, since sometimes different observations are in manifest contradiction. We take the attitude that, given the possible presence of systematic effects both in spectroscopic determinations of nonzero coupling variations and in the primordial \lise\ abundance, a viable model need not fit all data points. However, even if any given nonzero claimed variation is actually due to systematic error, we still expect the size of the error to be comparable to the size of the claimed variation. Thus, such claims are most conservatively interpreted as bounds on the absolute magnitude of variation. The surviving nonzero variation(s), in addition to the null bounds at other epochs, define a set of evolution factors which must be satisfied by any explicit model of evolution. 

For some scenarios we therefore also evaluate the evolution factors that are obtained by considering that some of the claimed observations of nonzero variation may instead be due to an underestimated systematic error. These alternative evolution factors are given in square brackets, corresponding to the following replacements:\\
Scenario 5, $\tilde{\gamma} = 42$: Neglecting \lise-abundance at BBN\\
Scenario 6, $\tilde{\gamma} = 70$: Neglecting \lise-abundance at BBN\\
Scenario 6, $\tilde{\gamma} = 25$: Replacing the $\mu$ measurements of \cite{Reinhold06} by the conservative upper bound of \cite{Wendt08}.\\
In the case where $\alpha$ alone varies, since the fit including \lise\ is poor we calculate a $2\sigma$ range using observational central values and errors of D and \hefo\ abundances as explained in \sect \ref{sec:VaryingAlphaAlone}.
\begin{table}

\hspace*{-1.5cm}
\begin{tabular}{|l|cccccc|}
\hline
\ \ Epoch        & 1    & 2     & 3    & 4    & 5      & 6  \\
\ \ \ \ \ $z_n$  & 0.14 &  0.53 &  1.6 &  3.8 & $10^3$ & $10^{10}$ \\
\hline
Scenario & $l_1\times 10^6$ & $l_2\times 10^6$ & $l_3\times 10^5$ & $l_4\times10^5$ &
	$l_5\times 10^4$ & $l_6\times 10^3$ \\
\hline
$\alpha$ only 
  & $-0.01 \pm 0.06$ & $-1.1 \pm 1.0 $  & $-0.26 \pm 0.10$ & $-0.85 \pm 0.37$&	
	$-150 \pm 350$ & $5 \pm 34$       \\
2 & $-0.1 \pm 0.1$   & $0.04 \pm 0.03$  & $-0.15 \pm 0.08$ & $0.10 \pm 0.03$ & 
	$0.9 \pm 14$   & $-0.37 \pm 0.20$ \\
3 & $ 4.1  \pm 4.8$  & $-1.5 \pm 1.2$   & $0.42 \pm 3.3$  & $-3.6 \pm 0.9$   & 
	$69 \pm 920$   & $14 \pm 8$       \\
4 & $ 3.9   \pm 8.5$ & $-3.4 \pm 2.7$   & $-8.4 \pm 5.1$  & $-8.7 \pm 2.1$   & 
	$31 \pm 450$   & $33 \pm 21$      \\
5, & $-0.02 \pm 0.18$ & $-0.24 \pm 0.18$ & $-0.25 \pm 0.10$ & $-0.61 \pm 0.13$ & 
	$0.6 \pm 8.6$  & $1.7 \pm 1.1$    \\
($\tilde{\gamma} = 42$) &&&&&& [$0.4 \pm 1.0$] \\
6, & $-0.02 \pm 0.12$ & $-0.10 \pm 0.07$ & $-0.17 \pm 0.07$ & $-0.44 \pm 0.10$ & 
	$0.3 \pm 5.0$  & $1.2 \pm 0.6$    \\
($\tilde{\gamma} = 70$) &&&&&& [$0.3 \pm 0.6$] \\
6, & $-0.12 \pm 0.18$ & $0.04 \pm 0.12$  & $-0.30 \pm 0.11$ & $0.29 \pm 0.08$  & 
	$0.7 \pm 10$   & $-1 \pm 0.3$     \\
($\tilde{\gamma}= 25$) &&&&[$-0.43 \pm 0.28$]&&\\
\hline
\end{tabular}
\caption[Redshifts and evolution factors for each epoch]{Redshifts and evolution factors for each epoch, for the scenarios defined in \sect \ref{sec:SixScenarios}. In the first row the values of $l_n$ give the fractional variation of $\alpha$; in Scenarios 2, 5 and 6 that of $\alphagut$; and in 3 and 4 that of $\vev{\phi}/\mgut$. Values in brackets give, for BBN ($l_6$) the evolution factors neglecting \lise; or for $l_4$, the evolution factor with the $\Delta \mu/\mu$ value of \cite{Reinhold06} substituted by that of \cite{Wendt08}.}
\label{tab:multiplierFactors}
\end{table}

\subsection{Monotonic evolution with unification}
\label{sec:monotony}
It seems natural to expect that variations of constants, if they occurred, evolve monotonically\footnote{In \sect \ref{sec:Growing} we will also consider scenarios of quintessence where the implied variation is non-monotonic.}. Looking on Figs.~\ref{fig:alphaonly} to \ref{fig:scen625} and Tab.~\ref{tab:multiplierFactors}, we can ask the question whether the claimed variations are consistent with monotonic variation within the specific GUT scenarios.
Here we briefly summarize whether the unified scenarios we consider can be consistent with a monotonic evolution of the single underlying varying parameter.

\paragraph{Varying $\alpha$ only}
\ \\
Although variation of $\alpha$ alone does not help to account for deviation of BBN abundances from standard theory, or for any nonzero variation of $\mu$, the cosmic history is interesting due to the significant nonzero value in Epochs 3 and 4. The Oklo bound in Epoch 1 restricts the present time variation to $3.7\times 10^{-17}\,{\rm y}^{-1}$ (assuming no acceleration of $\partial_t \alpha$). 

\newpage

\paragraph{Scenario 2}
\ \\
Scenario 2 favors a negative variation of $\alphagut$ at BBN, and a negative variation may also fit the M$\alpha$ results. However, the Reinhold $\mu$ measurement indicates a positive, but much smaller, variation. The R$\mu$ results dominate the weighted average for $l_4$ due to their small error on $\Delta \ln \alphagut$. The ratio $\Delta \ln \mu / \Delta \ln \alpha = 27$ makes this scenario unlikely to fit the reported signal of nonzero $\Delta \alpha$.

\paragraph{Scenario 3}
\ \\
In scenario 3 a positive variation of $\vev{\phi}/\mgut$ is favored by BBN. The high ratio \linebreak $\Delta \ln \mu / \Delta \ln \alpha \simeq -325$ makes the bounds obtained on a variation of $\mu$ strongly inconsistent with the claimed size of variation of $\alpha$. 
The Reinhold \etal values again dominate the results for $l_4$.

\paragraph{Scenario 4}
\ \\
In this scenario, the ratio $\Delta \ln \mu / \Delta \ln \alpha = -22$ is again large and makes any observation of significant nonzero $\Delta \ln \alpha$ unlikely. Both the M$\alpha$ and the R$\mu$ measurements point in opposite direction to BBN; however the two spectroscopic observations are also inconsistent with each other, within the scenario. Again, the R$\mu$ results dominate the determination of $l_4$ 
due to the small error.

\paragraph{Scenario 5, $\tilde{\gamma} = 42$}
\ \\
In this scenario the variation of $\alphagut$ favored by BBN is positive ($l_6 = (1.7 \pm 1) \times 10^3$), however both nonzero variations from spectroscopic data M$\alpha$ and R$\mu$ require negative variations. With $\Delta \ln \mu / \Delta \ln \alpha = -6$ the spectroscopic measurements appear consistent with each other. Hence one would require some non-monotonic evolution to fit nonzero variations both at BBN and at moderate $z$. 
In Table~\ref{tab:multiplierFactors} we have also evaluated $l_6$ using only the constraints given by D and \hefo\ (in brackets).

\paragraph{Scenario 6, $\tilde{\gamma} = 70$}
\ \\
As in the preceding scenario, BBN favors a positive variation in $\alphagut$, but M$\alpha$ and R$\mu$ favor negative. Again, Fig.~\ref{fig:scen670} may suggest a non-monotonic evolution. 
Fitting to BBN including \lise\ we would obtain $l_6 = (1.2 \pm 0.6) \times 10^{-3}$; \tab \ref{tab:multiplierFactors} also displays in brackets the value of $l_6$ obtained from D and \hefo\ bounds only.

\paragraph{Scenario 6, $\tilde{\gamma} = 25$}
\ \\
In this scenario, both BBN and the M$\alpha$ signal favor a negative variation of $\alphagut$, whereas the R$\mu$ observations point towards a positive variation. Following the argument of Wendt {\it et al}.~\cite{Wendt08}, we substitute the R$\mu$ value by the null constraint $|\Delta \mu / \mu| \le 2.5 \times 10^{-5}$ \cite{Wendt08} to obtain the bracketed value of $l_4$ in Table~\ref{tab:multiplierFactors}. In this scenario the evolution factors show a crossover from negligible variation at low redshift, to strong and monotonically increasing negative variation at $z \approx 2$. 

\newpage

\subsection{Tension between the \lise\ problem and variation of $\mu$}
\label{sec:TensionLithiumMu}
Measurements of the primordial \lise\ abundance show that the BBN abundance needs to decrease below the standard value to fit the observations, whereas the Reinhold $\mu$ measurement indicates $\mu$ to increase at $z\simeq 3$. We find that for all our unification scenarios the sign of the dependence on the fundamental parameter is the same for $\mu$ and \lise. Moreover, the coefficients of this dependence are nearly identical up to a common factor; hence the induced variations for $\mu$ and \lise\ point in the same direction, in contradiction to the tendency inferred from the observations. For example, for scenario 5 we find
\begin{align}
 \Delta \ln \mu &= (23.2 - 0.65 \tilde{\gamma}) \Delta \ln \alphagut, \nonumber \\ 
 \Delta \ln \mbox{\lise} &= (1692 - 49 \tilde{\gamma}) \Delta \ln \alphagut.
\end{align}
These expressions change sign at $\tilde{\gamma} = 35.7$ and $34.5$, respectively. For a monotonic evolution, there is no possibility to have both a significant variation of $\mu$ and a variation of opposite sign in the \lise\ abundance. (In the regime $\tilde{\gamma} \approx 35$ there is no $2 \sigma$ fit to BBN.) A similar result can be found for scenario 6 (including the SUSY partner mass dependence, which shows the same sort of degeneracy). Note that scenario 2 and 3 are just limiting cases of scenarios 5 and 6.

The main reason for this behavior is that variations of \lise\ and $\mu$ are dominated by the variations of $\hat{m}/\lqcd$ and $m_e/\lqcd$, respectively, with the same sign of prefactor. This degeneracy can be broken if $m_e$ varies differently from the quark masses, a possibility that we do not consider in this thesis. For our scenarios with constant $\hat{m}/m_e$, the conflict between a monotonic time evolution and the $\mu$- and \lise-observations is reflected in the opposite signs of $l_4$ and $l_6$.

\begin{figure}[tb]
\begin{center}
 \includegraphics[width=13.5cm]{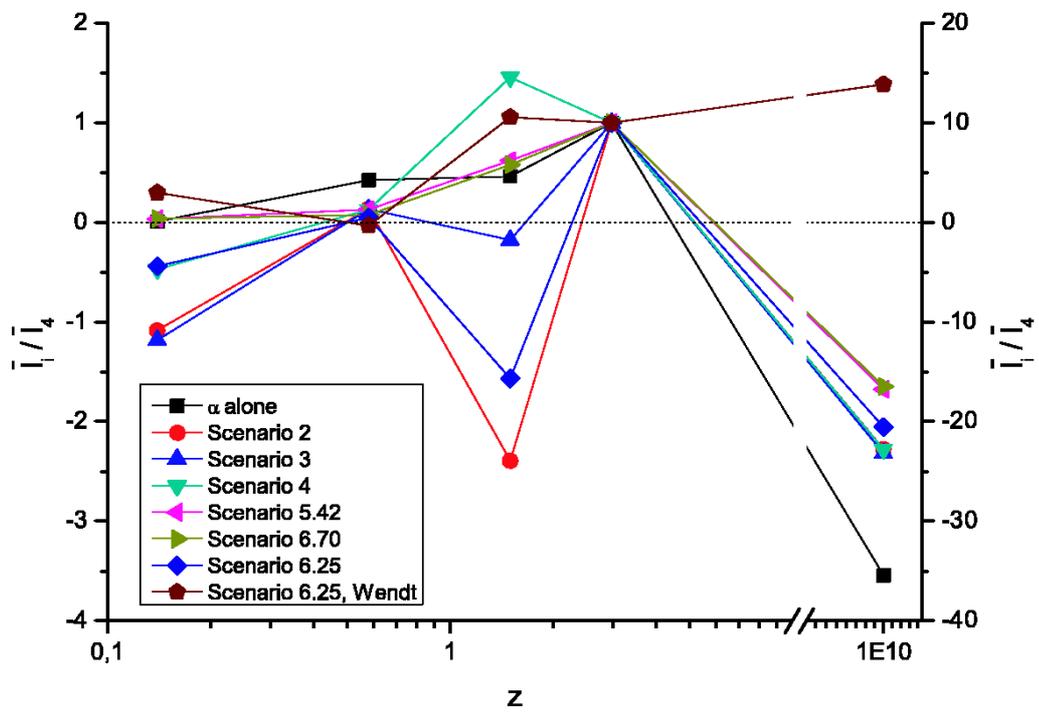}
\end{center}
\caption[Normalized evolution factors $\bar{l}_i/\bar{l}_4$ for each scenario]{Normalized evolution factors $\bar{l}_i/\bar{l}_4$ for each scenario, where $\bar{l}_i\equiv l_i/\ln(1+z_i)$.}
\label{fig:lionl4}
\end{figure}

This observational tension for monotonic behavior is clearly depicted in Fig.~\ref{fig:lionl4}, 
where we plot simultaneously the averaged observational values of evolution factors $l_i/\ln(1+z_i)$, normalized to $l_4/\ln(1+z_4)$. For Scenario 6, $\tilde{\gamma}=25$, we also display the result obtained by substituting the Wendt \etal value of $\mu$ variation for that of \cite{Reinhold06}.
The factor $\ln(1+z_i)$ is introduced as a convenient normalization to avoid compressing the scale of variations excessively in recent epochs.\footnote{In quintessence-like theories, if the scalar field contributes a constant fraction of the total energy density of the Universe, as in so-called ``tracker'' models, the evolution of the field is typically also proportional to $\ln(1+z)$. This is an additional motivation for our normalization.} For the purpose of a quick inspection we have omitted the error bars, which are of course necessary for a quantitative interpretation.

\chapter{Probing quintessence models}
\label{chap:ProbingQuintessence}

In this chapter we will describe how measurements of varying constants can be used to constrain models of quintessence under the assumption of grand unification. For that purpose we will use the measurements which we listed in \chap \ref{chap:ExpTestofVariations} in the six unified scenarios studied in the preceding chapter to constrain the free parameters of the quintessence models described in \chap \ref{chap:Quintessence}. Our aim is to see to what extent such models can be consistent with the behavior of time variations that we have outlined.

\section{Crossover quintessence}

As we have seen in \sect \ref{sec:CrossoverQuintessence}, crossover quintessence models yield a monotonic evolution of the cosmon and hence also a monotonic variation of constants, assuming a constant coupling $\delta$ to the fundamental varying parameter, 
\beq \label{eq:DalphaXgrowing2}
	\Delta \ln \alphagut(z) = \delta( \vp(z) - \vp(0)).
\eeq
We discussed the viability of monotonic evolution in \sect \ref{sec:monotony} where a first judgment can be made by inspection of Fig.~\ref{fig:alphaonly} to \ref{fig:scen625} for the various unification scenarios, or by inspection of Table~\ref{tab:multiplierFactors}.

To allow us to easily compare with the observational results, we observe that in Table~\ref{tab:multiplierFactors} the constraints for $l_5$ are considerably weaker than those for the remaining evolution factors. Furthermore, the one sigma range for $l_4$ is nonzero for all scenarios.
Hence we shall compare observational and theoretical values for the ratios $l_1/l_4$, $l_2/l_4$, $l_3/l_4$ and $l_6/l_4$. We note the opposite sign of $l_3$ and $l_4$ for scenario 2, 3 and scenario 6 with $\tilde{\gamma}=25$, which disfavors any monotonic evolution for these scenarios. The averaged observational values in each epoch are given in Table~\ref{tab:RatioMultiplierFactors}.
\begin{table}
\center
\begin{tabular}{|c|cccc|}
\hline
Scenario & $l_1/l_4$ & $l_2/l_4$ & $l_3/l_4$ & $l_6/l_4$\\
\hline
0   & $  0.00 \pm 0.01$ & $0.13 \pm 0.13$ & $0.31  \pm 0.18$ & $-592 \pm 4031$\\
2   & $ -0.10 \pm 0.12$ & $0.04 \pm 0.04$ & $-1.59 \pm 0.88$ & $-380 \pm 228$ \\
3   & $ -0.11 \pm 0.13$ & $0.04 \pm 0.03$ & $-0.12 \pm 0.91$ & $-387 \pm 241$ \\
4   & $ -0.04 \pm 0.10$ & $0.04 \pm 0.03$ & $0.97  \pm 0.64$ & $-381 \pm 259$ \\
5, $\tilde{\gamma} = 42$  & $ 0.00 \pm 0.03$ & $0.04 \pm 0.03$ & $0.41 \pm 0.19$ &
$-280 \pm 191$ \\
without BBN               &          ''       &       ''        &        ''        &
$-66 \pm 165$  \\
6, $\tilde{\gamma} = 70$  & $ 0.00  \pm 0.03$ & $0.02 \pm 0.02$ & $0.39  \pm 0.17$ &
$-275 \pm 150$ \\
without BBN               &          ''       &       ''        &        ''        &
$-69 \pm 139$  \\
6, $\tilde{\gamma} = 25$  & $ -0.04 \pm 0.06$ & $0.02 \pm 0.04$ & $-1.04 \pm 0.49$ &
$-343 \pm 142$ \\
with Wendt               & $ 0.03  \pm 0.05$ & $-0.01\pm 0.03$ & $0.70 \pm 0.52$  & 
$231 \pm 163$  \\
\hline
\end{tabular}
\caption{Ratios of the evolution factors from observations.}
\label{tab:RatioMultiplierFactors}
\end{table}
Note that the coupling $\delta$ drops out of the ratios $l_i/l_j$. Considering these ratios allows us to probe quintessence directly without knowing the absolute size of the coupling, only assuming its (approximate) constancy over the relevant range of evolution. In view of its monotonic evolution, crossover quintessence cannot give negative ratios $l_i/l_j$. Hence it cannot be a good fit to the lithium abundance within the unification scenarios 2 to 6 that we consider. This reflects the tension between the \lise\ problem and a positive variation of $\mu$ discussed in \sect \ref{sec:TensionLithiumMu}. 

For a simulation of the crossover quintessence model, we follow the procedure described in \sect \ref{sec:CrossoverQuintessence}. By specifying the present densities of matter, radiation and dark energy and the model parameters $w_{h0}$ and $z_+$, we can trace back the evolution of quintessence and the other components of our Universe. In Tab.~\ref{tab:TheoreticalRatios} we display the ratios $l_i/l_4$ expected from crossover quintessence for various parameters $w_{h0}$ and $z_{+}$, which can be compared with the observed ratios displayed in Tab.~\ref{tab:RatioMultiplierFactors}. Considering the lithium abundance to be affected by astrophysical systematics in scenarios 5 and 6 ($\tilde{\gamma} = 70$), or using the null result for $\Delta \ln \mu$ at intermediate redshift for scenario 6 ($\tilde{\gamma} = 25$ ``with Wendt''), we find that some crossover quintessence models indeed yield the observed order of magnitude for the ratios $l_i/l_4$. We conclude that crossover quintessence could, in principle, reconcile a coupling variation of the claimed size in epochs 3 and 4 with the bounds from late cosmology, \ie epochs 1 and 2. This is due to the ``slowing down'' of the cosmon evolution, as noted in \cite{Wetterich03}. Values of the present equation of state $w_{h0}$ quite close to $-1$ would be required, however. In other words, an observation of coupling variations would put strong bounds on the dynamics of the cosmon field\footnote{The current WMAP5 bound on the equation of state parameter of quintessence is $w = -0.972 \pm 0.06$ \cite{WMAP5}, but the derivation assumes a constant $w$ over a certain range of redshift, while our $w_{h0}$ gives the equation of state at the present time.} and provide for an independent source of information about the properties of dark energy.
\begin{table}
\center
\begin{tabular}{|cc|cccc|}
\hline
$w_{h0}$ & $z_{+}$ & $l_1/l_4$ & $l_2/l_4$ & $l_3/l_4$ & $l_6/l_4$ \\
\hline
-0.95    & 3 & 0.12 & 0.38 & 0.67 & 26\\
-0.99    & 3 & 0.09 & 0.28 & 0.55 & 38\\
-0.9999  & 3 & 0.02 & 0.10 & 0.37 & 54\\
-0.95    & 7 & 0.15 & 0.47 & 0.80 & 13\\
-0.99    & 7 & 0.15 & 0.47 & 0.80 & 24\\
-0.9999  & 7 & 0.08 & 0.27 & 0.52 & 125\\
-0.999999& 7 & 0.02 & 0.09 & 0.34 & 177\\
\hline
\end{tabular}
\caption[Ratios of evolution factors from crossover quintessence.]{Ratios of evolution factors from crossover quintessence. The $l_i$ are evaluated by averaging over the variations evaluated at the same redshift as the data in each epochs (weighting by the number of absorption systems if appropriate).}
\label{tab:TheoreticalRatios}
\end{table}

Note that observational probes of dark energy would {\em not}\/ give results for $w_{h0}$ that coincide with the values that we take in our model. Such probes do not actually measure the present-day equation of state, rather they extrapolate $w_0$ from past epochs under some parameterization.

\newpage

\section[Models with growing neutrinos]{Models with growing neutrinos and oscillating \\variation}
\label{sec:Growing}

In this section we investigate the two models of growing neutrinos which were laid out in \sect \ref{sec:GrowingNeutrinoModels}. These models do not obey the proportionality of all coupling variations for all redshifts and do not show a monotonic evolution of the cosmon field. A systematic analysis of all such models seems difficult, hence we concentrate on the specific examples given in \cite{Amendola08} and \cite{Wetterich08}.

The combined variation is different for each of the unified scenarios. For all unified parameters except $\vev{\phi}/\mgut$ we still have a proportionality for the variations at all $z$, since the variations are proportional to $\vp$. The variations due to the direct coupling \eqref{eq:DalphaXgrowing} can be described by our method of evolution factors $l_n$, even though the $l_n$ need not be strictly monotonic due to the oscillations in $\vp$. However, for $\vev{\phi}/\mgut$ we now have one variation linearly proportional to the variation of $\vp$, \eqn \eqref{eq:DalphaXgrowing} and an additional one with a nonlinear dependence, \eqn (\ref{eq:phiRofz}-\ref{eq:RofzphiWett}). A simple treatment with common evolution factors for all variations will no longer be applicable. Due to the additional $\vp$-dependence of $\vev{\phi}/\mgut$ we may have separate evolution factors for $\vev{\phi}/\mgut$, different from the (common) evolution factors for the other couplings. 

For example, the ``linear contribution'' \eqref{eq:DalphaXgrowing} may dominate at BBN and induce a positive $l_6$ common for all couplings. In the range $z<10$ the ``non-linear contribution'' \eqref{eq:phiRofz} could be more important, leading to effectively negative $l_{3,4}$ for $\vev{\phi}/\mgut$. (Such an effect could, in principle, relieve the tension between \lise\ and a positive $\mu$-variation at high $z$ which was explained in \sect \ref{sec:TensionLithiumMu}.) In practice, we calculate $\Delta \ln \alphagut$ and $\Delta \ln \frac{\vev{\phi}}{\mgut}$ at each epoch directly from the model, and extract the varying couplings and observables as explained in \chap \ref{chap:VariationsFromBBNtoTodayinGUT}.
Then we may search for a set of parameters $\delta$, $R_0$ which minimizes the $\chi^2$ for all measured variations. 

\subsection{The stopping growing neutrino model}

The stopping growing neutrino model which was introduced in \sect \ref{sec:StoppingGrowingModel} has an oscillation in $\vev{\phi}$ that grows both in frequency and amplitude at late times as $\vp$ approaches its asymptotic value. Such oscillations must not be too strong as measurements between $z=2$ and today would measure a high rate of change. The oscillation may be made arbitrarily small by choosing small $R_0$. The restrictions from the low-z epochs are actually so strong that to a good approximation the non-linear contribution $\sim R_0$ can be neglected. However, the linear variation \eqref{eq:DalphaXgrowing} is independent of $R_0$. It can be described by our method of evolution factors and yields for the set of parameters given in \cite{Wetterich08} ($\vp_t \approx 27.6$, $\alpha=10$, $\epsilon = -0.05$) the ratios 
\bea
  l_1/l_4 &=& 0.008, \nonumber\\
  l_2/l_4 &=& 0.09, \nonumber \\
  l_3/l_4 &=& 0.44, \nonumber \\
  l_6/l_4 &=& 175.
\eea
Comparing this with the numbers given in Table~\ref{tab:RatioMultiplierFactors} shows that this model naturally yields evolution factors which are of the correct order of magnitude. We emphasize that no new parameter has been introduced for this purpose.

\subsection{Global fit to the scaling growing neutrino model}
Each growing neutrino model contains a few parameters that determine the cosmological evolution of the cosmon $\vp$, and the Higgs v.e.v.\ $\vev{\phi}$. Two coupling parameters give respectively the relative strength of variation of $\alphagut$ with $\vp$, and the relative strength of the additional variation of the Higgs v.e.v.\ due to the varying triplet. For each example of cosmological evolution we may calculate the observables directly in terms of the two coupling parameters and make a global fit for their values. In performing the fit we take the 125 systems of the Murphy \etal $\alpha$ determination \cite{Murphy03.2} within Epochs 3 and 4 (\eqn \eqref{eq:Murphysplit2}) and further split them into 5 subsamples each with 25 absorption systems, since the data set extends over a wide range of redshift where there may be significant oscillations.

For the global fits, we take the scaling growing neutrino model (\sect \ref{sec:ScalingGrowingModel}) with $\beta = -52$, $\alpha = 10$, $m_{\nu,0} = 2.3$eV. With zero variation at all times (no degrees of freedom), we find $\chi^2 = 3.25$ including \lise\ at BBN, and $\chi^2 = 2.40$ neglecting \lise. The results of the best fits with varying couplings are given in Table~\ref{tab:chi2origAmendola}.
\begin{table}[h]
\centering
\begin{tabular}{|l|cc|cc|}
\hline
Scenario        & $\delta \times 10^4$ & $R_0$ & $\chi^2$ & $\Delta \chi^2$\\
\hline 
2               &-0.019 & 0.045& 3.09 & 0.16 \\
2 without Li7   &-0.040 & 0    & 2.09 & 0.31 \\
3               & 1.64  & 0    & 2.85 & 0.40 \\
3 without Li7   & 1.55  & 0    & 2.03 & 0.37 \\
4               & 3.79  & 0    & 2.85 & 0.40 \\
4 without Li7   & 3.80  & 0    & 1.98 & 0.42 \\
5.42            & 0.30  & 0    & 1.96 & 1.29 \\
5.42 without Li7& 0.24  & 0    & 1.87 & 0.53 \\
6.70            & 0.20  & 0    & 1.93 & 1.32 \\
6.70 without Li7& 0.16  & 0    & 1.87 & 0.53 \\
6.25            & 0.18  & 0.090& 2.73 & 0.52 \\
6.25 without Li7&-0.055 & 0.061& 2.20 & 0.20 \\
\hline
\end{tabular}
\caption[Fit to the scaling growing neutrino model]{Fitting parameters and minimal $\chi^2$ values for the different unification scenarios for best fit to the scaling growing neutrino model \cite{Amendola08}. The last column gives the increase in $\chi^2$ produced when $\delta$ and $R_0$ are forced to vanish, \ie for zero variation.}\label{tab:chi2origAmendola}
\end{table}

It turns out that $\chi^2$ cannot be reduced by more than $1.3$ in the fit including \lise\ and $0.53$ in the fit neglecting \lise, which we do not consider as convincing evidence for coupling variations within this model. We have investigated some other choices of parameters for the cosmological evolution and also the stopping growing neutrino model, without a substantial change in the overall situation. In view of the unsettled status of the observational data it seems premature to make a systematic scan in parameter space. Our investigation demonstrates, however, how a clear positive signal for a coupling variation could restrict the parameter space for quintessence models.

Most of the additional variation of the Higgs v.e.v.\ occurs at later epochs, $z<2$, thus recent observational bounds rule out any significant additional growth in $\vev{\phi}$. We considered fitting the observational values excluding BBN, as a function of the model parameters $\delta$ and $R_0$, and we find always that the value of $R_0$ at the minimum of $\chi^2$ is unobservably small.

\section{Tests of the weak equivalence principle}
\label{sec:WEP}

In Sec.~\ref{sec:ViolationOfEquivalencePrinciple} we have explained that variations of constants will influence the outcome of tests of the weak equivalence principle (WEP). Besides variations of couplings, the cosmon coupling to atoms also influences the outcome of tests of the weak equivalence principle, as it may also produce local gravitational effects which violate the WEP due to their interactions with matter. In \sect \ref{sec:PresentBounds} we use both aspects to set further constraints on a possible time variation of couplings. 
Test bodies with different composition have in general different couplings to the cosmon, and will hence experience different accelerations towards a common source. Usually, deviations from the universality of free fall are measured in terms of the E{\" o}tv{\" o}s parameter $\eta$ \cite{Wetterich02.1,Dent06}
\beq \label{eqn:etadef}
	\eta^{b-c} \equiv \frac{2|a_b-a_c|}{|a_b+a_c|} \; ,
\eeq
where $a_{b,c}$ are the accelerations towards the source of the two test masses. The experiment setting the currently tightest limits on $\eta$ \cite{Schlamminger07} has the result
\beq \label{eqn:Schlameta}
	\eta = (0.3\pm 1.8)\times 10^{-13}
\eeq
for test bodies of Be ($A=9$, $Z=4$) and Ti ($A=48$, $Z=22$) composition, where the gravitational source is taken to be the Earth. 

In contrast to the direct observations of time varying couplings, tests of the universality of free fall do not determine directly the values of fundamental `constants' or their possible variations. However, given our basic assumption of a slow time variation, driven by a light scalar degree of freedom, the current limits on composition-dependent long range forces put bounds on the scalar couplings to different constituents of matter. In our language, they measure or constrain the coefficients $\beta_k$ at $z=0$, which relate the evolution factors and the changes in the cosmon field (see \eqn \eqref{eq:defln1}),
\beq \label{eqn:betaDef}
	\Delta \ln G_{k} = d_{k} l = \beta_{k} \Delta \vp(z_n) \; .
\eeq
These constraints then imply bounds on the time variation of constants: Differentiating with respect to time, \eqn \eqref{eqn:betaDef} becomes
\beq
	\frac{\dot{G}_{k}}{G_k} = \beta_{k} \dot{\vp} \; .
\eeq
Applying the conservative bound \cite{Dent06} $\dot{\vp}/H_0 \le 0.7$, we derive
\beq
 \dot{\vp} \le \dot{\vp}_{\rm max} \simeq 5 \times 10^{-11} {\rm y}^{-1} \; .
\eeq
$\beta_k$ can be derived as a function of $\eta$, utilizing a model of nuclear masses and binding energies and using GUT relations to reduce the number of free parameters (see \cite{DSW08.2} for details on this treatment). It turns out that WEP violation places significant bounds on the present-day values of scalar couplings, as will be shown in the next section.

The differential acceleration $\eta$ for two bodies with equal mass but different composition and therefore different ``cosmon charge'' is related to the present time variation of couplings and cosmological parameters via \cite{DSW08.2}
\beq \label{eqn:DifferentialAccelerationParam}
 \eta \simeq 3.8 \times 10^{-12} \left( \frac{\dot{\alpha}/\alpha}{10^{-15}{\rm y}^{-1}} \right)^2 
 \frac{ F
}{\Omega_h (1 + w_h)}\, .
\eeq
Here $\dot{\alpha}/\alpha$ is the present relative variation of the fine structure constant in units of year$^{-1}$, $w_h$ is the present dark energy equation of state and $\Omega_h \approx 0.73$ the present dark energy fraction. The ``unification factor'' $F$ encodes the information on the specific relations between variations in fundamental parameters as implied by the different GUT scenarios, and the composition of the test bodies. The factor $F$ is displayed in Tab.~\ref{Tab:FforUnifiedScenarios} for our different unified scenarios and for the Be-Ti test masses of \cite{Schlamminger07}. For typical test mass compositions we find $1 \leq F\leq \rm{few}\times 10^2$. The very large value of $F$ for Scenario 3 reflects an accidental cancellation of terms which we do not consider to be typical.
\begin{table}
\center
\begin{tabular}{|c||c|c|c|c|c|c|}
\hline
Scenario & 2 & 3 & 4 & 5, $\tilde{\gamma}\! =\! 42$ & 6, $\tilde{\gamma}\! =\! 70$ & 6, $\tilde{\gamma}\! =\! 25$ \\ 
\hline 
$F$ (Be-Ti)& 95    & -9000  & -165   & -25   & -26   & 41 \\
\hline
\end{tabular}
\caption[Values of $F$ for the unified scenarios studied in this thesis]{Values of $F$ for a WEP experiment using Be-Ti masses for the unified scenarios studied in this thesis.} 
\label{Tab:FforUnifiedScenarios}
\end{table}

Once $F$ is fixed, the relation \eqref{eqn:DifferentialAccelerationParam} allows for a direct comparison between the sensitivity of measurements of $\eta$ versus the measurements of $\dot{\alpha}/\alpha$ from laboratory experiments, or bounds from recent cosmological history, such as from the Oklo natural reactor or the isotopic composition of meteorites.

\section[Bounds on present-day variation]{Bounds on present-day variation}
\label{sec:PresentBounds}

Within our theoretical framework there exist three distinct ways to bound or measure the present-day rate of variation of fundamental parameters. The first is a direct measurement, of the type probed by atomic clock experiments (see \sect \ref{sec:clocks}). If one or more nonzero variations are found in this way, bounds on unified models may immediately be set. The second method is by combining information on the size of scalar field couplings from WEP tests (Section~\ref{sec:WEP}) with a cosmological upper bound on the kinetic energy of scalar fields \cite{Wetterich02.1,Dent06}. Such bounds on scalar couplings will depend on the choice of unified model and in general will be independent of those derived from atomic clocks. Thirdly, under the assumption of a monotonic variation (that also does not significantly accelerate with time), we may convert any ``historic'' bound on the net variation of a fundamental parameter since a given epoch into a bound on the present rate of variation:
\beq \label{historic}
	|\dot{G}_k| \leq (t_0-t_n)^{-1} \left| G_k(t_0)-G_k(t_n)\right| \equiv \frac{|\Delta G_k|}{\Delta t},\qquad t_n < t_0.
\eeq
Here $t_0$ denotes the present, and any nonzero rate of variation should have the appropriate sign, \ie $\dot{G}_k$ has the opposite sign to $\Delta G_k$ referring to some past epoch.

For any given unified model of time variations, the three bounds on present-day evolution will have different sensitivity. Therefore if one method gives a nonzero variation we would (in some cases) be able to distinguish between models due to the fact that the other bounds are still consistent with zero. To give a simple example, the direct detection of a nonzero time variation in atomic clocks near the present upper bound would immediately rule out a large class of models that cannot account for such a variation without leading to WEP violation above current bounds; and would also rule out models in which such nonzero variations extrapolated to past epochs $t_n<t_0$ would exceed observational bounds.

However, this chain of inference does not function equally well in all directions. A nonzero finding of differential acceleration violating the WEP would indicate nontrivial scalar couplings, but need not imply nonzero time variation since the rate of change of the scalar is not bounded from below. Also, a nonzero variation at some past epoch $t_n$ would not necessarily imply a lower bound to the present-day rate of variation or size of scalar couplings, since the variation could have slowed substantially since then (either due to nonlinear scalar evolution or a nonlinear coupling function). Only with the assumption of a reasonably smooth and monotonic variation of the scalar field and its coupling functions, one can find, for any given unified model, where the first signals of present-day or recent variation are expected to appear. 

At present these methods give null results up to redshifts of about $0.8$, but if a nonzero time variation exists, we can determine for each unified scenario which observational method is most sensitive. Thus if a nonzero signal of late time variation arises it may be used to distinguish between models. 
We assume for this purpose an approximately linear variation over recent cosmological times, thus measurements of absolute variation at nonzero redshift $z$ imply time derivatives
\beq
  \frac{d \ln X}{dt} \simeq \frac{\Delta \ln X (z)}{t_0 - t(z)}.
\eeq
Here $X$ is the fundamental varying parameter: we consider first $X\equiv \alpha$, if only the fine structure varies; in scenario 1 $X\equiv \gnewton m_N^2$, in scenarios 2, 5 and 6 $X \equiv \alphagut$ and in scenarios 3 and 4 $X \equiv \vev{\phi}/\mgut$. Then Table~\ref{tab:PresentVariationErrors} gives the precision of bounds on time derivatives for the unified scenarios we consider, except scenario 1 (varying $\gnewton$) which is probed by a quite different set of measurements. As explained in \sect \ref{sec:Oklo}, we take the Oklo bound as applying directly to the variation of $\alpha$, and increase its uncertainty by a factor 3 to account for possible cancellations when other parameters also vary. We present the recent Rosenband \etal \cite{Rosenband08} Al/Hg ion clock bound separately to illustrate to what extent it improves over previous atomic clock results.

Extending this method beyond $z \approx 0.5$ becomes questionable. One could use linearity in $\ln(1+z)$ instead of $t$, but even this improvement may lead to unreliable extrapolations for models with a particular dynamics of the scalar field, as crossover quintessence.
\begin{table}
\center
\begin{tabular}{|lc|cccccc|}
\hline
Scenario & $X$              & Clocks         & Al/Hg  & WEP & Oklo $\alpha$ & Meteorite & Astro
\\
\hline
$\alpha$ only & $\alpha$    & 0.13 ($\alpha$)& 0.023 & 6.2   & 0.033 & 0.32   & 0.44 ($y$) \\
%1        & $G_{\rm N}m_N^2$ &  \\
2        & $\alpha_X$       & 0.074 ($\mu$)  & 0.027 & 0.007 & 0.12  & 0.015 & 0.006 (NH$_3$) \\
2S       & $\alpha_X$       & 0.12  ($\mu$)  & 0.044 & 0.012 & 0.19  & 0.026 & 0.010 (NH$_3$) \\
3        & $\vev{\phi}/M_X$ & 2.6 ($\mu$)    & 12.4  & 0.33  & 54    & 0.53  & 0.22  (NH$_3$) \\
4        & $\vev{\phi}/M_X$ & 6.2 ($\mu$)    & 1.78  & 0.35  & 7.7   & 1.2   & 0.51  (NH$_3$)
\\
5, $\tilde{\gamma} = 42$ & $\alpha_X$ & 0.32 ($\alpha$) & 0.024  & 0.013 & 0.11  & 0.069 & 0.035
(NH$_3$) \\
6, $\tilde{\gamma} = 70$ & $\alpha_X$ & 0.21 ($\alpha$) & 0.016  & 0.008 & 0.070 & 0.049 & 0.025
(NH$_3$) \\
6, $\tilde{\gamma} = 25$ & $\alpha_X$ & 0.25 ($\mu$)    & 0.027  & 0.011 & 0.12  & 0.056 & 0.021
(NH$_3$) \\
\hline
\end{tabular}
\caption[Competing bounds on recent time variations in unified scenarios]{Competing bounds on recent ($z\leq0.8$) time variations in unified scenarios. For each scenario we give $1\sigma$ uncertainties of bounds on $d(\ln X)/dt$ in units $10^{-15}{\rm y}^{-1}$, where $X$ is the appropriate fundamental parameter. For the Oklo bound we multiply the uncertainty by a factor 3 except when only $\alpha$ varies. The column ``Clocks'' indicates whether $\alpha$ or $\mu$ gives the stronger bound; the recent Al/Hg limit \cite{Rosenband08} is given a separate column. The column ``Astro'' indicates which measurements of astrophysical spectra are currently most sensitive in each scenario.}
\label{tab:PresentVariationErrors}
\end{table}

%%%%%%%%%%%%%%%%%%%%%%%%%%%%%%%%%%%%%%%%%%%%%%%%%%%%%%%%%%%%%%%%%%%%%%%%%
%%%%%%%%%%%%%%%%%%%%%%     CONCLUSION     %%%%%%%%%%%%%%%%%%%%%%%%%%%%%%%
%%%%%%%%%%%%%%%%%%%%%%%%%%%%%%%%%%%%%%%%%%%%%%%%%%%%%%%%%%%%%%%%%%%%%%%%%

\chapter{Conclusion and outlook}
\label{chap:Conclusion}

We have developed a systematic method to relate variations of fundamental parameters of particle physics to the primordial isotope abundances produced by BBN. The main advantage of the method, which is laid out in part two of this thesis, is that we are able to vary every parameter independently, both at the level of fundamental Standard Model parameters and of nuclear physics parameters, thus we are not dependent on any particular theoretical model which enforces particular relations between the variations. 

We follow a two step approach, first extracting the nuclear parameter dependence (without major theoretical uncertainties) and in a second step relating this to variations of fundamental Standard Model parameters. We define two linear response matrices, where the first, $C$, encodes the change in predicted abundances produced by small variations away from the current values of nuclear physics parameters which enter the BBN integration code. These parameters comprise the gravitational constant, fine structure constant, neutron lifetime, electron, proton and neutron masses, and binding energies of $A\leq7$ nuclei. The dependences of nuclear reaction rates on these parameters are also implemented insofar as they are calculated within some effective theory. One notable result is that the \lise\ abundance depends heavily on the binding energies of \heth, \hefo\ and \bese.

We also investigated possible further effects of variations in nuclear reaction rates on predicted abundances by varying each rate ({\it i.\,e.}\ thermal integrated cross-section $\vev{\sigma v}$) separately by a temperature-independent factor. We find that the \hefo\ abundance is insensitive to nuclear rates, and only eight reactions could lead to significant variation of the D, \heth\ or \lise\ abundances. 

The second response matrix, $F$, relates variations in nuclear parameters to the fundamental parameters of particle physics, comprising the gravitational constant, fine structure constant, Higgs vacuum expectation value, electron mass, and the light (up and down) quark masses. At this point theoretical uncertainty enters into the relation between quark masses and nuclear binding energies. We parameterize the dependence of binding energies on the pion mass (and hence on light quark masses) by the deuteron binding, which has been treated by a systematic expansion in effective field theory. 

The resulting fundamental response matrix $R=CF$ allows us, first, to bound the variations of the six fundamental couplings individually, some bounds being at the percent level. Secondly, studying three exemplary unified scenarios, we can also bound correlated variations affecting many couplings at once. We find that one scenario allows us to fit observed D, \hefo\ and \lise\ abundances within $2\sigma$ bounds, given a variation $\Delta\alpha/\alpha = -2\times 10^{-4}$ away from the present value; another fits these observational abundances within almost 1$\sigma$ bounds, given a variation $\Delta\alpha/\alpha = 4\times 10^{-4}$. For this analysis we have left the linear matrix approach and make use of our BBN code which allows us to derive the full nonlinear dependence.\\

In a next step (part three of the thesis), we combine our findings for BBN with further measurements of the variation of fundamental constants from BBN to today. Within grand unified theories, this set of different varying parameters can be consistently reduced to a variation of a few ``unification parameters'', namely the unification scale $\mgut/\mplanck$, gauge coupling $\alphagut$, the Fermi scale $\vev{\phi}/\mgut$ and SUSY-breaking masses $\tilde{m}/\mgut$. We define various GUT-scenarios for varying couplings by the assumption of proportionality of fractional variations of the unification parameters.

Assuming that couplings really vary, this is a way of excluding such GUT scenarios by demanding consistency of the implied variations. The assumption of proportionality permits us to project all observations into constraints on a common evolution factor $l(z)$ for each scenario. We show that different GUT scenarios yield different time evolutions of $l(z)$ assuming that certain claimed measurements of varying constants are correct. We confirm that ``simple'' models which have only one fundamental parameter varying ($\alphagut$ or $\vev{\phi}/\mgut$) result in inconsistent variations. However, combined variations of these two parameters, as described in scenarios 5 and 6, lead to results more consistent with the possible quintessence-induced time variations of fundamental couplings. For instance, some models of crossover quintessence and models of growing neutrinos naturally yield ratios of evolution factors which are of the same size as those derived from measurements of varying couplings.

Still, we have not found a scenario with a monotonic time evolution $l(z)$ that makes the two main signals or hints of variation (\cite{Murphy03.2}, \cite{Reinhold06}) and BBN mutually consistent. A monotonic evolution requires either to discount one of the ``signals'' by substantially increasing its uncertainty, or to alter our assumptions by including additional time variation of some Yukawa couplings.\\

In a last step we demonstrate how a clear observation of time variation of fundamental couplings would not only strongly disfavor a constant dark energy, but also put important constraints on the time evolution of a dynamical dark energy or quintessence. We have shown that for a given unified scenario, the bounds on the time variation of various couplings in different cosmological epochs can strongly restrict the possible time evolution of the cosmon field and put very strong bounds on late-time dynamics of the cosmon field. Hence, once at least one irrefutable observation of some coupling variation at some redshift becomes available, our method provides a powerful tool for testing extensions of the Standard Model or, vice versa, allows us to control consistency of claimed variations under the demand of unification.

We have demonstrated this by an analysis that implicitly assumes a nonzero variation, considering both general features and specific quintessence models. However, we are aware that the actual values for the evolution factors $l(z)$ from this analysis may be premature, since the observational situation is unclear and on moving grounds. For example, taking the recent reanalysis of the variation of the proton to electron mass ratio $\mu$ in Ref.~\cite{King08} instead of the results in Ref.~\cite{Reinhold06} used in this thesis, would strongly influence the values of the evolution factors. We have demonstrated this in a somewhat different way by investigating the change in the evolution factors if some claimed observations of varying couplings are omitted.

\newpage
\section*{Outlook}
Progress in the field of BBN requires both observational and theoretical improvements. Both statistical and systematic errors in abundance measurements could be improved, for example observations to better determine the nature of systems where \hefo\ is measured \cite{Steigman05}, or stellar modeling to test possible solutions of the \lise\ problem. On the theoretical side the relation between quark masses and nuclear physics remains unclear beyond the level of the two-nucleon system: the largest uncertainty in our BBN bounds arises from the poorly known dependence of the binding energies on the fundamental couplings. BBN is already the most powerful probe of fundamental ``constants'' in the early Universe, and precision bounds may well be obtained, given continued efforts in observation and theory, to rule out or confirm the presence of a cosmological variation.

Our investigation has further shown how the variations of different couplings in the Standard Model may be compared. If the observational situation becomes clearer and at least one nonzero time variation is established, such methods may be used for new tests of the idea of grand unification and, even further, models of quintessence. The presented method could then easily be applied to constrain both theories beyond the Standard Model like grand unification and theories beyond the concordance model of cosmology like quintessence. Hence, it can establish relations between two a priori distinct theories, namely high-energy physics and general relativity, a capability which is so far dominated by theories with radically new concepts of physics like string theory.

\vspace{3cm}

%%%%%%%%%%%%%%%%%%%%%%%%%%%%%%%%%%%%%%%%%%%%%%%%%%%%%%%%%%%%%%%%%%%%%%%%%%%%%%%

\section*{Acknowledgements}
\addcontentsline{toc}{chapter}{Acknowledgements}
First of all I would like to thank Christof Wetterich, the supervisor of this thesis, for giving me the chance to work on the highly interesting questions of varying constants, cosmology and quintessence. His offer allowed me to stay at the beautiful Institute for Theoretical Physics in Heidelberg, a place which I will always remember. 

Furthermore, I would like to thank Thomas Dent for almost three years of deep and successful cooperation. His valuable hints and explanations opened the door to new insights and approaches which I would not have found that easily without him. 

The members of the Institute and especially the ``Dachzimmer'' deserve gratitude for the wonderful atmosphere. Thanks to Georg Robbers and Thomas Dent for proofreading this thesis. 

The final thanks go to the \textit{Studienstiftung des Deutschen Volkes}. I am greatly indebted to them for supporting me during my whole studies and PhD, which made important things possible in my life.

%%%%%%%%%%%%%%%%%%%%%%%%%%%%%%%%%%%%%%%%%%%%%%%%%%%%%%%%%%%%%%%%%%%%%
%%%%%%%%%%%%%%%%%%%%%%%%%%   APPENDIX   %%%%%%%%%%%%%%%%%%%%%%%%%%%%%

\appendix

\chapter{Conventions}
\section{Symbols and abbreviations}

\begin{table}[h!]
\begin{tabular}{l|l}
Symbol & Explanation \\
\hline
$\alpha$, $\alpha_{em}$ & fine structure constant\\
$\alphastrong$ & strong coupling constant\\
$\alphagut$  & grand unified coupling constant\\
$B_i$        & binding energy of nucleus $i$\\
$\delta_q$   & light quark mass difference\\
$\eta$       & baryon-to-photon ratio; E{\" o}tv{\" o}s parameter\\
$\gnewton$   & Newton's constant\\
$\lqcd$      & QCD scale\\
$\hat{m}$    & average light quark mass\\
$m_d$        & d-quark mass\\
$m_e$        & electron mass\\
$\mneutron$  & neutron mass\\
$m_{\pi}$    & average pion mass\\
$\mproton$   & proton mass\\
$m_s$        & strange quark mass\\
$m_u$        & u-quark mass\\
$\mnucleon$  & average nucleon mass, $\mnucleon = \frac{1}{2} ( \mneutron + \mproton)$\\
$\mgut$    & GUT scale\\
$\mplanck$ & Planck mass\\
$\mplanckred$ & reduced Planck mass\\
$M_X$        & GUT scale ($\equiv \mgut$) \\
$\mu$       & proton to electron mass ratio, $\mu \define m_p/m_e$\\
$\OmegaBaryon$  & baryon fraction\\
$Q_N$      & neutron proton mass difference, $Q_N \define \mneutron - \mproton$\\
\end{tabular}
\caption{List of symbols}
\end{table}

\begin{table}[h!]
\begin{tabular}{l|l}
Abbreviation & Explanation \\
\hline
BAO    & baryon acoustic oscillations\\
BBN    & \BigBangNucleosynthesis \\
EFT    & effective field theory\\
CMB    & cosmic microwave background\\
DOF    & degree of freedom\\
MSSM   & minimal supersymmetric standard model\\
QCD    & quantum chromodynamics\\
QED    & quantum electrodynamics\\
SBBN   & standard big bang nucleosynthesis\\
SM     & Standard Model (of particle physics)\\
SN     & supernovae\\
v.e.v. & vacuum expectation value\\
WEP    & weak equivalence principle\\
\end{tabular}
\caption{List of abbreviations}
\end{table}

%%%%%%%%%%%%%%%%%%%%%%%%%%%%%%%%%%%%%%%%%%%%%%%%%%%%%%%%%%%%%%%%%%%%%%%%%%%%%%%%%
%%%%%%%%%%%%%%%%%%%%%     LISTS OF TABLES AND FIGURES    %%%%%%%%%%%%%%%%%%%%%%%%
%%%%%%%%%%%%%%%%%%%%%%%%%%%%%%%%%%%%%%%%%%%%%%%%%%%%%%%%%%%%%%%%%%%%%%%%%%%%%%%%%

\listoftables
\addcontentsline{toc}{chapter}{List of tables}

\listoffigures
\addcontentsline{toc}{chapter}{List of figures}

%%%%%%%%%%%%%%%%%%%%%%%%%%%%%%%%%%%%%%%%%%%%%%%%%%%%%%%%%%%%%%%%%%%%%%%%%%%%%%%%%
%%%%%%%%%%%%%%%%%%%%%%%%%%      BIBLIOGRAPHY      %%%%%%%%%%%%%%%%%%%%%%%%%%%%%%%
%%%%%%%%%%%%%%%%%%%%%%%%%%%%%%%%%%%%%%%%%%%%%%%%%%%%%%%%%%%%%%%%%%%%%%%%%%%%%%%%%

\newpage
\addcontentsline{toc}{chapter}{Bibliography}

\nocite*    %outputs all references in bibliography
\bibliography{thesisStern}{}
\bibliographystyle{steffenStyle}

\end{document}